\documentclass[12pt,draftclsnofoot,peerreview,onecolumn,notitlepage]{IEEEtran}
\IEEEoverridecommandlockouts

\usepackage{cite}
\usepackage{ieeefig}
\usepackage{amsfonts}
\usepackage{epsfig,color}
\usepackage{graphicx}
\usepackage[caption=false,font=footnotesize]{subfig}
\usepackage{stfloats}
\usepackage{amsmath}
\usepackage{paralist}
\usepackage{color}

\newtheorem{theorem}{\bf Theorem}
\newtheorem{lemma}{\bf Lemma}
\newtheorem{corollary}{\bf Corollary}

\newtheorem{remark}{\bf Remark}
\newtheorem{definition}{\bf Definition}

\newcommand{\achsch}{\mathcal{I}\mathcal{A}\rm{-Scheme}}

\begin{document}
%
\title{\LARGE{The Degrees of Freedom Region and Interference Alignment for the MIMO Interference Channel with Delayed CSI} }

\author{Chinmay S.~Vaze %
        and~Mahesh K.~Varanasi
\thanks{This work was supported in part by NSF Grant CCF-0728955. The authors
are with the Department of Electrical, Computer, and Energy
Engineering, University of Colorado, Boulder, CO 80309-0425 USA
(e-mail: {Chinmay.Vaze, varanasi}@colorado.edu). } }

%



\maketitle

\begin{abstract}
The degrees of freedom (DoF) region of the $2$-user multiple-antenna or MIMO (multiple-input, multiple-output) interference channel (IC) is studied under fast fading and the assumption of {\em delayed} channel state information (CSI) wherein all terminals know all (or certain) channel matrices perfectly, but with a delay, and each receiver in addition knows its own incoming channels instantaneously. The general MIMO IC is considered with an arbitrary number of antennas at each of the four terminals. Dividing it into several classes depending on the relation between the numbers of antennas at the four terminals, the fundamental DoF regions are characterized under the delayed CSI assumption for {\em all} possible values of number of antennas at the four terminals. In particular, an outer bound on the DoF region of the general MIMO IC is derived. This bound is then shown to be tight for all MIMO ICs by developing interference alignment based achievability schemes for each class. A comparison of these DoF regions under the delayed CSI assumption is made with those of the idealistic `perfect CSI' assumption where perfect and instantaneous CSI is available at all terminals on the one hand and with the DoF regions of the conservative `no CSI' assumption on the other, where CSI is available at the receivers but not at all at the transmitters.
\end{abstract}



\newpage

\section{Introduction}

\IEEEPARstart{T}{he} interference channel (IC) is an archetype of a network for multiple unicast communication. This channel consists of two transmit-receive pairs that wish to communicate over a common noisy medium. Despite over three decades of research the capacity region of the IC remains unknown in general. Much progress has been made towards this goal however, in the form of capacity characterizations of special classes of the interference channel \cite{Carleial75,Benzel,Han_Kobayashi,Sato,Gamal_Costa,Annapureddy_Veeravalli,Motahari_Khandani,Shang2009} and capacity approximations for the scalar (single-antenna) Gaussian IC \cite{Etkin_Tse_Wang} and the vector or MIMO Gaussian IC \cite{Telatar-Tse-IC,Constant-Gap-Karmakar-MV-2011}, where the capacity region is characterized to within a constant gap that is independent of signal-to-noise ratio (SNR) and the channel parameters. These works assume perfect channel state information (CSI) at all terminals.

For the MIMO Gaussian IC, under the same perfect CSI assumption, the degrees of freedom (DoF) region -- which denotes the rate of growth with respect to $\log \rm{SNR}$ of the capacity region at high SNR -- was found in \cite{Jafar-Maralle, Chiachi-Jafar}. As shown there, the DoF region can be achieved via a combination of transmit/receive zero-forcing beamforming with time sharing, wherein the transmitters exploit the channel null spaces to beamform transmitted signals so as to cause minimal interference at the unintended receiver. However, such schemes can be sensitive to imperfections in CSI at the transmitters (CSIT). Indeed, it has been recently proved that in the extreme case of having no CSI whatsoever at the transmitters, the DoF region of the channel collapses to the extent that just receive zero-forcing beamforming (and time sharing) are DoF-region optimal \cite{Chiachi2, Zhu_Guo_noCSIT_DoF_2010, Vaze_Dof_final}.

The impact of lack of CSIT on the achievable rate emphasizes the need for transmitters to obtain CSI from the receivers. The receivers can learn their current incoming channel state via pilot transmissions with a high degree of accuracy\footnote{Throughout this paper, it is assumed that receivers know their incoming channel state perfectly and instantaneously.}, which in practice, can be broadcast back to the other terminals via a feedback link. As a result, CSI available at different terminals of the network may be neither perfect, due to the limited capacity of the feedback link, nor instantaneous, due to the delays involved in channel estimation and feedback. The first problem of having imperfect CSI has been relatively well investigated and it is now known in the variety of contexts that if the quality of CSI available at the terminals can be improved at a sufficiently fast rate with $\log \rm{SNR}$, then the perfect CSI DoF can be achieved even with imperfect CSIT (cf. \cite{Jindal, Vaze_fb_scaling_GBC, Rajesh_Varanasi_IA_2009}, and references therein). However, fundamental approaches to the study of the second problem of having just delayed CSI, a setting that is highly relevant in mobile environments with short channel coherence times, has only recently begun \cite{maddah_ali_tse_delayed_CSIT,Jafar_Shamai_retrospective_IA,Vaze-Varanasi-delay-MIMOBC}.

In particular, Maddah-Ali and Tse in \cite{maddah_ali_tse_delayed_CSIT} prove in the context of the MISO broadcast channel (BC) (ie., a BC with multiple-antenna transmitter and single antenna receivers) the surprising and interesting result that the presence of delayed CSI at the terminals can lead to a dramatic increase in the achievable rate relative to having no CSI to the extent that even the DoF of the channel increase. Moreover, this is shown to be possible even if the channel coefficients vary independently across each time slot so that the delayed CSI available to the terminals is actually outdated (or `stale'). As a case in point, in the MISO BC with $2$ transmit antennas, $2$ single-antenna users, and i.i.d. fast fading, the achievable sum-DoF improve from $1$ with no CSI to $\frac{4}{3}$ with delayed CSI. The achievable scheme in \cite{maddah_ali_tse_delayed_CSIT} is based on the idea that the interference encountered at a prior time instant by one of the users, which is a linear combination (LC) of the data symbols intended for some other user, can be perfectly evaluated by the transmitter at the current time instant using delayed CSI, and hence, can be subsequently transmitted to provide an opportunity to the interfered user to learn that interference while simultaneously providing a new LC to the user where these symbols are desired. Furthermore, the achievability scheme based on this idea has been shown there to be sum-DoF optimal for the class of MISO BCs in which there are at least as many transmit antennas as the number of users. This work was recently extended by the authors of this paper to the case of the MIMO BC (with multiple and arbitrary number of antennas as each terminal) in \cite{Vaze-Varanasi-delay-MIMOBC} where an outer-bound to the DoF region is obtained for the general $K$-user case, and was proved, through an interference alignment based achievability scheme, to be tight in the $2$-user case.

Meanwhile, Maleki et al in \cite{Jafar_Shamai_retrospective_IA} show how the so-called blind interference alignment scheme for the $(3,1,4,2)$ MIMO IC (i.e., with 3 and 1 antennas at transmitters one and two, and 4 and 2 antennas at the receivers one and two, respectively) obtained earlier for the staggered block fading model in \cite{Jafar_correlations} can be adapted to the same MIMO IC but under the delayed CSI model to achieve the $(1,1.5)$ DoF tuple, a point that lies outside the DoF region with no CSI \cite{Zhu_Guo_noCSIT_DoF_2010}. However, with no outer bound for that example, the question of whether this example can be improved is left open and so is the much more general question of characterizing the DoF region of the $(M_1,M_2,N_1,N_2)$ MIMO IC (with $M_i$ antennas at transmitter $i$ and $N_i$ antennas at receiver $i$ for $i\in\{1,2\}$). Two more schemes were also proposed in \cite{Jafar_Shamai_retrospective_IA} under the delayed CSI model for the $2$-user scalar X channel (with a sum DoF of 8/7) and the $3$-user scalar IC (with a sum DoF of 9/8) but the question of DoF-optimality in these cases also remains open as is the characterization of the DoF regions.

In this paper, we study the problem of characterizing the DoF region with delayed CSI of the general $(M_1,M_2,N_1,N_2)$ MIMO IC . In particular, we first obtain an outer-bound for the general case. Toward this end, Fano's inequality is used to upper-bound the rate achievable for each user by the mutual information between the signal received by that user and its desired message, and then a technique is developed to bound the DoF of a certain key weighted sum of these two mutual information terms to finally obtain an upper-bound on that weighted sum of the DoF achievable for the two users (see Sections \ref{sec: proof of thm: outer-bound DoF 2-user IC d-CSI} and \ref{sec: proof of thm: outer-bound DoF 2-user IC d-CSI L4}). Note that apart from the $2$-user MIMO IC considered here, the delayed-CSI DoF region is known only for the MISO BC \cite{maddah_ali_tse_delayed_CSIT} and its generalization to MIMO BC in \cite{Vaze-Varanasi-delay-MIMOBC}, which prompts us to contrast the techniques developed in this work to that in \cite{maddah_ali_tse_delayed_CSIT,Vaze-Varanasi-delay-MIMOBC}. In order to derive their outer bound, the authors in \cite{maddah_ali_tse_delayed_CSIT} first outer bound the capacity region by transforming the original BC through genie-aided side information into a physically-degraded BC with a transmitter that knows the past channel states and past channel outputs and then compute the DoF region of the resultant physically-degraded BC by using the result of \cite{Gamal_fb_capacity_degraded_BC} that the capacity of the physically-degraded BC with feedback is the same as that without feedback. However, the reliance on the result of \cite{Gamal_fb_capacity_degraded_BC} makes the technique in \cite{maddah_ali_tse_delayed_CSIT} not applicable to other channels with delayed CSI, such as the IC. In contrast, the general idea developed here can be used for dealing with a variety of channels with delayed CSI. For instance, using our technique, it is possible to provide an alternate proof for the outer-bound of \cite{maddah_ali_tse_delayed_CSIT} and \cite{Vaze-Varanasi-delay-MIMOBC} for the MISO and MIMO BCs without invoking the result of \cite{Gamal_fb_capacity_degraded_BC}. Moreover, the same technique could also be employed with the MIMO cognitive IC under the delayed CSI assumption. The cognitive IC is an interference channel in which one or more terminals can be simultaneously cognitive (cf. \cite{Devroye, Chiachi-Jafar, Vaze_Dof_Cognitive_IC_ISIT} and references therein).

Next, the general $(M_1,M_2,N_1,N_2)$ MIMO IC is divided into several classes of MIMO ICs depending on the relative numbers of antennas at the four terminals. Interference alignment based communication schemes specifically tailored for each class (when necessary) are given and shown to have DoF regions that coincide with the outer bound for all classes. Hence, the fundamental DOF region is obtained for the general $(M_1,M_2,N_1,N_2)$  MIMO IC without any restriction on the numbers of antennas. The MIMO IC, unlike the MISO or MIMO BC, consists of distributed, or non-cooperating, transmitters. Hence, neither transmitter can compute the past received signals (excluding noise terms) of the two receivers even with delayed CSI, unlike the case with the transmitter in a BC. However, since the transmit signal of any given transmitter is intended for just its paired receiver, each transmitter can compute the interference caused by its signal to the unintended receiver at previous time instants. Thus, interference alignment can be achieved by requiring the transmitter(s) to transmit the interference they created in the past, which provides additional information to the intended receiver about its desired message without causing any extra interference at their unpaired receiver. Interestingly, there is even a class of MIMO ICs -- in which the no-CSI DoF region is strictly smaller than the perfect-CSI DoF region -- for which the DoF region with delayed CSI is seen to coincide with the entire perfect-CSI DoF region.

Incidentally, as a result of this work, it can now be asserted that the $(1,1.5)$ DoF pair achievable using the retrospective interference alignment scheme of \cite{Jafar_Shamai_retrospective_IA} over the $(3,1,4,2)$ MIMO IC lies on the boundary of the DoF region (see Lemma \ref{lem: Case B.I d-CSI 2user IC} in Section \ref{sec: proof of thm: MIMO_2-user IC_d-CSI_inner-bound DoF_Case B.I} which derives the DoF region for a class of MIMO ICs that contains the $(3,1,4,2)$ MIMO IC).

A complete comparative characterization of the DoF regions of the MIMO IC under the delayed CSI assumption is given with the DoF regions under the idealized assumption of perfect CSI \cite{Jafar-Maralle, Chiachi-Jafar} on the one hand and with the DoF regions under the conservative assumption of no CSI \cite{Chiachi2, Zhu_Guo_noCSIT_DoF_2010, Vaze_Dof_final} on the other, thereby revealing a rich classification of MIMO ICs according to whether the no CSI DoF region is strictly contained by (or is equal to) the delayed CSI DoF region which in turn is strictly contained by (or is equal to) the perfect CSI DoF region.

The rest of the paper is organized as follows. In Section \ref{sec:model}, the MIMO IC model is defined. Section \ref{sec: main results dCSI 2user IC} contains the main results, namely, the outer and inner bounds on the DoF region with delayed CSI. Sections \ref{sec: proof of thm: outer-bound DoF 2-user IC d-CSI}-\ref{sec: proof of thm: MIMO_2-user IC_d-CSI_inner-bound DoF_Case B.III} provide the proofs of the main results. Section \ref{sec:conclusions} concludes the paper.

\section{The Channel Model}
\label{sec:model}
In this section, we describe the $(M_1,M_1,N_1,N_2)$ MIMO IC with fading under the assumption of delayed CSI. The two transmitters of the MIMO IC are denoted as T1 and T2, with transmitter $i$ having $M_i$ antennas and their corresponding receivers are denoted as R1 and R2, with receiver $i$ equipped with $N_i$ antennas. A given transmitter has a message only for its respective/paired receiver. However, its signal is received at the unintended receiver as interference. The input-output relationship at the $t^{th}$ channel use is given at R1 and R2, respectively, as
\begin{eqnarray}
 Y_1(t) = H_{11}(t) X_1(t) + H_{12}(t) X_2(t) + W_1(t), \\
Y_2(t) = H_{21}(t) X_1(t) + H_{22}(t) X_2(t) + W_2(t),
\end{eqnarray}
where $X_i(t)$ is the signals transmitted by the $i^{th}$ transmitter Ti; $H_{ij}(t) \in \mathbb{C}^{N_i \times M_j}$ denotes the channel matrix between $i^{th}$ receiver and $j^{th}$ transmitter; and $W_i(t) \sim \mathcal{C}\mathcal{N}(0,I_{N_i})$, for $i = 1,~ 2$, is the additive noise at receiver $i$. The power is constrained to be $P$, i.e., $\mathbb{E} ||X_i(t)||^2 \leq P$ $\forall$ $i,t$.

For simplicity, we assume the case of Rayleigh fading, i.e., all the entries of all channel matrices $\{H_{ij}(t)\}_{i,j}$ are independent, identically distributed (i.i.d.) zero-mean, unit variance complex normal (denoted $\mathcal{C}\mathcal{N}(0,1)$) random variables. Further, the channel and noise realizations are taken to be i.i.d. across time and they are independent of each other.

It is assumed that every receiver knows the channel matrices corresponding to itself perfectly and instantaneously, while all other terminals know them perfectly with a delay of one time unit. In particular, for each $i \in \{1,2\}$, the channel matrices $H_{i1}(t)$ and $H_{i2}(t)$ are known to the $i^{th}$ receiver at time $t$, whereas all other terminals (i.e., T1, T2, and the $j^{th}$ receiver with $j \not= i$) know these matrices at time $t+1$. We refer to this assumption about CSI as the ``delayed CSI" assumption.

The case where all terminals have perfect and instantaneous knowledge of all channel matrices is referred to as the ``perfect CSI" assumption. If the $i^{th}$ receiver knows the channel matrices $H_{i1}(t)$ and $H_{i2}(t)$ perfectly and instantaneously and if there is no additional CSI at any of the terminals, then we refer to the corresponding case as the ``no CSI" assumption. It must be noted that in all the three cases, all terminals are always assumed to know the distribution of the channel matrices.

Let $\mathcal{M}_1$ and $\mathcal{M}_2$ be two independent messages to be sent by T1 and T2, respectively, over a block length of $n$, where the message $\mathcal{M}_i$ is intended for the $i^{th}$ receiver. It is assumed that $\mathcal{M}_i$ is distributed uniformly over a set of cardinality $2^{nR_i(P)}$, when there is a power constraint of $P$ at the transmitters. If $\overline{H}(n) = \big\{ H_{11}(t), H_{12}(t), H_{21}(t), H_{22}(t) \big\}_{t=1}^n$ with $\overline{H}(0) = \phi$ and $\overline{Y}_i(n) = \{Y_i(t)\}_{i=1}^n$, then a coding scheme for blocklength $n$ consists of two encoding functions $f_i^{(n)} = \{f_{i,t}^{(n)}\}_{t=1}^n$, $i = 1,2$, such that
\[
X_i(t) = f_{i,t}^{(n)} \Big( \mathcal{M}_i, \overline{H}(t-1) \Big) ~  \forall t \in \{1,2,\cdots,n\}
\]
and two decoding functions such that
\[
\hat{\mathcal{M}}_i = g_i^{(n)} \Big( \overline{Y}_i(n), \overline{H}(n-1), H_{i1}(n), H_{i2}(n) \Big)  \mbox{ where } i \in \{1,2\}.
\]
A rate tuple $\big( R_1(P),R_2(P) \big)$ is said to be achievable if there exists a sequence of coding schemes such that probability of $\mathcal{M}_1 \not= \hat{\mathcal{M}}_1$ or $\mathcal{M}_2 \not= \hat{\mathcal{M}}_2$ tends to zero as $n \to \infty$. The capacity region $\mathcal{C}(P)$ is defined to be the set of all achievable rate tuples $\big( R_1(P),R_2(P) \big)$ when the power constraint at the transmitters is $P$. We then define the DoF region as
\[
\mathbf{D}^{\rm{d-CSI}} = \left\{ (d_1,d_2) \left| d_i \geq 0 \mbox{ and } \exists ~ \big( R_1(P),R_2(P) \big) \in \mathcal{C}(P) \mbox{ s.t. } d_i = \lim_{P \to \infty} \frac{R_i(P)}{\log_2 P} ~ i \in \{1,2\} \right. \right\}.
\]
The DoF regions corresponding to the cases of perfect and no CSI, denoted respectively as $\mathbf{D}^{\rm{p-CSI}}$ and $\mathbf{D}^{\rm{no-CSI}}$, can be defined in analogous fashion with the only difference being in the definitions of the encoding and decoding functions. In particular, when there is perfect CSI, we have
\[
X_i(t) = f_{i,t}^{(n)} \Big( \mathcal{M}_i, \overline{H}(t) \Big) \; \mbox{ and } \; \hat{\mathcal{M}}_i = g_i^{(n)} \big( \overline{Y}_i(n), \overline{H}(n) \Big) \, .
\]
When there is no CSI, we have
\[
X_i(t) = f_{i,t}^{(n)} \Big( \mathcal{M}_i \Big) \mbox{ and } \hat{\mathcal{M}}_i = g_i^{(n)} \Big( \overline{Y}_i(n), \{H_{i1}(t), H_{i2}(t)\}_{t=1}^n \Big).
\]
Clearly, $\mathbf{D}^{\rm{no-CSI}} \subseteq \mathbf{D}^{\rm{d-CSI}} \subseteq \mathbf{D}^{\rm{p-CSI}}$.

\section{The DoF Region of the IC with Delayed CSI} \label{sec: main results dCSI 2user IC}
In this section the main results regarding the DoF region of the MIMO IC are stated for which we need the following two definitions.

\begin{definition}
For a given $i \in \{1,2\}$, Condition $i$ is said to hold, whenever the inequality
\[
M_i > N_1 + N_2 - M_j > N_i > N_j > M_j > N_j \frac{N_j - M_j}{N_i - M_j}
\]
is true for $j \in \{1,2\}$ with $j \not= i$.
\end{definition}
Clearly, the two conditions are symmetric counterparts of each other (i.e., one can be obtained from the other by switching user indices $1$ and $2$). Moreover, the two conditions can not be true simultaneously, and Condition $i$ can not hold if $N_j \geq N_i$.

\begin{definition}
The region $\mathbf{D}_{\rm{outer}}^{\rm{d-CSI}}$ is defined as follows:
\begin{eqnarray*}
\hspace{1cm} \mathbf{D}_{\rm{outer}}^{\rm{d-CSI}} = \Big\{ (d_1,d_1) \Big| ~ L_{o1} \equiv 0 \leq d_1 \leq \min(M_1,N_1) \mbox{ and } L_{o2} \equiv 0 \leq d_2 \leq \min(M_2,N_2) , \Big. \Big. \\
&& {} \hspace{-15cm} L_1 \equiv \frac{d_1}{\min(N_1 + N_2, M_1)} + \frac{d_2}{\min(N_2,M_1)} \leq \frac{\min (N_2,M_1+M_2)}{\min(N_2,M_1)}; \\
&& {} \hspace{-15cm} L_2 \equiv \frac{d_1}{\min(N_1,M_2)} + \frac{d_2}{\min(N_1 + N_2,M_2)} \leq \frac{\min(N_1, M_1+M_2)}{\min(N_1,M_2)};  \\
&& {} \hspace{-15cm} L_3 \equiv  d_1 + d_2 \leq \min \big\{ M_1 + M_2, N_1+N_2, \max(M_1,N_2), \max(M_2,N_1) \big\}; \\
&& {} \hspace{-15cm} \mbox{if Condition } 1 \mbox{ holds, } L_4 \equiv d_1 + d_2 \frac{N_1 + 2 N_2 - M_2}{N_2} \leq N_1 + N_2;  \\
&& {}  \Big. \hspace{-15cm} \mbox{if Condition } 2 \mbox{ holds, } L_5 \equiv d_2 + d_1 \frac{N_2 + 2 N_1 - M_1}{N_1} \leq N_1 + N_2 \Big\}.
\end{eqnarray*}
\end{definition}
In the sequel, the first two ``single-user" bounds on $d_1$ and $d_2$ appearing in the above definition are referred to as $L_{o1}$ and $L_{o2}$, respectively; while the remaining five bounds on the weighted sum of $d_1$ and $d_2$ are referred to respectively as $L_1$, $L_2$, $\cdots$, $L_5$.

The following theorem gives an outer-bound to the DoF region.
\begin{theorem}[Outer-bound] \label{thm: outer-bound DoF 2-user IC d-CSI}
The DoF region of the MIMO IC with delayed CSI is outer-bounded by $\mathbf{D}_{\rm{outer}}^{\rm{d-CSI}}$, i.e.,
\[ \mathbf{D}^{\rm{d-CSI}} \subseteq \mathbf{D}_{\rm{outer}}^{\rm{d-CSI}}.\]
\end{theorem}
\begin{IEEEproof}
If $(d_1,d_2) \in \mathbf{D}^{\rm{d-CSI}}$, then $d_i \leq \min(M_i,N_i)$ for each $i$ because the DoF of the point-to-point MIMO channel with $M$ transmit and $N$ receive antennas are upper-bounded by $\min(M,N)$ \cite{Telatar}. Now, note that if $(d_1,d_2) \in \mathbf{D}^{\rm{p-CSI}} $, then $d_1$ and $d_2$ satisfy bound $L_3$ \cite{Chiachi-Jafar} (see also \cite[Section III-C]{Vaze_Dof_final}), and hence, if $(d_1,d_2) \in \mathbf{D}^{\rm{d-CSI}} \subseteq \mathbf{D}^{\rm{p-CSI}} $, then the bound $L_3$ must hold. It now remains to prove the bounds $L_1$, $L_2$, $L_4$ and $L_5$. To this end, we observe that bounds $L_1$ and $L_2$ are symmetric counterparts of each other, and so are the bounds $L_4$ and $L_5$. Hence, it is sufficient to prove bounds $L_1$ and $L_4$. The proofs of $L_1$ and $L_4$ as being the outer-bounds are provided in Sections \ref{sec: proof of thm: outer-bound DoF 2-user IC d-CSI} and \ref{sec: proof of thm: outer-bound DoF 2-user IC d-CSI L4}, respectively.
\end{IEEEproof}

The following theorem asserts that the above outer-bound is tight for all possible values of $(M_1,M_2,N_1,N_2)$.
\begin{theorem}[The DoF region] \label{thm: MIMO_2-user IC_d-CSI_inner-bound DoF}
The DoF region of the MIMO IC with delayed CSI is equal to $\mathbf{D}_{\rm{outer}}^{\rm{d-CSI}}$. In other words,
\[    \mathbf{D}^{\rm{d-CSI}} = \mathbf{D}_{\rm{outer}}^{\rm{d-CSI}}.\]
\end{theorem}
\begin{IEEEproof}
It has been proved in the previous theorem that $\mathbf{D}^{\rm{d-CSI}} \subseteq \mathbf{D}_{\rm{outer}}^{\rm{d-CSI}}$. It will be shown here that $\mathbf{D}_{\rm{outer}}^{\rm{d-CSI}} \subseteq \mathbf{D}^{\rm{d-CSI}} $, i.e., the region $\mathbf{D}_{\rm{outer}}^{\rm{d-CSI}}$ is achievable. In the remainder of the proof it is assumed, without loss of generality, that $ N_1 \geq N_2 $ (thus Condition $2$ can not hold).

We now divide the analysis into three main cases which are further subdivided to obtain a total of seven cases, which are analyzed individually in their respective sections (stated beside the definition of the case). Note that the assumption of $N_1 \geq N_2$ applies to every case.
\begin{itemize}
\item Case 0: $N_2 \geq M_1$ -- See Section \ref{sec: proof of thm: MIMO_2-user IC_d-CSI_inner-bound DoF_Case A.I}.
\item Case A: $M_1 > N_2$ and $M_2 \geq N_2$:
\begin{itemize}
\item Case A.I: $M_2 \geq N_1$ -- See Section \ref{sec: proof of thm: MIMO_2-user IC_d-CSI_inner-bound DoF_Case A.I}. \footnote{Thus, under Case A.I, the inequality $M_2 \geq N_1$ holds, in addition to the inequality $N_1 \geq N_2$ and the inequalities that are true with Case A.}
\item Case A.II: $M_2 < N_1$ -- See Section \ref{sec: proof of thm: MIMO_2-user IC_d-CSI_inner-bound DoF_Case A.II}.
\end{itemize}
\item Case B: $M_1 > N_2$ and $N_2 > M_2$:
\begin{itemize}
\item Case B.0: $N_1 = N_2$ -- See Section \ref{sec: proof of thm: MIMO_2-user IC_d-CSI_inner-bound DoF_Case B.I}.
\item Case B.I: $N_1 > N_2$ and $N_1 \geq M_1$ -- See Section \ref{sec: proof of thm: MIMO_2-user IC_d-CSI_inner-bound DoF_Case B.I}.
\item Case B.II: $M_1 > N_1 > N_2$ and Condition $1$ does not hold -- See Section \ref{sec: proof of thm: MIMO_2-user IC_d-CSI_inner-bound DoF_Case B.II}.
\item Case B.III: $M_1 > N_1 > N_2$ and Condition $1$ holds -- See Section \ref{sec: proof of thm: MIMO_2-user IC_d-CSI_inner-bound DoF_Case B.III}.
\end{itemize}
\end{itemize}

It turns out that for some values of the 4-tuple ($M_1, M_2, N_1, N_2$), we have $\mathbf{D}^{\rm{no-CSI}} = \mathbf{D}_{\rm{outer}}^{\rm{d-CSI}}$, which implies the achievability of the outer-bound, and hence the theorem. However, if $\mathbf{D}^{\rm{no-CSI}} \not= \mathbf{D}_{\rm{outer}}^{\rm{d-CSI}}$, we need to develop a new achievability scheme, which is done here using the idea of interference alignment, to prove that $\mathbf{D}^{\rm{d-CSI}} = \mathbf{D}_{\rm{outer}}^{\rm{d-CSI}}$.  The reader is referred to Sections \ref{sec: proof of thm: MIMO_2-user IC_d-CSI_inner-bound DoF_Case A.I} - \ref{sec: proof of thm: MIMO_2-user IC_d-CSI_inner-bound DoF_Case B.III} for detailed proofs.
\end{IEEEproof}

From the above two theorems, we obtain two corollaries corresponding to two different sets of assumptions about the availability of CSI. Consider first the case wherein it is assumed that both the receivers have perfect and instantaneous knowledge of all channel matrices while the transmitters (as before) know the channel matrices perfectly, but with a delay of one time unit. We refer to this case as ``delayed CSI at the transmitters" and denote the corresponding DoF region by $\mathbf{D}^{\rm{d-CSI-T}}$. The following corollary characterizes this DoF region.
\begin{corollary}
The DoF region of the MIMO IC with delayed CSI at the transmitters is equal to $\mathbf{D}_{\rm{outer}}^{\rm{d-CSI}}$, i.e.,
\[
\mathbf{D}^{\rm{d-CSI-T}} = \mathbf{D}^{\rm{d-CSI}}.
\]
\end{corollary}
\begin{IEEEproof}
Since $\mathbf{D}_{\rm{outer}}^{\rm{d-CSI}} = \mathbf{D}^{\rm{d-CSI}} \subseteq \mathbf{D}^{\rm{d-CSI-T}}$, the region $\mathbf{D}_{\rm{outer}}^{\rm{d-CSI}}$ is an inner-bound to $\mathbf{D}^{\rm{d-CSI-T}}$. Therefore, it is sufficient to prove that $\mathbf{D}_{\rm{outer}}^{\rm{d-CSI}}$ is also an outer-bound. To this end, note that the bounds $L_{o1}$, $L_{o2}$, $L_3$ are outer-bounds, even in the present case. Moreover, from Sections \ref{sec: proof of thm: outer-bound DoF 2-user IC d-CSI} and \ref{sec: proof of thm: outer-bound DoF 2-user IC d-CSI L4}, we observe that the bounds $L_1$ and $L_4$ have been derived by assuming that both the receivers have perfect and instantaneous CSI of all channel matrices, and hence, they are applicable to the IC with delayed CSI, as well.
\end{IEEEproof}

Consider next a different assumption about CSI knowledge. The transmitters, instead of having delayed CSI of all channel matrices, have delayed CSI of only the cross channel matrices. That is, at time $t$, the $i^{th}$ transmitter knows perfectly the channel matrix $H_{ji}(t-1)$ with $j \not= i$, but does not have any knowledge of the realizations of the other three channel matrices. The receivers, on the other hand, have delayed CSI (as in the last section), in addition to the instantaneous CSI of their incoming channel matrices. We refer to this case as ``delayed CSI of cross channel matrices" and denote the corresponding DoF region by $\mathbf{D}^{\rm{d-CSI-c}}$, which is characterized by the following corollary.
\begin{corollary}
The DoF region of the MIMO IC with delayed CSI of cross channel matrices is equal to $\mathbf{D}_{\rm{outer}}^{\rm{d-CSI}}$, i.e.,
\[
\mathbf{D}^{\rm{d-CSI-c}} = \mathbf{D}^{\rm{d-CSI}}.
\]
\end{corollary}
\begin{IEEEproof}
Since $\mathbf{D}_{\rm{outer}}^{\rm{d-CSI}} = \mathbf{D}^{\rm{d-CSI}} \supseteq \mathbf{D}^{\rm{d-CSI-c}}$, the region $\mathbf{D}_{\rm{outer}}^{\rm{d-CSI}}$ is an outer-bound to $\mathbf{D}^{\rm{d-CSI-c}}$. It is thus sufficient to show that it is achievable even if the transmitters have just delayed CSI of cross channel matrices. To this end, note that all achievability schemes developed in Sections \ref{sec: proof of thm: MIMO_2-user IC_d-CSI_inner-bound DoF_Case A.I}-\ref{sec: proof of thm: MIMO_2-user IC_d-CSI_inner-bound DoF_Case B.III} (which prove the achievability of $\mathbf{D}_{\rm{outer}}^{\rm{d-CSI}}$ with delayed CSI) make use of only  delayed CSI of cross channel matrices at the transmitters. Hence, the schemes developed in these sections are directly applicable to the case of delayed CSI of cross channel matrices, which proves the achievability part.
\end{IEEEproof}

\begin{table}[t] \centering
\begin{tabular}{|c|c|c|c|c|} \hline
Case & Definition of the Case                      & Bounds & Reference & Remarks \\
&  $(N_1 \geq N_2)$ & Active &  & \\ \hline \hline
0 & $N_2 \geq M_1$ & $L_3$ & Lemma \ref{lem: Case 0 d-CSI 2user IC} &  $\mathbf{D}^{\rm{no-CSI}} = \mathbf{D}^{\rm{d-CSI}} = \mathbf{D}^{\rm{p-CSI}}$ \\ \hline \hline
\multicolumn{5}{|c|}{Case A: $M_1 > N_2$ and $M_2 \geq N_2$}\\ \hline
A.I.  & $M_2 \geq N_1$  & $L_{\{1,2\}}$ & Lemma \ref{lem: Case A.I d-CSI 2user IC} & \\ \cline{1-3} \cline{5-5}
1 & $M_1 \leq N_1$ & $L_1$ & See & $\mathbf{D}^{\rm{no-CSI}} = \mathbf{D}^{\rm{d-CSI}} \subset \mathbf{D}^{\rm{p-CSI}}$ \\
2 & $M_1>N_1$ and $N_2 = M_2$ & $L_2$ or $L_3$ & Section & $\mathbf{D}^{\rm{no-CSI}} = \mathbf{D}^{\rm{d-CSI}} = \mathbf{D}^{\rm{p-CSI}}$ \\
3 & $M_1>N_1$ and $N_2 < M_2$ & $L_{\{1,2\}}$ & \ref{sec: proof of thm: MIMO_2-user IC_d-CSI_inner-bound DoF_Case A.I}. & $\mathbf{D}^{\rm{no-CSI}} \subset \mathbf{D}^{\rm{d-CSI}} \subset \mathbf{D}^{\rm{p-CSI}}$ \\ \hline
A.II. & $M_2 < N_1$ & $L_{\{1,3\}}$ & Lemma \ref{lem: Case A.II d-CSI 2user IC} & \\ \cline{1-3} \cline{5-5}
1 & $N_1 \geq M_1$ & $L_1$ & See Section &  $\mathbf{D}^{\rm{no-CSI}} = \mathbf{D}^{\rm{d-CSI}} \subset \mathbf{D}^{\rm{p-CSI}}$ \\
2 & $N_1 < M_1$ & $L_{\{1,3\}}$ & \ref{sec: proof of thm: MIMO_2-user IC_d-CSI_inner-bound DoF_Case A.II}. & $\mathbf{D}^{\rm{no-CSI}} \subset \mathbf{D}^{\rm{d-CSI}} \subset \mathbf{D}^{\rm{p-CSI}}$ \\ \hline \hline
\multicolumn{5}{|c|}{Case B: $M_1 > N_2$ and $N_2 > M_2$}\\ \hline
B.0 & $N_1 = N_2$ & $L_3$ & Lemma \ref{lem: Case B.0 d-CSI 2user IC} & $\mathbf{D}^{\rm{no-CSI}} = \mathbf{D}^{\rm{d-CSI}} = \mathbf{D}^{\rm{p-CSI}}$ \\ \hline
B.I. & $N_1 > N_2$ and $N_1 \geq M_1$ & $L_1$ & Lemma \ref{lem: Case B.I d-CSI 2user IC}  &  $\mathbf{D}^{\rm{no-CSI}} \subset \mathbf{D}^{\rm{d-CSI}} \subset \mathbf{D}^{\rm{p-CSI}}$ \\ \hline
B.II. & $M_1 > N_1> N_2$ and Condition $1$ does not hold & $L_{\{1,3\}}$ & Lemma \ref{lem: Case B.II d-CSI 2user IC} & \\ \cline{1-3} \cline{5-5}
1 & $M_2 \leq m$ & $L_3$ & See Section & $\mathbf{D}^{\rm{no-CSI}} \subset \mathbf{D}^{\rm{d-CSI}} = \mathbf{D}^{\rm{p-CSI}}$ \\
2 & $M_1 = M_1'$ and $M_2 > m$ & $L_{\{1,3\}}$ & \ref{sec: proof of thm: MIMO_2-user IC_d-CSI_inner-bound DoF_Case B.II}. & $\mathbf{D}^{\rm{no-CSI}} \subset \mathbf{D}^{\rm{d-CSI}} \subset \mathbf{D}^{\rm{p-CSI}}$ \\ \hline
B.III. & $M_1 > N_1> N_2$ and Condition $1$ holds & $L_{\{1,3,4\}}$ & Lemma \ref{lem: Case B.III d-CSI 2user IC} & \\  \cline{1-3} \cline{5-5}
1 & $M_1 \geq N_1 + N_2 - m $ & $L_{\{3,4\}}$ & See Section & $\mathbf{D}^{\rm{no-CSI}} \subset \mathbf{D}^{\rm{d-CSI}} \subset \mathbf{D}^{\rm{p-CSI}}$ \\
2 & $M_1< N_1 + N_2 - m$ & $L_{\{1,3,4\}}$ & \ref{sec: proof of thm: MIMO_2-user IC_d-CSI_inner-bound DoF_Case B.III}. & $\mathbf{D}^{\rm{no-CSI}} \subset \mathbf{D}^{\rm{d-CSI}} \subset \mathbf{D}^{\rm{p-CSI}}$ \\ \hline
\end{tabular}
\newline \newline $ M_1' = \min(M_1,N_1+N_2-M_2)$ $~ ~ ~ m = N_2 \frac{M_1'-N_1}{M_1'-N_2}$ \newline
Condition $1$ can equivalently be stated as $M_1 > M_1' > N_1 > N_2 > M_2 > m$; See Lemma \ref{lem: eqvt form Condition 1 d-CSI 2user IC} in Appendix \ref{app: lemmas to determine the shape d-CSI 2user IC}.
\caption{Summary of Results on the DoF region of the IC with Delayed CSI: $N_1 \geq N_2$.}
\label{table: summary of results d-CSI 2user IC}
\end{table}

\begin{table}[h] \centering
\begin{tabular}{|c|c|} \hline
Situation & Cases \\ \hline
$\mathbf{D}^{\rm{no-CSI}} = \mathbf{D}^{\rm{d-CSI}} = \mathbf{D}^{\rm{p-CSI}}$ & 0, A.I.2, B.0 \\ \hline
$\mathbf{D}^{\rm{no-CSI}} = \mathbf{D}^{\rm{d-CSI}} \not= \mathbf{D}^{\rm{p-CSI}}$ & A.I.1, A.II.1 \\ \hline
$\mathbf{D}^{\rm{no-CSI}} \not= \mathbf{D}^{\rm{d-CSI}} = \mathbf{D}^{\rm{p-CSI}}$ & B.II.1 \\ \hline
$\mathbf{D}^{\rm{no-CSI}} \not= \mathbf{D}^{\rm{d-CSI}} \not= \mathbf{D}^{\rm{p-CSI}}$ & A.I.3, A.II.2, B.I, B.II.2, B.III.1,B.III.2 \\ \hline
\end{tabular}
\caption{Comparison of DoF Regions of the IC with No, Delayed, and Perfect CSI: $N_1 \geq N_2$.} \label{table: comparison of DoF regions d-CSI IC}
\end{table}

\begin{figure}[h] \centering
\includegraphics[bb=0bp 350bp 540bp 680bp,clip, height=2in, width=3.5in]{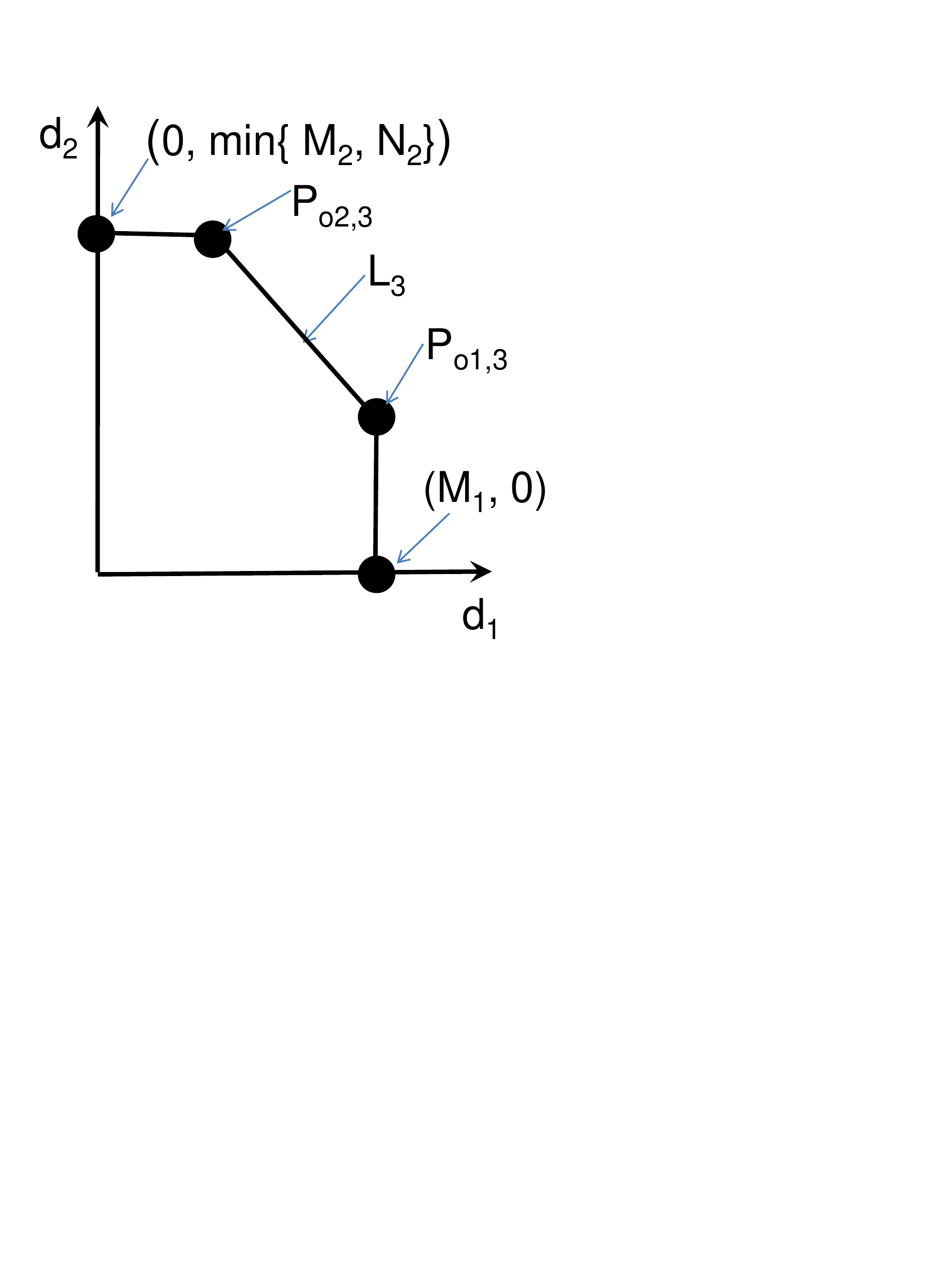}
\caption{Typical Shape of the DoF Region of the $2$-user MIMO IC with Delayed CSI: Case 0} \label{fig: typical shape d-CSI 2user IC Case 0}
\end{figure}

\begin{figure}[h] \centering
\includegraphics[bb=0bp 150bp 540bp 710bp,clip, height=3.2in, width=3.2in]{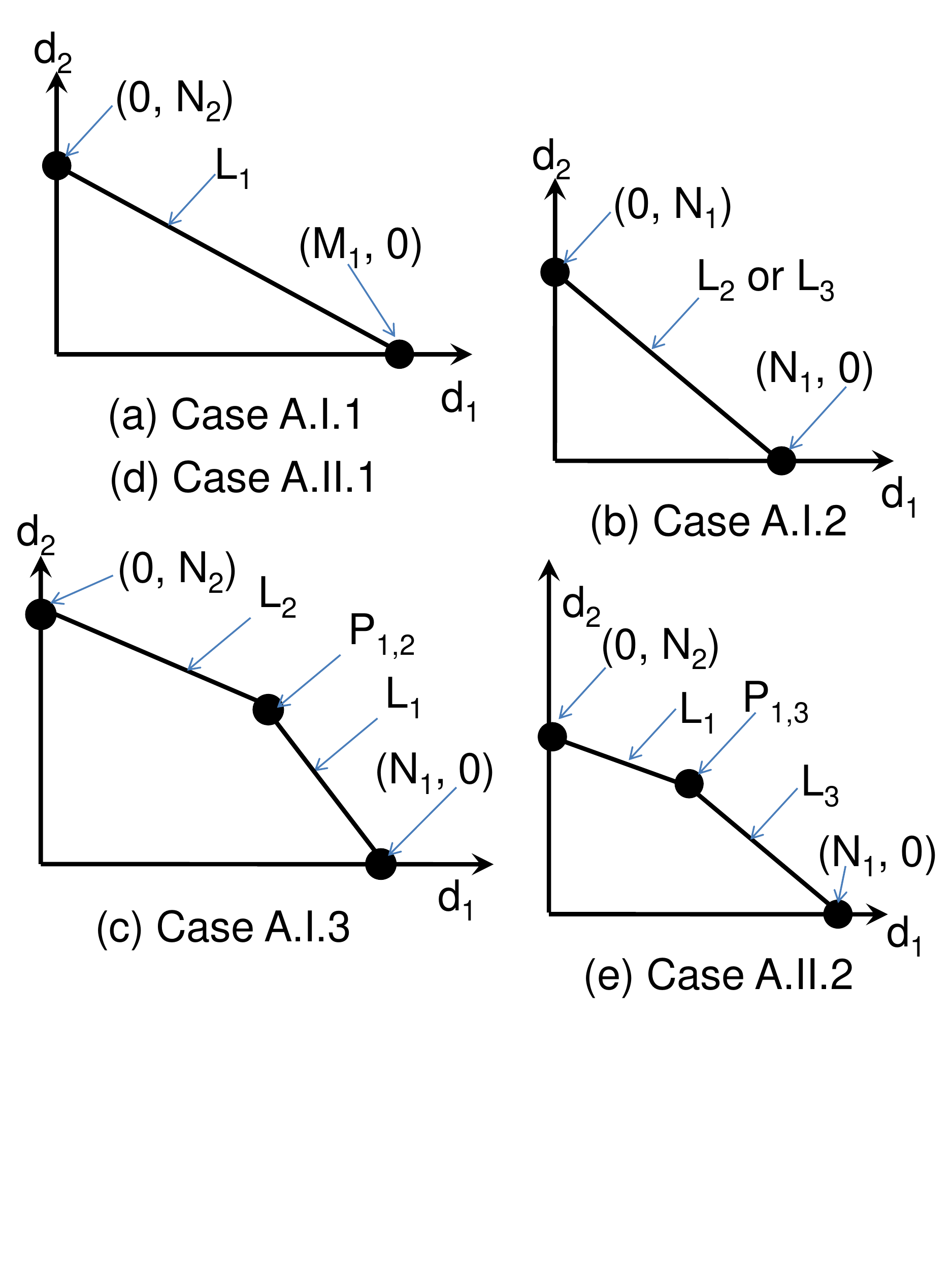}
\caption{Typical Shape of the DoF Region of the $2$-user MIMO IC with Delayed CSI: Case A} \label{fig: typical shape d-CSI 2user IC Case A}
\end{figure}

\begin{figure}[h] \centering
\includegraphics[bb=0bp 180bp 540bp 720bp,clip, height=3.5in, width=3.5in]{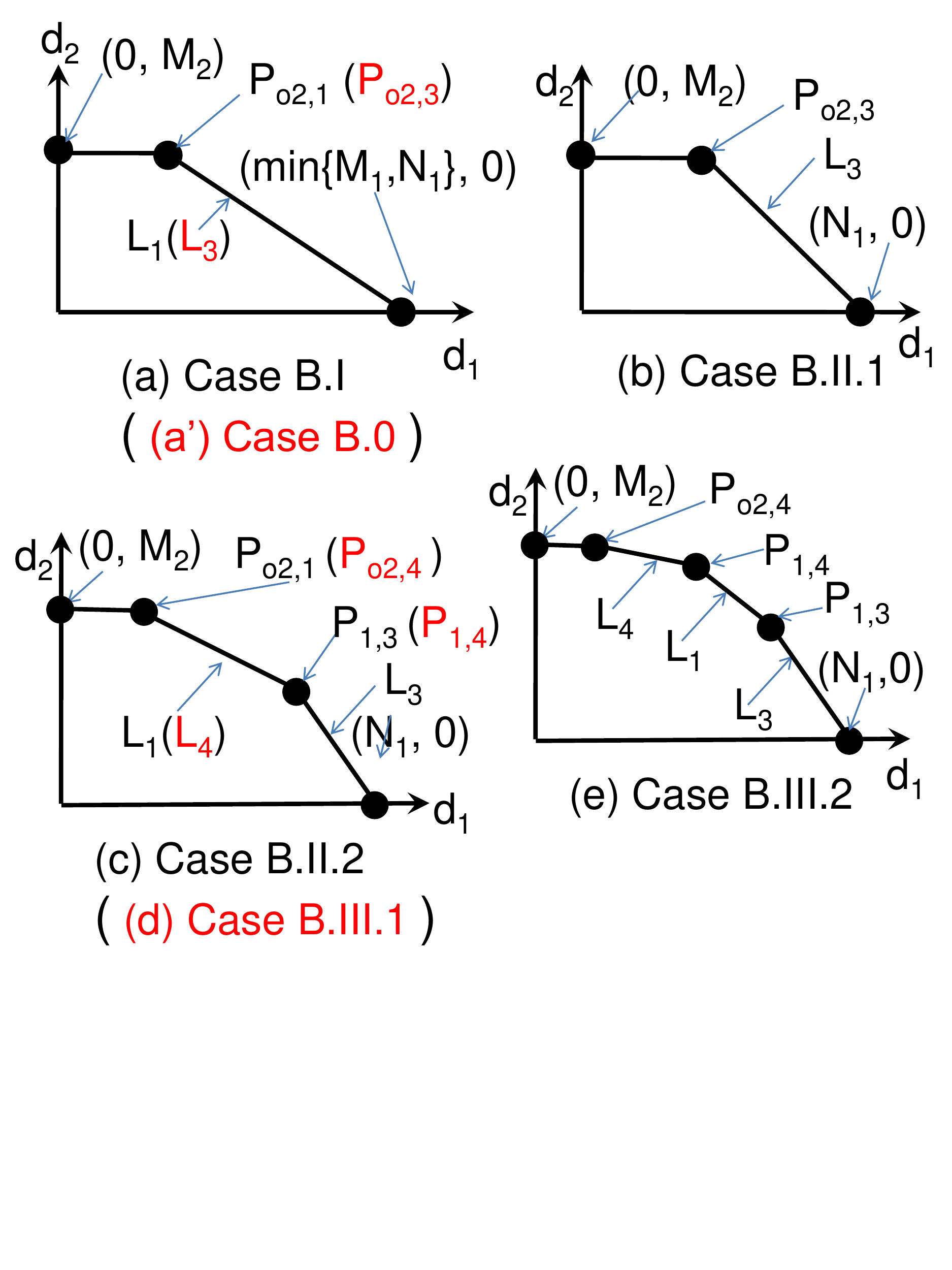}
\caption{Typical Shape of the DoF Region of the $2$-user MIMO IC with Delayed CSI: Case B} \label{fig: typical shape d-CSI 2user IC Case B}
\end{figure}

\begin{figure}[h] \centering
\includegraphics[bb=0bp 185bp 540bp 650bp,clip, height=3.5in, width=3.5in]{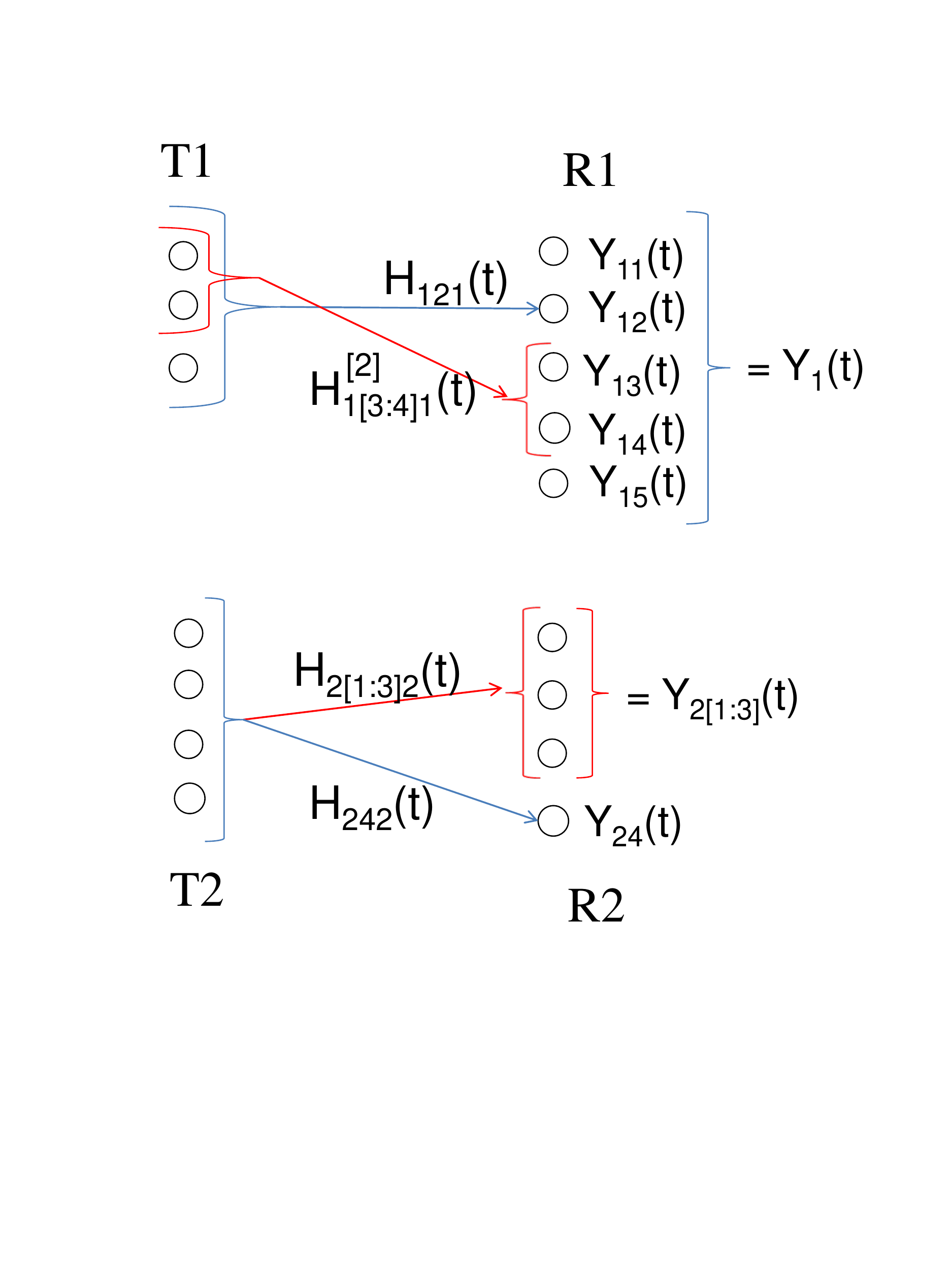}
\caption{Illustrating the Notation Used in This Paper} \label{fig: notation_used}
\end{figure}

\subsection{Summary of Results}
We know from the previous section that $\mathbf{D}^{\rm{d-CSI}} = \mathbf{D}_{\rm{outer}}^{\rm{d-CSI}}$. The outer-bound $\mathbf{D}_{\rm{outer}}^{\rm{d-CSI}}$ is defined in terms of two single-user bounds and five bounds labeled as $L_1$, $L_2$, $\cdots$, $L_5$. In this section, we discuss how the shape of the DoF region changes with the number of antennas at the four terminals. To this end, we need to determine which of the bounds in the description of the outer-bound are essential (i.e., not redundant or not implied by the other bounds). Thus, it is convenient to have the following definition.

Let $S$ be a subset of $\{1,2,3,4,5\}$ and $S^c$ be the complement of $S$, i.e., $S^c = \{1,2,3,4,5\} \backslash S$,
\begin{definition}
The bounds $L_{S}$ are said to be active if the single-user bounds $L_{o1}$ and $L_{o2}$, and $\{L_i\}_{i \in S}$ imply bound(s) $\{L_j\}$ $\forall$ $j \in S^c$ (bound $L_{3+i}$ is trivially implied by the other when Condition $1$ does not hold).
\end{definition}
If bound(s) $L_{S}$ are active for some set $S \subseteq \{1,2,3\}$, then bound(s) $\{L_j\}_{j \in S^c}$ can be dropped from the description of $\mathbf{D}_{\rm{outer}}^{\rm{d-CSI}}$. Moreover, if bounds $L_{\{S\}}$ are active with $S = {i}$, then we simply say that $L_i$ is active.

Now, the shape of the DoF region can be determined by identifying the bounds that are active. Again, without loss of generality, it is assumed that $N_1 \geq N_2$ (the results for $N_2 > N_1$ can be obtained by switching the order of the two users).

The summary of our results on the DoF region of the MIMO IC with delayed CSI is presented in Table \ref{table: summary of results d-CSI 2user IC}. Various details can be easily obtained from this table. For example, under Case A.I, the inequality $M_2 \geq N_1$ holds, in addition to the inequalities that define Case A. Moreover, as proved by Lemma \ref{lem: Case A.I d-CSI 2user IC}, in this case, bounds $L_{\{1,2\}}$ are active in general. But, more specifically, under Case A.I.1, only bound $L_1$ is active (see the proof of Lemma \ref{lem: Case A.I d-CSI 2user IC} in Section \ref{sec: proof of thm: MIMO_2-user IC_d-CSI_inner-bound DoF_Case A.I}) and $\mathbf{D}^{\rm{no-CSI}} = \mathbf{D}^{\rm{d-CSI}} \not= \mathbf{D}^{\rm{p-CSI}}$; whereas under Case A.I.2, we have $\mathbf{D}^{\rm{no-CSI}} = \mathbf{D}^{\rm{d-CSI}} = \mathbf{D}^{\rm{p-CSI}}$, i.e., $L_3$ is active. In this manner information about all other cases can be obtained from Table \ref{table: summary of results d-CSI 2user IC}.

In Figs. \ref{fig: typical shape d-CSI 2user IC Case 0}-\ref{fig: typical shape d-CSI 2user IC Case B}, we depict the DoF region of the IC with delayed CSI for the cases considered in Table \ref{table: summary of results d-CSI 2user IC}. Here, we make use of two facts: First, under Cases A and B, $d_1 = \min(M_1,N_1) \Rightarrow d_2 = 0$; see Lemma \ref{lem: d1max then d2=0 d-CSI 2user IC}. Second, under Case A, $d_2 = \min(M_2,N_2) \Rightarrow d_1 = 0$, which is not true with Case B; see Lemma \ref{lem: d2max then d1=0 d-CSI 2user IC}.

\subsection{Comparison of the DoF Regions with No, Delayed, and Perfect CSI}
For easy reference, we state the DoF regions of the IC with perfect and no CSI in the following remark.
\begin{remark} \label{rem: No and perfect CSI DoF regions}
The perfect-CSI DoF region of the IC is equal to the region $\mathbf{D}_{\rm{outer}}^{\rm{d-CSI}}$ with bounds $L_1$, $L_2$, $L_4$, and $L_5$ dropped \cite{Jafar-Maralle} (see also \cite[Subsection III-C]{Vaze_Dof_final}). Further, the no-CSI DoF region of the IC with $N_1 \geq N_2$ is given by \cite{Vaze_Dof_final, Chiachi2, Zhu_Guo_noCSIT_DoF_2010, D.Guo}
\begin{eqnarray*}
\lefteqn{ \mathbf{D}^{\rm{no-CSI}} = \Bigl\{ (d_1,d_2) \Big| 0 \leq d_i \leq \min(M_i,N_i) \mbox{ for } i=1,~2 \Big. \Bigr. }\\
&& {} \Biggl. \frac{d_1}{\min(M_1,N_1)} + \frac{d_2}{\min(M_1,N_2)} \leq \frac{\min(N_2,M_1+M_2)}{\min(M_1,N_2)} \Biggr\} .
\end{eqnarray*}
\end{remark}

Table \ref{table: summary of results d-CSI 2user IC} makes it easy to compare the DoF regions under the three cases of no, delayed, and perfect CSI. For example, we see that $\mathbf{D}^{\rm{no-CSI}} = \mathbf{D}^{\rm{d-CSI}}$ under Cases 0, A.I.1, A.I.2, A.II.1, and B.0.  Moreover, out of these, under Cases 0, A.I.2, and B.0, $\mathbf{D}^{\rm{no-CSI}}$ equals $\mathbf{D}^{\rm{d-CSI}}$ because the perfect-CSI DoF region can be achieved even without CSI. In other words, under Cases A.I.1 and A.II.1, the availability of delayed CSI can not improve the no-CSI DoF region, even though the no CSI and perfect CSI DoF regions are not equal. These two cases can collectively be described by the inequalities $N_1 \geq M_1 > N_2$ and $M_2 \geq N_2$.

Consider now the comparison of the delayed-CSI DoF region with the perfect-CSI DoF region. The two are equal under Cases 0, A.I.2, B.0, and B.II.1. Again, out of these, only under Case B.II.1, we have $\mathbf{D}^{\rm{d-CSI}} = \mathbf{D}^{\rm{p-CSI}} \not= \mathbf{D}^{\rm{no-CSI}}$. These observations are summarized in Table \ref{table: comparison of DoF regions d-CSI IC}.

\section{Proof of Theorem \ref{thm: outer-bound DoF 2-user IC d-CSI}: $L_1$ is an Outer-Bound} \label{sec: proof of thm: outer-bound DoF 2-user IC d-CSI}

Before starting the proof, we introduce the notation used in the remainder of the paper. \newline \emph{Notation: } The set of four channel matrices at time $t$ is denoted by $H(t)$, i.e., $H(t) = \big\{ H_{ij}(t) \big\}$ where $i,j \in \{1,2\}$. For integers $n_1$ and $n_2$, if $n_1 \leq n_2$, $[n_1:n_2] =  \{n_1,n_1+1, \cdots, n_2\}$; whereas if $n_1 > n_2$, then $[n_1: n_2]$ denotes the empty set. For a random variable $X(t)$, $X([n_1:n_2]) = \{X(t)\}_{t=n_1}^{n_2}$ if $n_1 \leq n_2$, whereas $X([n_1:n_2])$ denotes an empty set if $n_1 > n_2$. Further, for $n \geq 1$, $\overline{X}(n) = X([1:n])$. For the received signal $Y_i(t)$ and the channel matrix $H_{ik}(t)$, the $j^{th}$ entry and the $j^{th}$ row are denoted respectively by $Y_{ij}(t)$ and $H_{ijk}(t)$. Further, whenever $n_1 \leq n_2$ and $n_3 \leq n_4$, $Y_{i[n_1:n_2]}(t) = \{Y_{ij}(t)\}_{j=n_1}^{n_2}$, $Y_{i[n_1:n_2]}([n_3:n_4]) = \big\{ \{Y_{ij}(t)\}_{j=n_1}^{n_2} \big\}_{t=n_3}^{n_4}$, $H_{i[n_1:n_2]j}(t)$ is the channel matrix from $j^{th}$ transmitter to channel outputs $Y_{i[n_1:n_2]}(t)$ (see Fig. \ref{fig: notation_used}); however, if $n_1 > n_2$ and/or $n_3>n_4$, then $Y_{i[n_1:n_2]}(t)$ and $Y_{i[n_1:n_2]}([n_3:n_4])$ denote empty sets. Moreover, for $n \geq 1$, $\overline{Y}_{i[n_1:n_2]}(n) = Y_{i[n_1:n_2]}([1:n])$. Finally, $o(\log_2 P)$ denotes any real-valued function $x(P)$ of $P$ such that $\lim_{P \to \infty} \frac{x(P)}{\log_2 P} = 0$.

As discussed earlier, the aim is to prove that $L_1$ is a valid outer-bound. It turns out that to prove this bound, we need an inequality which is stated in the following lemma.
\begin{lemma} \label{cor: main inequality d-CSI IC outer-bound}
Let $m_1 = \min(M_1,N_1+N_2)$ and $m_2 = \min(M_1,N_2)$. Then, for each n, we have
\begin{eqnarray*}
\frac{1}{m_2} h \Big( \overline{Y}_2(n) \Big| \mathcal{M}_2, \overline{H}(n) \Big) \hspace{-9pt} & \geq & \hspace{-9pt} \frac{1}{m_1} h \Big( \overline{Y}_1(n), \overline{Y}_2(n) \Big| \mathcal{M}_2, \overline{H}(n) \Big) + n \cdot o(\log_2 P), \mbox{ or equivalently,} \\
\frac{1}{m_2} I \Big( \mathcal{M}_1; \overline{Y}_2(n) \Big| \mathcal{M}_2, \overline{H}(n) \Big) \hspace{-9pt} & \geq & \hspace{-9pt} \frac{1}{m_1} I \Big(\mathcal{M}_1; \overline{Y}_1(n), \overline{Y}_2(n) \Big| \mathcal{M}_2, \overline{H}(n) \Big) + n \cdot o(\log_2 P),
\end{eqnarray*}
where the term $o(\log_2 P)$ is constant with $n$.
\end{lemma}
In fact, the proof of bound $L_1$ is an application of the above lemma. In the following section, bound $L_1$ is obtained using the above lemma. Later, in Section \ref{subsec: proof of outer-bound d-CSI 2-user IC lemmas}, Lemma \ref{cor: main inequality d-CSI IC outer-bound} is proved. Section \ref{subsec: remarks on the proof of the outer-bound d-CSI 2user IC} contains some remarks on the proof.

\subsection{Proof of Bound $L_1$} \label{subsec: proof of outer-bound d-CSI 2-user IC main proof}

We first outer-bound the capacity region of the given MIMO IC with delayed CSI as follows. Assume that both the receivers know all channel matrices perfectly and instantaneously. Further, R1 is assumed have non-causal knowledge of the message $\mathcal{M}_2$ and instantaneous knowledge of the signal $Y_2(t)$. These assumptions can only enhance the capacity region of the given IC. We apply Fano's inequality \cite{CT} under these assumptions to upper bound the rates achievable for the two users as follows:
\begin{eqnarray}
n R_2 & \leq & I \Big( \mathcal{M}_2;\overline{Y}_2(n) \Big| \overline{H}(n) \Big) + n \epsilon_n \nonumber \\
      & = & h \Big( \overline{Y}_2(n) \Big| \overline{H}(n) \Big) - h \Big( \overline{Y}_2(n) \Big| \mathcal{M}_2, \overline{H}(n) \Big) + ~ n \epsilon_n \\
      & = & \underbrace{I \Big( \mathcal{M}_1, \mathcal{M}_2; \overline{Y}_2(n) \Big| \overline{H}(n) \Big)}_{= t_1} - \underbrace{I \Big( \mathcal{M}_1 ; \overline{Y}_2(n) \Big| \mathcal{M}_2, \overline{H}(n) \Big)}_{=t_2} + ~ n \epsilon_n  \mbox{ and } \label{eq: bound on r2 d-CSI IC}\\
n R_1 & \leq & \underbrace{I \Big( \mathcal{M}_1;\overline{Y}_1(n), \overline{Y}_2(n) \Big| \mathcal{M}_2, \overline{H}(n) \Big)}_{=t_3} + ~ n \epsilon_n   \label{eq: bound on r1 d-CSI IC}
\end{eqnarray}
where $\epsilon_n \to 0$ as $n \to \infty$.

Let $f (\cdot)$ denote the operation:
\[
f ( x ) = \lim_{P \to \infty} \left\{ \frac{1}{\log_2 P} \cdot \lim_{n \to \infty} ~ \frac{x}{n} \right\}
\]
whenever the limits exist. Note here that the operator $f(\cdot)$ preserves inequalities, i.e., if $x \leq y$, then $f(x) \leq f(y)$.

We now apply $f (\cdot)$ to both the sides of equations (\ref{eq: bound on r2 d-CSI IC}) and (\ref{eq: bound on r1 d-CSI IC}) to obtain
\begin{equation}
d_2 \leq f \big( t_1 \big) - f \big( t_2 \big) \mbox{ and } d_1 \leq f \big( t_3 \big), \label{eq: bounds on d1 and d2 d-CSI 2user IC}
\end{equation}
where we make use of the facts that $f \big( n R_i \big) = d_i$ for $i=1,~2$, $f \big( n \epsilon_n \big) = 0$.

Consider the first term $f \big( t_1 \big)$, which stands for the DoF available at R2 per time slot (or the total number of dimensions of the receive-signal space of R2 per time slot). Since the DoF of the point-to-point MIMO channel are limited by the minimum of the number of transmit and receive antennas, we get
\begin{eqnarray}
f \big( t_1 \big) \leq \min(M_1+M_2,N_2). \label{eq: bound on term 1 d-CSI IC outer-bound}
\end{eqnarray}

Consider the term $f \big( t_2 \big)$. Given the message $\mathcal{M}_2$ and the channel matrices $\overline{H}(n)$, the transmit signal $\overline{X}_2(n)$ is deterministic. Hence, the term $t_2$ equals the amount of information R2 can get about $\mathcal{M}_1$ after decoding its intended message $\mathcal{M}_2$. That is, the term $t_2$ is a measure of the interference caused by the transmission of T1 at R2, and $f \big( t_2 \big)$ is equal to the DoF occupied by the interference at R2 per time slot. Thus, the first inequality in (\ref{eq: bounds on d1 and d2 d-CSI 2user IC}) says that the number of DoF that can be achieved for the second user equals the total number of DoF available at R2 per time slot minus the number of DoF occupied by the interference. Moreover, using Lemma \ref{cor: main inequality d-CSI IC outer-bound} and the second inequality in (\ref{eq: bounds on d1 and d2 d-CSI 2user IC}), we get
\begin{equation}
f \big( t_2 \big) \geq \frac{m_2}{m_1} f \big( t_3 \big) \geq \frac{m_2}{m_1} d_1, \label{eq: immediate applicn of cor: main inequality d-CSI IC outer-bound}
\end{equation}
which shows that if $d_1$ DoF are to be achieved for the first user when there is delayed CSI, interference will occupy at least $\frac{m_2}{m_1}$ times $d_1$ DoF at R2, regardless of the achievability scheme used. This lower-bound on the DoF of the interference allows us to obtain an upper-bound on $d_2$ as follows:
\begin{equation}
d_2 \leq \min(N_2, M_1+M_2) - \frac{m_2}{m_1} d_1,
\end{equation}
where we make use of inequalitites (\ref{eq: bounds on d1 and d2 d-CSI 2user IC}), (\ref{eq: bound on term 1 d-CSI IC outer-bound}), and (\ref{eq: immediate applicn of cor: main inequality d-CSI IC outer-bound}), and the above bound in turn yields us
\[
\frac{d_1}{m_1} + \frac{d_2}{m_2} \leq \frac{\min(N_2,M_1+M_2)}{m_2},
\]
which is the bound $L_1$. \IEEEQED

\subsection{Proof of Lemma \ref{cor: main inequality d-CSI IC outer-bound}} \label{subsec: proof of outer-bound d-CSI 2-user IC lemmas}

We will derive a series of lemmas and a corollary using which Lemma \ref{cor: main inequality d-CSI IC outer-bound} will be proved. Let us first focus on the first inequality, using which the second inequaltiy will be proved later. The first inequality relates the two differential entropy terms involving random variables $\overline{Y}_1(n)$ and $\overline{Y}_2(n)$, which denote the signal received by the two receivers over the blocklength of $n$. In the following two lemmas, it is shown that although the received signals $Y_1(t)$ and $Y_2(t)$ are $N_1$ and $N_2$ dimensional, respectively, only the first $m_1-m_2$ and $m_2$ entries of them are relevant as far as the current DoF analysis is concerned.

\begin{lemma} \label{lem: extracting essential part of term1 main inequality d-CSI IC outer-bound}
If $m_2 = \min(M_1,N_2)$, then we have following:
\[
h \Big( \overline{Y}_2(n) \Big| \mathcal{M}_2, \overline{H}(n) \Big) \geq h \Big( \overline{Y}_{2[1:m_2]}(n) \Big| \mathcal{M}_2, \overline{H}(n) \Big) + n \cdot o(\log_2 P),
\]
where the term $o(\log_2 P)$ is constant with $n$.
\end{lemma}
\begin{IEEEproof}
This can be proved as follows:
\begin{eqnarray}
\lefteqn{ h \Big( \overline{Y}_2(n) \Big| \mathcal{M}_2, \overline{H}(n) \Big) } \nonumber \\
& = & h \Big( \overline{Y}_{2[1:m_2]}(n) \Big| \mathcal{M}_2, \overline{H}(n) \Big) + h \Big( \overline{Y}_{2[m_2+1:N_2]}(n) \Big| \overline{Y}_{2[1:m_2]}(n), \mathcal{M}_2, \overline{H}(n) \Big) \nonumber \\
& \geq &  h \Big( \overline{Y}_{2[1:m_2]}(n) \Big| \mathcal{M}_2, \overline{H}(n) \Big) + h \Big( \overline{Y}_{2[m_2+1:N_2]}(n) \Big| \overline{X}_1(n), \overline{X}_2(n), \overline{Y}_{2[1:m_2]}(n), \mathcal{M}_2, \overline{H}(n) \Big) \label{eq: lem1 d-CSI IC outer-bound step1} \\
& =  & h \Big( \overline{Y}_{2[1:m_2]}(n) \Big| \mathcal{M}_2, \overline{H}(n) \Big) + h \Big( \overline{W}_{2[m_2+1:N_2]}(n) \Big)\label{eq: lem1 d-CSI IC outer-bound step2}\\
& = & h \Big( \overline{Y}_{2[1:m_2]}(n) \Big| \mathcal{M}_2, \overline{H}(n) \Big)  + n \cdot o(\log_2 P), \label{eq: lem1 d-CSI IC outer-bound step3}
\end{eqnarray}
where inequality (\ref{eq: lem1 d-CSI IC outer-bound step1}) follows because conditioning reduces entropy; the next equality (\ref{eq: lem1 d-CSI IC outer-bound step2}) holds because translation does not change differential entropy and because noise is independent of the channel matrices, the transmit signals, and the messages; and the last equality is true since noise random variables are i.i.d. across time and their statistics are constant with $P$, which also explains why the term $o(\log_2 P)$ is constant with $n$.
\end{IEEEproof}

The following lemma deals with the second differential term involved in the first inequality of Lemma \ref{cor: main inequality d-CSI IC outer-bound}.

\begin{lemma} \label{lem: extracting essential part of term2 main inequality d-CSI IC outer-bound}
If $m_1 = \min(M_1,N_1+N_2)$, then
\[
h \Big( \overline{Y}_1(n), \overline{Y}_2(n) \Big| \mathcal{M}_2, \overline{H}(n) \Big) \leq h \Big( \overline{Y}_{1[1:m_1-m_2]}(n), \overline{Y}_{2[1:m_2]}(n) \Big| \mathcal{M}_2, \overline{H}(n) \Big) + n \cdot o(\log_2 P).
\]
where the term $o(\log_2 P)$ is constant with $n$.
\end{lemma}
\begin{IEEEproof}
This lemma holds trivially if $M_1 \geq N_1+N_2$. We thus consider the case where $M_1 < N_1 + N_2$. Given $\mathcal{M}_2$ and $\overline{H}(n)$, the transmit signal $X_2(t)$ is deterministic $\forall$ $t \in [1:n]$. Define $Y_i'(t) = Y_i(t) - H_{i2}(t) X_2(t)$ and then define quantities $\overline{Y}_i'(n)$, $\overline{Y}_{2[1:m_2]}'(n)$, etc. in an analogous fashion. Since translation does not change differential entropy, we get
\[
h \Big( \overline{Y}_{1[1:m_1-m_2]}(n), \overline{Y}_{2[1:m_2]}(n) \Big| \mathcal{M}_2, \overline{H}(n) \Big) = h \Big( \overline{Y}_{1[1:m_1-m_2]}'(n), \overline{Y}_{2[1:m_2]}'(n) \Big| \mathcal{M}_2, \overline{H}(n) \Big) =: q.
\]
Denote by $Q(n)$ random variables $\{ \mathcal{M}_2, \overline{H}(n) \}$. Then we have
\begin{eqnarray*}
\lefteqn{ h \Big( \overline{Y}_1(n), \overline{Y}_2(n) \Big| Q(n) \Big) =  h \Big( \overline{Y}_1'(n), \overline{Y}_2'(n) \Big| Q(n) \Big) } \\
\hspace{-5cm} & = & q + h \Big( \overline{Y}_{1[m_1-m_2+1:N_1]}'(n), \overline{Y}_{2[m_2+1:N_2]}'(n) \Big| Q(n), \overline{Y}_{1[1:m_1-m_2]}'(n), \overline{Y}_{2[1:m_2]}'(n) \Big) \\
\hspace{-5cm} & \leq & q + \sum_{t=1}^n q_t ~ \mbox{with } q_t = h \Big( Y_{1[m_1-m_2+1:N_1]}'(t), Y_{2[1:m_2]}'(t) \Big| H(t), Y_{1[1:m_1-m_2]}'(t), Y_{2[1:m_2]}'(t) \Big),
\end{eqnarray*}
where the inequality follows because the conditioning reduces entropy. Now, given the signal received at $m_1$ receive antennas, namely, $Y_{1[1:m_1-m_2]}'(t)$ and $Y_{2[1:m_2]}'(t)$, a noisy version of $X_1(t)$ can be constructed via channel inversion with probability $1$ (c.f. \cite[Section II-C]{Vaze_Dof_final}). Hence, the term $q_t$ just equals  the differential entropy of some noise random variables. Since noise as well as channel matrices are i.i.d. across time and the statistics of both are constant with respect to $P$, the terms $q_t$ are equal for all $t$ and $ q_t = o(\log_2 P)$, irrespective of $n$.
\end{IEEEproof}

\begin{remark}
Since all the differential entropy or the mutual information terms considered in this section are conditioned on $\mathcal{M}_2$ and $\overline{H}(n)$, it is assumed henceforth in this section that $X_2(t) = 0$, $\forall$ $t$ because $H_{ij2}(t) X_2(t)$ can always be subtracted from the received signal $Y_{ij}(t)$.
\end{remark}

Consider the following lemma, which is the crucial step in establishing Lemma \ref{cor: main inequality d-CSI IC outer-bound}.

\begin{lemma} \label{lem:symmetry outer-bound IC d-CSI}
For a given $t \in [1:n]$, $i \in [1 : m_2-1]$, and $j = i+1$, we have
\[
h \Big( Y_{2i}(t) \Big| \overline{Y}_{2[1:m_2]}(t-1), \mathcal{M}_2,\overline{H}(t), Y_{2[1:i-1]}(t) \Big) \geq h \Big( Y_{2j}(t) \Big| \overline{Y}_{2[1:m_2]}(t-1), \mathcal{M}_2,\overline{H}(t), Y_{2[1:i]}(t) \Big).
\]
\end{lemma}
\begin{IEEEproof}
Consider the first term
\begin{eqnarray}
\lefteqn{ h \Big( Y_{2i}(t) \Big| \overline{Y}_{2[1:m_2]}(t-1), \mathcal{M}_2,\overline{H}(t), Y_{2[1:i-1]}(t) \Big) } \nonumber\\
& = & h \Big( Y_{2i}(t) \Big| \overline{Y}_{2[1:m_2]}(t-1), Y_{2[1:i-1]}(t), \mathcal{M}_2, \overline{H}(t-1), H_{2[1:i]1}(t) \Big) \label{eq: step1 lem:symmetry outer-bound IC d-CSI}  \\
& = & \mathbb{E}_{H_{2i1}(t) = a} ~  h \Big( Y_{2i}(t) \Big| \overline{Y}_{2[1:m_2]}(t-1), Y_{2[1:i-1]}(t), \mathcal{M}_2, \overline{H}(t-1), H_{2[1:i-1]1}(t), H_{2i1}(t) = a \Big) \label{eq: step2 lem:symmetry outer-bound IC d-CSI} \\
& = & \mathbb{E}_{H_{2j1}(t) = a} ~  h \Big( Y_{2j}(t) \Big| \overline{Y}_{2[1:m_2]}(t-1), Y_{2[1:i-1]}(t), \mathcal{M}_2, \overline{H}(t-1), H_{2[1:i-1]1}(t), H_{2j1}(t) = a \Big) \label{eq: step3 lem:symmetry outer-bound IC d-CSI} \\
& \geq &  h \Big( Y_{2j}(t) \Big| \overline{Y}_{2[1:m_2]}(t-1), \mathcal{M}_2,\overline{H}(t), Y_{2[1:i]}(t) \Big), \label{eq: step4 lem:symmetry outer-bound IC d-CSI}
\end{eqnarray}
where equality (\ref{eq: step1 lem:symmetry outer-bound IC d-CSI}) holds because all the concerned random variables are independent of $H_{2[i+1:N_2]1}(t)$; the equation (\ref{eq: step2 lem:symmetry outer-bound IC d-CSI}) follows from the definition of conditional differential entropy; the next equality (\ref{eq: step3 lem:symmetry outer-bound IC d-CSI}) holds because conditioned of random variables $\big\{  \overline{Y}_{2[1:m_2]}(t-1), ~ Y_{2[1:i-1]}(t), ~ \mathcal{M}_2,~ \overline{H}(t-1), \mbox{ and } H_{2[1:i-1]1}(t) \big\}$, the joint distribution of the pair of random variables $\big\{ H_{2i1}(t), ~ X(t) \big\}$ is identical to that of the pair $\big\{ H_{2j1}(t), ~ X(t) \big\}$; and the final inequality (\ref{eq: step4 lem:symmetry outer-bound IC d-CSI}) holds since conditioning reduces entropy.
\end{IEEEproof}

Consider the equality in (\ref{eq: step3 lem:symmetry outer-bound IC d-CSI}), which proves that the the signals received $Y_{2i}(t)$ and $Y_{2j}(t)$ at the $i^{th}$ and $j^{th}$ antenna, respectively, of R2 have equal differential entropy, when conditioned on the channel matrices $\overline{H}(t)$, the message $\mathcal{M}_2$, the past channel outputs $\overline{Y}_2(t-1)$, and the present channel outputs $Y_{2[1:i-1]}(t)$ at some other receive antennas. We refer to this property as the statistical equivalence of the channel outputs, which essentially says that given the past and present channel outputs, the signals received at any two antennas of the system provide equal amount of information about $\mathcal{M}_1$ (when T2 is silent). This property provides a basis for us to relate the differential entropies of the signals received by R1 and R2, and indeed it is the important point of this proof. See Section \ref{subsec: remarks on the proof of the outer-bound d-CSI 2user IC} for further discussion.

The following corollary uses the above lemma to obtain inequalities in more useful form.
\begin{corollary} \label{cor: applicn of lem:symmetry outer-bound IC d-CSI}
For a given $t \in [1:n]$, given $i \in [1: m_1-m_2-1]$, $j = i+1$, $V_1(t) = \big\{ \overline{Y}_2(t-1), \mathcal{M}_2, \overline{H}(t) \big\}$, and $V_2(t) = \big\{ \overline{Y}_{2[1:m_2]}(t), \overline{Y}_{1[1:m_1-m_2]}(t-1), \mathcal{M}_2, \overline{H}(t) \big\}$, we have the following inequalities:
\begin{eqnarray*}
h \Big( Y_{2[1:m_2]}(t) \Big| V_1(t) \Big) & \geq & m_2 \cdot h \Big( Y_{2m_2}(t) \Big| V_1(t), Y_{2[1:m_2-1]}(t) \Big)
\end{eqnarray*}
\begin{eqnarray*}
h \Big( Y_{2m_2}(t) \Big| V_1(t), Y_{2[1:m_2-1]}(t) \Big) & \geq & h \Big( Y_{11}(t) \Big| V_2(t) \Big),
\end{eqnarray*}
\begin{eqnarray*}
h \Big( Y_{1i}(t) \Big| V_2(t), Y_{1[1:i-1]}(t) \Big) & \geq & h \Big( Y_{1j}(t) \Big| V_2(t), Y_{1[1:i]}(t) \Big), \mbox{ and}
\end{eqnarray*}
\begin{eqnarray*}
(m_1-m_2) \cdot h \Big( Y_{11}(t) \Big| V_2(t) \Big) & \geq & h \Big( Y_{1[1:m_1-m_2]}(t) \Big| V_2(t) \Big).
\end{eqnarray*}
\end{corollary}
\begin{IEEEproof}
By applying the previous lemma, we obtain
\begin{eqnarray*}
\lefteqn{ h \Big( Y_{2m_2}(t) \Big| V_1(t), Y_{2[1:m_2-1]}(t) \Big)  \leq  h \Big( Y_{2(m_2-1)}(t) \Big| V_1(t), Y_{2[1:m_2-2]}(t) \Big) }\\
&& {} \leq  h \Big( Y_{2(m_2-2)}(t) \Big| V_1(t), Y_{2[1:m_2-3]}(t) \Big) \leq \cdots \leq  h \Big( Y_{2k}(t) \Big| V_1(t), Y_{2[1:k-1]}(t) \Big) \\
&& {} \leq  \cdots \leq h \Big(Y_{22}(t) \Big| V_1(t), Y_{21}(t) \Big) \leq h \Big( Y_{21}(t) \Big| V_1(t) \Big),
\end{eqnarray*}
which implies that
\begin{eqnarray*}
\lefteqn{  h \Big( Y_{2[1:m_2]}(t) \Big| V_1(t) \Big)  =  \sum_{i=1}^{m_2} h \Big( Y_{2i}(t) \Big| V_1(t), Y_{2[1:i-1]}(t) \Big)} \\
&& {} \geq  \sum_{i=1}^{m_2} h \Big( Y_{2m_2}(t) \Big| V_1(t), Y_{2[1:m_2-1]}(t) \Big) = m_2 \cdot h \Big( Y_{2m_2}(t) \Big| V_1(t), Y_{2[1:m_2-1]}(t) \Big).
\end{eqnarray*}
The next inequality is based on the idea of statistical equivalence of the signal received at two of the receive antennas. It can be proved as follows:
\begin{eqnarray*}
h \Big( Y_{2m_2}(t) \Big| \overline{Y}_2(t-1), \mathcal{M}_2, \overline{H}(t), Y_{2[1:m_2-1]}(t) \Big)
& = & h \Big( Y_{11}(t) \Big| \overline{Y}_2(t-1), \mathcal{M}_2, \overline{H}(t), Y_{2[1:m_2-1]}(t) \Big) \\
& \geq & h \Big( Y_{11}(t) \Big| V_2(t) \Big) ,
\end{eqnarray*}
where the last inequality follows because conditioning reduces entropy.
The third inequality is immediate from the previous lemma, whereas the last inequality can be proved in a manner analogous to the proof of the first inequality of this corollary.
\end{IEEEproof}

The nextlemma uses the above corollary.
\begin{lemma} \label{lem: ultimate applicn lem:symmetry outer-bound IC d-CSI}
The following inequality holds:
\[
\frac{1}{m_2} h \Big( \overline{Y}_{2[1:m_2]}(n) \Big| \mathcal{M}_2, \overline{H}(n) \Big) \geq \frac{1}{m_1} h \Big( \overline{Y}_{1[1:m_1-m_2]}(n), \overline{Y}_{2[1:m_2]}(n) \Big| \mathcal{M}_2, \overline{H}(n) \Big).
\]
\end{lemma}
\begin{IEEEproof}
To simplify the notation, with abuse of notation, we omit in this proof the subscripts $[1:m_2]$ and $[1:m_1-m_2]$ appearing in the above differential entropy terms. That is, we write $Y_{2[1:m_2]}(t)$, $Y_{1[1:m_1-m_2]}(t)$, $\overline{Y}_{2[1:m_2]}(n)$, and $\overline{Y}_{1[1:m_1-m_2]}(n)$ respectively as $Y_2(t)$, $Y_1(t)$, $\overline{Y}_2(n)$, and $\overline{Y}_1(n)$. Then
\begin{eqnarray*}
h \Big( \overline{Y}_2(n) \Big| \mathcal{M}_2, \overline{H}(n) \Big) & = &
\sum_{t=1}^n h \Big( Y_2(t) \Big| \overline{Y}_2(t-1), \mathcal{M}_2, \overline{H}(t) \Big) \mbox{ and } \\
h \Big( \overline{Y}_1(n), \overline{Y}_2(n) \Big| \mathcal{M}_2, \overline{H}(n) \Big)  & = & \sum_{t=1}^n  h \Big( Y_1(t), Y_2(t) \Big| \overline{Y}_1(t-1), \overline{Y}_2(t-1),  \mathcal{M}_2, \overline{H}(t) \Big),
\end{eqnarray*}
where we make use of the fact that random variables $\overline{Y}_1(t)$, $\overline{Y}_2(t)$, $\mathcal{M}_2$, and $\overline{H}(t)$ are independent of $H([t+1:n])$.
Hence it is sufficient to prove that for each $t \in [1:n]$,
\[
m_1 \cdot  h \Big( Y_2(t) \Big| \overline{Y}_2(t-1), \mathcal{M}_2, \overline{H}(t) \Big) \geq m_2 \cdot h \Big( Y_1(t), Y_2(t) \Big| \overline{Y}_1(t-1), \overline{Y}_2(t-1),  \mathcal{M}_2, \overline{H}(t) \Big).
\]
Moreover, since $h \Big( Y_1(t), Y_2(t) \Big| \overline{Y}_1(t-1), \overline{Y}_2(t-1),  \mathcal{M}_2, \overline{H}(t) \Big)$
\[
= h \Big( Y_2(t) \Big| \overline{Y}_1(t-1), \overline{Y}_2(t-1),  \mathcal{M}_2, \overline{H}(t) \Big) + h \Big( Y_1(t) \Big| \overline{Y}_1(t-1), \overline{Y}_2(t),  \mathcal{M}_2, \overline{H}(t) \Big)
\]
and
\[
h \Big( Y_2(t) \Big| \overline{Y}_2(t-1), \mathcal{M}_2, \overline{H}(t) \Big) \geq
h \Big( Y_2(t) \Big| \overline{Y}_1(t-1), \overline{Y}_2(t-1),  \mathcal{M}_2, \overline{H}(t) \Big),
\]
it is enough to establish the following inequality:
\[
(m_1 - m_2) \cdot  h \Big( Y_2(t) \Big| \overline{Y}_2(t-1), \mathcal{M}_2, \overline{H}(t) \Big) \geq m_2 \cdot h \Big( Y_1(t) \Big| \overline{Y}_1(t-1), \overline{Y}_2(t),  \mathcal{M}_2, \overline{H}(t) \Big).
\]
To this end, consider the following:
\begin{eqnarray*}
\lefteqn{ (m_1 - m_2) \cdot  h \Big( Y_2(t) \Big| \overline{Y}_2(t-1), \mathcal{M}_2, \overline{H}(t) \Big) } \\
&& {} \geq (m_1 - m_2) \cdot m_2 \cdot h \Big( Y_{2m_2}(t) \Big| \overline{Y}_2(t-1), \mathcal{M}_2, \overline{H}(t), Y_{2[1:m_2-1]}(t) \Big) \\
&& {} \geq m_2 \cdot (m_1 - m_2) \cdot h \Big( Y_{11}(t) \Big| \overline{Y}_2(t), \overline{Y}_1(t-1), \mathcal{M}_2, \overline{H}(t) \Big) \\
&& {} \geq  m_2 \cdot h \Big( Y_1(t) \Big| \overline{Y}_1(t-1), \overline{Y}_2(t),  \mathcal{M}_2, \overline{H}(t) \Big),
\end{eqnarray*}
where the three inequalities follow respectively from the first, second, and the fourth inequalities of Corollary \ref{cor: applicn of lem:symmetry outer-bound IC d-CSI}.
\end{IEEEproof}

Using the above lemma, Lemma \ref{cor: main inequality d-CSI IC outer-bound} can now be proved as follows.

\begin{IEEEproof}[Proof of Lemma \ref{cor: main inequality d-CSI IC outer-bound}]
Consider the first inequality. Recall that the differential entropy term on the left hand side of the above inequality is lower-bounded (within $o(\log_2 P)$) via Lemma \ref{lem: extracting essential part of term1 main inequality d-CSI IC outer-bound}, whereas the one on the right hand side is upper-bounded (within $o(\log_2 P)$) via Lemma \ref{lem: extracting essential part of term2 main inequality d-CSI IC outer-bound}. The inequality now follows from Lemma \ref{lem: ultimate applicn lem:symmetry outer-bound IC d-CSI} by noting that the sum or the difference of two $o(\log_2 P)$ yields another $o(\log_2 P)$ term.

The second inequality follows from the first by noting that the differential entropy of the received signals conditioned on the channel matrices and the messages is equal to that of some noise terms, which is of the order of $n \cdot o(\log_2 P)$.
\end{IEEEproof}

\subsection{Comments on the Proof of Lemma \ref{cor: main inequality d-CSI IC outer-bound}} \label{subsec: remarks on the proof of the outer-bound d-CSI 2user IC}

Let us revisit the equality in (\ref{eq: step3 lem:symmetry outer-bound IC d-CSI}), which proves the statistical equivalence of the channel outputs at two of the receive antennas, namely, $Y_{2i}(t)$ and $Y_{2j}(t)$, given $\overline{H}(t)$, $\mathcal{M}_2$, $\overline{Y}(t-1)$ and $Y_{2[1:i-1]}(t)$. The random variables $\big\{  \overline{Y}_2(t-1), ~ Y_{2[1:i-1]}(t), ~ \mathcal{M}_2,~ \overline{H}(t-1), ~ H_{2[1:i-1]1}(t) \big\}$ are dependent on the transmit signal $X_1(t)$. However, conditioned on this set of random variables, the transmit signal $X_1(t)$ is independent of $H_{2i1}(t)$ as well as $H_{2j1}(t)$, and moreover, the channel vectors are $H_{2i1}(t)$ and $H_{2j1}(t)$ are themselves i.i.d.  Hence, conditioned on the above set of random variables, the situation at the $i^{th}$ antenna of R2 is identical to that at its $j^{th}$ antenna, and hence, they would provide equal amount of information about the message. This observation is the key to the equality in (\ref{eq: step3 lem:symmetry outer-bound IC d-CSI}), which is important for proving Lemma \ref{cor: main inequality d-CSI IC outer-bound}.

Note here that the equality of the two differential entropy terms claimed in (\ref{eq: step3 lem:symmetry outer-bound IC d-CSI}) holds because the random variables, on which these two terms are conditioned, involve only the past channel outputs and the present channel outputs at some other receive antennas of the system, in addition to the channel matrices and the message $\mathcal{M}_2$. In fact, the same result may not hold if the two differential entropy terms are conditioned on some future channel outputs as well. This is because these future channel outputs are correlated with the signal transmitted (by T1) during some future time slot, which, due to the availability of delayed CSI, can depend on the present channel outputs $Y_{2i}(t)$ and $Y_{2j}(t)$ (and hence on the channel vectors $H_{2i1}(t)$ and $H_{2j1}(t)$) in any arbitrary fashion. Hence, conditioned on the future channel outputs, the vectors $H_{2i1}(t)$ and $H_{2j1}(t)$ may no longer be identically distributed, which otherwise is the case. This implies that the equality in (\ref{eq: step3 lem:symmetry outer-bound IC d-CSI}) may not hold.

Recall that the outer-bound derived in \cite{Vaze_Dof_final} for the no-CSI case also rests on the statistical equivalence of the channel outputs. Particularly, in the case of no CSI, the channel outputs corresponding to any two receive antennas would have equal differential entropy, when conditioned on the same set of random variables, because the channel inputs are independent of all channel matrices. In other words, when there is no CSI, the statistical equivalence holds as long as the two differential entropies are conditioned on the same set of random variables, regardless of what these random variables are. In this sense, as compared to the case of delayed CSI, the property of statistical equivalence holds more generally under no CSI.

As mentioned earlier, an outer-bound to the DoF region of the $K$-user MISO BC with delayed CSI was derived in \cite{maddah_ali_tse_delayed_CSIT}, which was later generalized by the authors in \cite{Vaze-Varanasi-delay-MIMOBC} to the case of MIMO BC. These outer-bounds have been found to be tight in some special cases (see \cite{maddah_ali_tse_delayed_CSIT, Vaze-Varanasi-delay-MIMOBC}) for details). These bounds have been derived by making use of the result of \cite{Gamal_fb_capacity_degraded_BC} which states that feedback can not improve the capacity region of the physically-degraded BC. However, using the idea developed here for proving bound $L_1$, it is possible to derive the outer-bound derived in \cite{Vaze-Varanasi-delay-MIMOBC} without making use of the result of \cite{Gamal_fb_capacity_degraded_BC}. 

\section{Proof of Theorem \ref{thm: outer-bound DoF 2-user IC d-CSI}: $L_4$ is an Outer-Bound} \label{sec: proof of thm: outer-bound DoF 2-user IC d-CSI L4}

We first outer-bound the capacity region of the given MIMO IC with delayed CSI by assuming that each receiver knows all channel matrices perfectly and instantaneously. Under this assumption, using Fano's inequality, we upper-bound the rates achievable for the two users as follows:
\begin{eqnarray}
n R_2 & = & \underbrace{ I \Big( \mathcal{M}_1, \mathcal{M}_2; \overline{Y}_2(n) \Big| \overline{H}(n) \Big) }_{=t_1}- \underbrace{I \Big( \mathcal{M}_1 ; \overline{Y}_2(n) \Big| \mathcal{M}_2, \overline{H}(n) \Big) }_{=t_2}+ ~ n \epsilon_n \mbox{ and}  \label{eq: bound on r2 d-CSI IC part2} \\
n R_1 & \leq & \underbrace{I \Big( \mathcal{M}_1; \overline{Y}_1(n) \Big| \overline{H}(n) \Big)}_{= t_3} ~ + ~ n \epsilon_n, \label{eq: bound on r1 d-CSI IC part2}
\end{eqnarray}
where $\epsilon_n \to 0$ as $n \to \infty$. Using these bounds, we get
\begin{eqnarray}
d_2 & \leq & N_2 - f(t_2) \label{eq: bound on d2 d-CSI IC part2} \\
d_1 & \leq & f(t_3) = f \Big\{ ~ I \Big( \mathcal{M}_1;\overline{Y}_1(n) \Big| \overline{H}(n) \Big) ~ \Big\},  \label{eq: bound on d1 d-CSI IC part2}
\end{eqnarray}
where we use equation (\ref{eq: bound on term 1 d-CSI IC outer-bound}). In the following, we obtain a tight upper-bound on $f(t_3)$ so that the desired bound $L_4$ can be derived.

To this end, define a unitary matrix $U_{12}(t)$ such that the last $N_1-M_2$ rows of $U_{12}(t)H_{12}(t)$ consist only of zeros. Such a unitary matrix can be obtained from the singular-value decomposition \cite{Horn-Johnson} of $H_{12}(t)$, and therefore, $U_{12}(t)$ is a deterministic function of $H_{12}(t)$. Define $Y_1'(t) = U_{12}(t) Y_1(t)$. Note that $X_2(t)$ affects only the first $M_2$ entries of $Y'(t)$. Since a unitary transformation can not affect the mutual information, we have
\begin{eqnarray}
\lefteqn{ \hspace{-0.9cm} d_1 \leq f \Big\{ I \Big( \mathcal{M}_1;\overline{Y}_1'(n) \Big| \overline{H}(n) \Big) \Big\} }\nonumber \\
&& {} \hspace{-1.1cm} = f \Big\{ I \Big( \mathcal{M}_1;\overline{Y}_{1[1:M_2]}'(n) \Big| \overline{Y}_{1[M_2+1:N_1]}'(n), \overline{H}(n) \Big) \Big\} +  f \Big\{ I \Big( \mathcal{M}_1;\overline{Y}_{1[M_2+1:N_1]}'(n) \Big| \overline{H}(n) \Big) \Big\}. \label{eq: bound on d1 advanced dCSI MIMO IC part2}
\end{eqnarray}
We upper-bound each of the two terms appearing above through the following two lemmas.

\begin{lemma} \label{lem: first part of Y1 dCSI IC part2}
We have
\begin{equation}
f \Big\{ I \Big( \mathcal{M}_1;\overline{Y}_{1[1:M_2]}'(n) \Big| \overline{Y}_{1[M_2+1:N_1]}'(n), \overline{H}(n) \Big) \Big\} \leq M_2 - d_2. \label{eq: lem: first part of Y1 dCSI IC part2}
\end{equation}
\end{lemma}
\begin{IEEEproof}
See Section \ref{subsec: proof of lem: first part of Y1 dCSI IC part2}.
\end{IEEEproof}

\begin{lemma} \label{lem: second part of Y1 dCSI IC part2}
We have
\begin{eqnarray}
\lefteqn{ \hspace{-1cm} f \Big\{ I \Big( \mathcal{M}_1;\overline{Y}_{1[M_2+1:N_1]}'(n) \Big| \overline{H}(n) \Big) \Big\} \leq f \Big\{ I \Big( \mathcal{M}_1;\overline{Y}_{1[M_2+1:N_1]}'(n) \Big| \mathcal{M}_2, \overline{H}(n) \Big) \Big\} } \label{eq: lem: second part of Y1 step1 dCSI IC part2}\\
&& {} = f \Big\{ I \Big( \mathcal{M}_1;\overline{Y}_{1[M_2+1:N_1]}(n) \Big| \mathcal{M}_2, \overline{H}(n) \Big) \Big\} \label{eq: lem: second part of Y1 step2 dCSI IC part2} \\
&& {} \leq f \Big\{ I \Big( \mathcal{M}_1;\overline{Y}_{1[M_2+1:N_1]}(n), \overline{Y}_2 \Big| \mathcal{M}_2, \overline{H}(n) \Big) \Big\} \label{eq: lem: second part of Y1 step3 dCSI IC part2} \\
&& {} \leq \frac{N_1 + N_2 - M_2}{N_2}  f \Big\{ I \Big( \mathcal{M}_1; \overline{Y}_2 \Big| \mathcal{M}_2, \overline{H}(n) \Big) \Big\} =  \frac{N_1 + N_2 - M_2}{N_2} f(t_2). \label{eq: lem: second part of Y1 step4 dCSI IC part2}
\end{eqnarray}
\end{lemma}
\begin{IEEEproof}
See Section \ref{subsec: proof of lem: second part of Y1 dCSI IC part2}.
\end{IEEEproof}

Using inequalities (\ref{eq: bound on d1 advanced dCSI MIMO IC part2}), (\ref{eq: lem: first part of Y1 dCSI IC part2}), and (\ref{eq: lem: second part of Y1 step4 dCSI IC part2}), we obtain
\[
d_1 \leq (M_2 - d_2) + \frac{N_1 + N_2 - M_2}{N_2} f(t_2) \Rightarrow f(t_2) \geq \frac{N_2}{N_1 + N_2 - M_2} (d_1 + d_2 - M_2).
\]
We substitute this lower-bound on $f(t_2)$ into equation (\ref{eq: bound on d2 d-CSI IC part2}), we get
\begin{eqnarray*}
\lefteqn{ \frac{d_2}{N_2} \leq 1 - \frac{1}{N_1 + N_2 - M_2} (d_1 + d_2 - M_2 ) } \\
&& {} \Rightarrow \frac{d_1}{N_1 + N_2 - M_2} + d_2 \left\{ \frac{1}{N_2} + \frac{1}{N_1 + N_2 - M_2} \right\} \leq 1 + \frac{M_2}{N_1+N_2-M_2} \\
&& {} \Rightarrow \frac{d_1}{N_1 + N_2 - M_2} + d_2 \frac{N_1 + 2N_2 - M_2}{N_2 ( N_1 + N_2 - M_2)} \leq \frac{N_1 + N_2}{N_1 + N_2 - M_2},
\end{eqnarray*}
which is the desired bound $L_4$.

\subsection{Proof of Lemma \ref{lem: first part of Y1 dCSI IC part2}} \label{subsec: proof of lem: first part of Y1 dCSI IC part2}

To prove this lemma, we make use of the techniques developed in \cite{Jafar-Maralle}. We first define to following quantities: Let the matrix formed by retaining the first $M_2$ rows of $U_{12}(t) H_{12}(t)$ be $Z(t)$ (it is thus $M_2 \times M_2$ in size); define
\[
\alpha(t) = \min \left\{ \frac{1}{\sigma_{\max}^2 [Z(t)]}, ~ \frac{1}{\sigma_{\max}^2 [ H_{22}(t)]} \right\},
\]
where $\sigma_{\max}[A]$ denotes the largest singular-value of $A$; let $W_b(t)$ be an $M_2$-dimensional noise vector that is distributed as
\[
W_b(t) \sim \mathcal{C}\mathcal{N} \left( 0, Z(t) \Big\{[Z(t)]^* Z(t) \Big\}^{-1} [Z(t)]^* - \alpha(t) Z(t) [Z(t)]^* \right)
\]
with its realizations being i.i.d. across time; and finally, set $Y_{1[1:M_2]}^{\ddagger}(t) =  Y_{1[1:M_2]}'(t) - W_b(t)$.

Now consider the following:
\begin{eqnarray}
\lefteqn{ \hspace{-1cm} f \Big\{ I \Big( \mathcal{M}_1;\overline{Y}_{1[1:M_2]}'(n) \Big| \overline{Y}_{1[M_2+1:N_1]}'(n), \overline{H}(n) \Big) \Big\} \leq f \Big\{ I \Big( \mathcal{M}_1;\overline{Y}^{\ddagger}_{1[1:M_2]}(n) \Big| \overline{Y}_{1[M_2+1:N_1]}'(n), \overline{H}(n) \Big) \Big\} } \nonumber \\
&& {} = f \Big\{ I \Big( \mathcal{M}_1, \mathcal{M}_2; \overline{Y}_{1[1:M_2]}^{\ddagger}(n) \Big| \overline{Y}_{1[M_2+1:N_1]}'(n), \overline{H}(n) \Big) \Big\}  \nonumber  \\
&& {} \hspace{5cm} -  f \Big\{ I \Big( \mathcal{M}_2; \overline{Y}_{1[1:M_2]}^{\ddagger}(n) \Big| \mathcal{M}_1,  \overline{Y}_{1[M_2+1:N_1]}'(n), \overline{H}(n) \Big) \Big\} \nonumber \\
&& {} \leq M_2 - f \Big\{ I \Big( \mathcal{M}_2;\overline{Y}_{1[1:M_2]}^{\ddagger}(n) \Big| \mathcal{M}_1, \overline{H}(n) \Big) \Big\}, \label{eq: penultimate step proof of lem: first part of Y1 dCSI IC part2}
\end{eqnarray}
where the first inequality holds because, within the partial order of positive semi-definite matrices, the covariance matrix of $W_b(t)$ is smaller than the identity matrix \cite{Jafar-Maralle}; while the last inequality follows from the following reasons: \begin{inparaenum}[(a)] \item the number of DoF of the point-to-point MIMO channel are limited by the number of receive antennas \cite{Telatar}, and \item conditioned on $\mathcal{M}_1$ and $\overline{H}(n)$, $X_1(n)$ is deterministic, which implies that $\mathcal{M}_2$ and $\overline{Y}_{1[1:M_2]}^{\ddagger}(n)$ are independent of $\overline{Y}_{1[M_2+1:N_1]}'(n)$, conditioned on $\mathcal{M}_1$ and $\overline{H}(n)$. \end{inparaenum}

Now, using the arguments developed in \cite{Jafar-Maralle} (see Steps 3-5 in the proof of Theorem 1 therein), it can be shown that
\[
f \Big\{ I \Big( \mathcal{M}_2;\overline{Y}_{1[1:M_2]}^{\ddagger}(n) \Big| \mathcal{M}_1, \overline{H}(n) \Big) \Big\} \geq f \Big\{ I \Big( \mathcal{M}_2; \overline{Y}_2(n) \Big| \mathcal{M}_1, \overline{H}(n) \Big) \Big\} \geq d_2 .
\]
Substituting the above lower-bound into (\ref{eq: penultimate step proof of lem: first part of Y1 dCSI IC part2}), we obtain the desired inequality.

\subsection{Proof of Lemma \ref{lem: second part of Y1 dCSI IC part2}} \label{subsec: proof of lem: second part of Y1 dCSI IC part2}

The first inequality in (\ref{eq: lem: second part of Y1 step1 dCSI IC part2}) follows by noting that $\mathcal{M}_1$ and $\mathcal{M}_2$ are independent.

To prove the next equality in (\ref{eq: lem: second part of Y1 step2 dCSI IC part2}), consider the following argument. Conditioned on $\mathcal{M}_2$ and $\overline{H}(n)$, $\overline{X}_2(n)$ is deterministic. Further, since translation does not change differential entropy \cite{CT}, conditioned on $\mathcal{M}_2$ and $\overline{H}(n)$, we may let $X_2(t) = 0$ $\forall t \leq n$. Thus, in the following, it may be assumed, without loss of generality, that
\begin{equation}
Y_1(t) = H_{11}(t) X_1(t) + W_1(t) \mbox{ and } Y_1'(t) = U_{12}(t) H_{11}(t) X_1(t) + U_{12}(t) W_1(t). \label{eq: third-last step proof of lem: second part of Y1 dCSI IC part2}
\end{equation}

Let $H_{11}'(t) = U_{12}(t) H_{11}(t)$. Since $U_{12}(t)$ is independent of $H_{11}(t)$ and $H_{11}(t)$ is i.i.d. Rayleigh faded, we have $H_{11}'(t) \sim H_{11}(t)$. For similar reasons, $U_{12}(t) W_1(t) \sim W_1(t)$. Now, let $H'(t)$ be a collection of channel matrices $H_{11}'(t)$, $H_{12}(t)$, $H_{21}(t)$, and $H_{22}(t)$. Then $H(t) \sim H'(t)$.

Since $U_{12}(t)$ is a deterministic function of $H_{12}(t)$, a one-to-one and onto mapping exists between $H(t)$ and $H'(t)$, which gives us the following equality:
\begin{equation}
f \Big\{ I \Big( \mathcal{M}_1;\overline{Y}_{1[M_2+1:N_1]}'(n) \Big| \mathcal{M}_2, \overline{H}(n) \Big) \Big\} = f \Big\{ I \Big( \mathcal{M}_1;\overline{Y}_{1[M_2+1:N_1]}'(n) \Big| \mathcal{M}_2, \overline{H}'(n) \Big) \Big\}. \label{eq: second-last step proof of lem: second part of Y1 dCSI IC part2}
\end{equation}
The quantity on the right hand side of the above equation is a function of the joint distribution of random variables
\[
\Big\{ \mathcal{M}_1,\mathcal{M}_2,\overline{H}'(n),\overline{X}_1(n), \overline{Y}_1'(n) \Big\},
\]
which, in fact, is equal to that of the random variables
\[
\Big\{ \mathcal{M}_1,\mathcal{M}_2,\overline{H}(n),\overline{X}_1(n), \overline{Y}_1(n) \Big\}
\]
because $H'(t) \sim H(t)$, $U_{12}(t) W_1(t) \sim W_1(t)$, and equation (\ref{eq: third-last step proof of lem: second part of Y1 dCSI IC part2}) holds. This equality of the joint distributions implies that
\begin{equation}
f \Big\{ I \Big( \mathcal{M}_1;\overline{Y}_{1[M_2+1:N_1]}'(n) \Big| \mathcal{M}_2, \overline{H}'(n) \Big) \Big\} = f \Big\{ I \Big( \mathcal{M}_1; \overline{Y}_{1[M_2+1:N_1]}(n) \Big| \mathcal{M}_2, \overline{H}(n) \Big) \Big\}. \label{eq: last step proof of lem: second part of Y1 dCSI IC part2}
\end{equation}
The desired equality in (\ref{eq: lem: second part of Y1 step2 dCSI IC part2}) can now be obtained by combining equations (\ref{eq: second-last step proof of lem: second part of Y1 dCSI IC part2}) and (\ref{eq: last step proof of lem: second part of Y1 dCSI IC part2}).

The inequality (\ref{eq: lem: second part of Y1 step3 dCSI IC part2}) holds trivially. The next inequality (\ref{eq: lem: second part of Y1 step4 dCSI IC part2}) is a simple application of Lemma \ref{cor: main inequality d-CSI IC outer-bound}, and the final equality follows from the definition of $t_2$.

\section{Proof of Theorem \ref{thm: MIMO_2-user IC_d-CSI_inner-bound DoF}: Cases 0 and A.I} \label{sec: proof of thm: MIMO_2-user IC_d-CSI_inner-bound DoF_Case A.I}

Consider first Case 0 for which we have the following lemma.
\begin{lemma} \label{lem: Case 0 d-CSI 2user IC}
Under Case 0, $\mathbf{D}^{\rm{p-CSI}} = \mathbf{D}^{\rm{d-CSI}}$, which implies that bound $L_3$ is active.
\end{lemma}
\begin{IEEEproof}
Under Case 0, the inequality $N_1 \geq N_2 \geq M_1$ is true (recall $N_1 \geq N_2$ throughout). Then by \cite[Remark 17]{Vaze_Dof_final}, we have $\mathbf{D}^{\rm{no-CSI}} = \mathbf{D}^{\rm{p-CSI}}$, which yields $
\mathbf{D}^{\rm{no-CSI}} = \mathbf{D}^{\rm{d-CSI}} = \mathbf{D}^{\rm{p-CSI}} = \mathbf{D}_{\rm{outer}}^{\rm{d-CSI}}$. This also implies that $L_3$ is active.
\end{IEEEproof}

The remainder of this section will now deal with Case A.I, for which the defining inequalities can be stated as
\[
M_2 \geq N_1 \geq N_2 \mbox{ and } M_1>N_2,
\]
where we account for $N_1 \geq N_2$, the inequalities $M_1 > N_2$ and $M_2 \geq N_2$ that hold with Case A, and the one that defines Case A.I. The following lemma proves that the proposed outer-bound is tight and also helps determine its shape.

\begin{lemma} \label{lem: Case A.I d-CSI 2user IC}
Under Case A.I, bounds $L_{\{1,2\}}$ are active and the outer-bound $\mathbf{D}_{\rm{outer}}^{\rm{d-CSI}}$ is achievable. More specifically, under Case A.I,
\begin{enumerate}[1.]
\item if $M_1 \leq N_1$, then $L_1$ is active and $\mathbf{D}^{\rm{no-CSI}} = \mathbf{D}^{\rm{d-CSI}} \subset \mathbf{D}^{\rm{p-CSI}}$;
\item if $M_1 > N_1$ and $N_2 = M_2$, then $L_2$ is active and $\mathbf{D}^{\rm{no-CSI}} = \mathbf{D}^{\rm{d-CSI}} = \mathbf{D}^{\rm{p-CSI}}$;
\item if $M_1 > N_1$ and $N_2 < M_2$ then $L_{\{1,2\}}$ is active and $\mathbf{D}^{\rm{no-CSI}} \subset \mathbf{D}^{\rm{d-CSI}} \subset \mathbf{D}^{\rm{p-CSI}}$.
\end{enumerate}
\end{lemma}
\begin{IEEEproof}
Consider the region $\mathbf{D}_{\rm{outer},1}^{\rm{d-CSI}}$ obtained from the region $\mathbf{D}_{\rm{outer}}^{\rm{d-CSI}}$ by simply ignoring the bound $L_3$. Clearly, $\mathbf{D}^{\rm{d-CSI}} \subseteq \mathbf{D}_{\rm{outer}}^{\rm{d-CSI}} \subseteq \mathbf{D}_{\rm{outer},1}^{\rm{d-CSI}}$. Hence, to prove this lemma, it is sufficient to establish the achievability of $\mathbf{D}_{\rm{outer},1}^{\rm{d-CSI}}$, which is the goal of this proof.

Under Case A.I, bounds $L_1$ and $L_2$ are given by
\[
L_1 \equiv \frac{d_1}{\min(N_1+N_2,M_1)} + \frac{d_2}{N_2} \leq 1 \mbox{ and } L_2 \equiv \frac{d_1}{N_1} + \frac{d_2}{\min(N_1 + N_2, M_2)} \leq 1, \mbox{ respectively}.
\]
We can now have three different scenarios, depending upon whether only one of $L_1$ and $L_2$ is active or both are active:
\begin{itemize}
\item \underline{Case A.I.1: $M_1 \leq N_1 \Rightarrow M_2 \geq N_1 \geq M_1 >N_2$:} \newline Since $\min(M_1,N_1+N_2) = M_1 \leq N_1$ and $N_2 \leq \min(M_2,N_1+N_2)$, $L_1$ implies $L_2$ (i.e., if $L_1$ is true, $L_2$ is also true), or $L_1$ is active. Moreover, it can be verified that $\mathbf{D}_{\rm{outer},1}^{\rm{d-CSI}} = \mathbf{D}^{\rm{no-CSI}}$ (see Remark \ref{rem: No and perfect CSI DoF regions}), which implies the achievability of $\mathbf{D}_{\rm{outer},1}^{\rm{d-CSI}}$ and the fact that $\mathbf{D}^{\rm{no-CSI}} = \mathbf{D}^{\rm{d-CSI}}$. Hence, the lemma holds in this case.
\item \underline{Case A.I.2: $\{M_1 > N_1 \mbox{ and } N_2 = M_2\} \Rightarrow M_1> N_1 = M_2 = N_2$:} \newline Since $N_2 = M_2$ and $M_2 \geq N_1 \geq N_2$, we have the inequality $M_1> N_1 = M_2 = N_2$. Since $\min(M_1,N_1+N_2) > N_1$ and $N_2 = M_2$, $L_2$ is active. It can be verified that in this case, the region $\mathbf{D}_{\rm{outer},1}^{\rm{d-CSI}}$ with $L_2$ active  coincides with $\mathbf{D}^{\rm{no-CSI}}$ as well as $\mathbf{D}^{\rm{p-CSI}}$. Hence, we get $\mathbf{D}^{\rm{no-CSI}} = \mathbf{D}^{\rm{d-CSI}} = \mathbf{D}^{\rm{p-CSI}}$ (see also \cite[Remark 17]{Vaze_Dof_final}).
\item \underline{Case A.I.3: $\{M_1 > N_1 \mbox{ and } N_2 < M_2\}$:} \newline In this case, we have
    \[
    M_1 > N_1, N_2; \mbox{ and } M_2 \geq N_1 \geq N_2 \mbox{ with either } M_2 \not= N_1 \mbox{ or } N_1 \not= N_2.
    \]
    Here, both the bounds $L_{\{1,2\}}$ are (strictly) active because $\min(N_1+N_2,M_1) > N_1$ but $N_2 < \min(M_2,N_1+N_2)$, and hence no bound can imply the other. The typical shape of $\mathbf{D}_{\rm{outer},1}^{\rm{d-CSI}}$ is as shown in Fig. \ref{fig: typical shape d-CSI 2user IC Case A}(c), whence we deduce that if the point $P_{1,2}$ (i.e., the point of intersection of $L_1$ and $L_2$) is known to be achievable, the entire outer-bound $\mathbf{D}_{\rm{outer},1}^{\rm{d-CSI}}$ can be achieved via time sharing. Thus, to prove the lemma, it is sufficient to show the achievability of point $P_{1,2}$.
\end{itemize}

\emph{\underline{An achievability scheme for point $P_{1,2}$:} }

Point $P_{1,2}$ is given by
\[
P_{1,2} \equiv \left( \frac{N_1 \cdot M_1' \cdot (M_2'-N_2)}{N_1(M_2'-N_2) + M_2' ( M_1' - N_1)}, ~ \frac{N_2 \cdot M_2' \cdot (M_1'-N_1)}{N_1(M_2'-N_2) + M_2' ( M_1' - N_1)} \right),
\]
where\footnote{In this paper, the some variables like $M_1'$, $M_2'$, $t'$, etc. have been reused in various sections with different definitions. In each section, follows only the respective definition.} $M_i' = \min(M_i,N_1+N_2)$ for $i=1,2$. It will be shown that over $N_1(M_2'-N_2) + M_2' ( M_1' - N_1)$ time slots, we can achieve
\[
N_1 \cdot M_1' \cdot (M_2'-N_2) ~ \mbox{ and } ~ N_2 \cdot M_2' \cdot (M_1'-N_1)
\]
DoF for the two users, respectively. The achievability scheme consists of three phases.

\underline{Phase One:} This comprises of the initial $t_1 = N_1(M_2'-N_2)$ time slots over which T2 remains silent. On the other hand, T1 transmits $M_1'$ i.i.d. complex Gaussian data symbols intended for R1 in each time slot; and thus, a total of $N_1 M_1' (M_2'-N_2)$ data symbols are sent to R1. We denote these symbols by $\{u_{1i}(j)\}$, where $i \in [1: M_1']$, $j \in [1:t_1]$, and $u_{1i}(j)$ are i.i.d. (across $i$ and $j$) according to $\mathcal{C}\mathcal{N}\Big( 0,\frac{P}{(N_1+N_2)^2} \Big)$ distribution. The signal received by R1 is given by
\[
Y_1(t) = \tilde{H}_{11}^{[M_1']}(t) \begin{bmatrix} u_{11}^*(t) & u_{12}^*(t) & \cdots & u_{1M_1'}^*(t) \end{bmatrix}^* + W_1(t), ~ ~ \forall ~ t \in [1: t_1],
\]
where $\tilde{H}_{ij}^{[x]}(t)$ denotes the matrix obtained obtained from $H_{ij}(t)$ by retaining only the first $x$ columns of it. Thus, for a given $t \in [1:t_1]$, R1 observes $N_1$ linear combinations (LCs) of input data symbols; and these LCs are almost surely linearly independent because $M_1' > N_1$ and the Rayleigh-faded channel matrices are almost surely full rank. Since the receiver needs to have one interference-free LC per data symbol to be decoded (provided all LCs are linearly independent of each other) for being able to decode the desired symbols, R1 needs to receive $M_1'-N_1$ extra LCs of $\{u_{1i}(t)\}_i$; and as we would see soon, these LCs are present at R2.

R2, over a given time slot of this phase, observes $N_2$ LCs of data symbols sent to R1. More importantly, any $M_1'-N_1 (< N_2)$ LCs out of these would almost surely be linearly independent of each other and also of $N_1$ LCs observed at that time by R1 (again, because of full-rank property of Rayleigh-faded matrices). Therefore, if R1 is given the signal received at (say) the first $M_1'-N_1$ antennas of R2 over the duration of this phase, it can decode the all desired data symbols. We show that this can indeed be accomplished by making use of delayed CSI available to all terminals. With this motivation, let us focus on the signal received by R2 at its first $M_1'-N_1$ antennas. At time $t \in [1:t_1]$, it is given by
\begin{eqnarray*}
Y_{2[1:M_1'-N_1]}(t) & = & I_{2[1:M_1'-N_1]}(t) + W_{2[1:M_1'-N_1]}(t), \mbox{ with}\\
I_{2j}(t) & = & \tilde{H}_{2j1}^{[M_1']}(t) \begin{bmatrix} u_{11}^*(t) & u_{12}^*(t) & \cdots & u_{1M_1'}^*(t) \end{bmatrix}^* ~ \forall j \in [1:M_1'-N_1],
\end{eqnarray*}
where $\tilde{H}_{2j1}^{[M_1']}(t)$ is a row vector obtained from another row vector $H_{2j1}(t)$ by retaining only its first $M_1'$ entries (see also Fig. \ref{fig: notation_used} for notation). Here, $\{I_{2j}(t)\}$ is a LC of $\{u_{1i}(t)\}_i$ whose coefficients are decided by the row vector $\tilde{H}_{2j1}^{[M_1']}(t)$. If suppose R1 knows the LCs $I_{2[1:M_1'-N_1]}(t)$ as well as the channel matrix $H_{2[1:M_1'-N_1]}^{[M_1']}(t)$, then via channel inversion it can almost surely compute
\begin{equation}
\begin{bmatrix} H_{11}^{[M_1']}(t) \\ H_{2[1:M_1'-N_1]}^{[M_1']}(t) \end{bmatrix}^{-1} \begin{bmatrix} Y_1(t) \\ I_{2[1:M_1'-N_1]}(t) \end{bmatrix} = X_{1[1:M_1']}(t) + (\rm{noise ~ terms}) \label{eq: channel inversion dCSI 2user IC Case A.I.3}
\end{equation}
to recover the input symbols $\{u_{1i}(t)\}_{i=1}^{M_1'}$ (where the noise terms have been ignored since they can not affect the DoF result). Now, R1 by the virtue of delayed CSI, knows the channel matrices $\{H_{21}(t)\}_{t=1}^{t_1}$ at the end of this phase. Hence, it is sufficient for it to learn the values of the LCs $\{I_{2j}(t)\}_{j,t}$. Moreover, at the end of Phase One, T1 knows perfectly the values of LCs $\{I_{2j}(t)\}_{j,t}$.

The next phase is analogous to this one during which T2 transmits the $N_2 M_2' (M_1'-N_1)$ number of data symbols to R2. R2, just like the case of R1, does not observe a sufficient number of LCs required to decode the desired data symbols and requires some of the LCs observed by R1. Over the third or the last phase, the transmitters signal such that both the receivers learn the remaining set of LCs that they need to. During the last phase, interference alignment is achieved by making use of the fact that the LCs that are required by the $i^{th}$ receiver, $i=1,2$, have already been observed by the $j^{th}$ receiver, $j \not= i$; and hence, when these LCs are being delivered to the $i^{th}$ receiver, they do not cause any additional interference to the other receiver. See also Remark \ref{rem: how IA is achieved dCSI 2user IC}.

\underline{Phase Two:} Over this phase of duration $t_2 = M_2' ( M_1' - N_1)$ time slots, T1 remains silent, while T2 is transmitting $M_2'$ i.i.d. complex Gaussian symbols, per time slot, intended for R2. Let these symbols be $\{u_{2i}(j)\}$, where $i \in [1: M_2']$, $j \in [1:t_2]$, and $u_{2i}(j)$ are i.i.d. (across $i$ and $j$) according to $\mathcal{C}\mathcal{N} \Big(0, \frac{P}{(N_1+N_2)^2} \Big)$ distribution. Then, if $t' = t-t_1$, the signal received by R2 is given by
\[
Y_2(t) = \tilde{H}_{22}^{[M_2']}(t) \begin{bmatrix} u_{21}^*(t') & u_{22}^*(t') & \cdots & u_{2M_2'}^*(t') \end{bmatrix}^* + W_2(t) ~ ~ ~ \forall ~ t \in [t_1+1: t_1+t_2].
\]
Thus, to decode symbols $\{u_{2i}(t)\}_{i=1}^{M_2'}$, R2 needs another $(M_2'-N_2)$ LCs of them\footnote{Throughout, whenever we say that a certain receiver needs few more LCs for decoding the desired data symbols, we always mean that these new LCs are linearly independent of each other and also of the LCs this receiver has already observed.}. As argued before, these LCs are present at the first $M_2'-N_2$ antennas of R1, where the signal received at time $t \in [t_1+1:t_1+t_2]$ is given by
\begin{eqnarray*}
Y_{1[1:M_2'-N_2]}(t) & = & I_{1[1:M_2'-N_2]}(t') + W_{2[1:M_2'-N_2]}(t), \mbox{ where } t' = t-t_1 \mbox{ and} \\
I_{1j}(t') & = & H_{1j2}^{[M_2']}(t) \begin{bmatrix} u_{21}^*(t') & u_{22}^*(t') & \cdots & u_{2M_2'}^*(t') \end{bmatrix}^* ~ \forall j \in [1:M_2'-N_2].
\end{eqnarray*}
As in the case of R1, R2 can decode its desired data symbols, if the values of LCs $\{I_{ij}(t')\}$ for $j \in [1: M_2'-N_2]$ and $t' \in [1:t_2]$ can be conveyed to it because at the end of this phase, R2 will know the channel-dependent coefficients that produce the LCs $\{I_{ij}(t')\}_{j,t'}$ from the input symbols.

\underline{Phase Three:} This is the last phase and takes the remaining
\[
N_1(M_2'-N_2) + M_2' ( M_1' - N_1) - t_1 - t_2 = (M_2'-N_2) (M_1'-N_1) = t_3
\]
time slots. Over this phase, T1 and T2 signal such that the values of $N_1 t_3$ LCs, $\big\{ \{ I_{2j}(t) \}_{j=1}^{M_1'-N_1} \big\}_{t=1}^{t_1}$, are communicated to R1, while those of $N_2 t_3$ LCs, $\big\{ \{I_{1i}(t')\}_{i=1}^{M_2'-N_1} \big\}_{t'=1}^{t_2}$, are delivered to R2. Once this is accomplished, each receiver gets the one LC per desired data symbols (and all these LCs would be linearly independent) and hence can recover the input symbols.

First, let us partition the total of $N_1 t_3$ LCs, $\{I_{2j}(t)\}_{j,t}$, into $t_3$ disjoint subsets of cardinality $N_1$ each\footnote{Treat here the LC $I_{2j}(t)$ as a random variable and $\{I_{2j}(t)\}_{j,t}$ as a set of $N_1 t_3$ random variables. This step partitions the set $\{I_{2j}(t)\}_{j,t}$ of random variables into $t_3$ disjoint subsets.}. After partitioning, let us relabel these LCs as $\big\{ I^{[2]}_j(k) \big\}$, where $j \in [1:N_1]$ and $k \in [1: t_3]$. Similarly, partition $N_2 t_3$ LCs, $\{I_{1i}(t')\}_{i,t'}$, into $t_3$ subsets of cardinality $N_2$ each, and after partitioning, let us relabel them as $\big\{ I^{[1]}_i(k) \big\}$, where $i \in [1: N_2]$ and $k \in [1: t_3]$. This procedure of partitioning the set of LCs is deterministic and is known to all terminals.

At any time $t \in [t_1+t_2+1: t_1+t_2+t_3]$ and $\bar{t} = t- t_1-t_2$, T1 and T2 transmit the LCs belonging to the set $\{ I_j^{[2]}(\bar{t}) \}_j$ and $\{ I_i^{[1]}(\bar{t}) \}_i$, respectively. Thus, the transmit signals are formed as follows:
\[
X_1(t) = \begin{bmatrix} I^{[2]}_1(\bar{t}) \\ I^{[2]}_2(\bar{t}) \\ \vdots \\ I^{[2]}_{N_1}(\bar{t}) \\ 0_{M_1-N_1} \end{bmatrix} \mbox{ and } X_2(t) = \begin{bmatrix} I^{[1]}_1(\bar{t}) \\ I^{[1]}_2(\bar{t}) \\ \vdots \\ I^{[1]}_{N_2}(\bar{t}) \\ 0_{M_2-N_2} \end{bmatrix},
\]
where $0_x$ denotes the all-zero column vector of size $x$.

Consider now the decoding procedure at R1. The signal received by it is given by
\[
Y_1(t) = H_{11}^{[N_1]}(t) \begin{bmatrix} I^{[2]}_1(\bar{t}) \\ I^{[2]}_2(\bar{t}) \\ \vdots \\ I^{[2]}_{N_1}(\bar{t}) \end{bmatrix} + H_{12}^{[N_2]}(t) \begin{bmatrix} I^{[1]}_1(\bar{t}) \\ I^{[1]}_2(\bar{t}) \\ \vdots \\ I^{[1]}_{N_2}(\bar{t}) \end{bmatrix} + W_1(t), ~ \forall t \in [t_1+t_2+1: t_1+t_2+t_3].
\]
Since R1 knows LCs $\big\{I^{[1]}_i(k) \big\}$ $\forall$ $i,k$ as well as the channel matrices $H_{12}(t)$, it can compute\footnote{R1 knows noisy versions of LCs $\big\{I^{[1]}_i(k) \big\}$. However, the presence or absence of noise does not alter a DoF result. It is in this sense that we say that R1 knows these LCs.}
\begin{equation}
Y_1'(t) = Y_1(t) - H_{12}^{[N_2]}(t) \begin{bmatrix} I^{[1]}_1(\bar{t}) \\ I^{[1]}_2(\bar{t}) \\ \vdots \\ I^{[1]}_{N_2}(\bar{t}) \end{bmatrix} =  H_{11}^{[N_1]}(t) \begin{bmatrix} I^{[2]}_1(\bar{t}) \\ I^{[2]}_2(\bar{t}) \\ \vdots \\ I^{[2]}_{N_1}(\bar{t}) \end{bmatrix} + W_1(t) ~ \forall \bar{t} \in \in [1:t_3] \label{eq: subtracting contribution dCSI 2user IC Case A.I.3}
\end{equation}
by subtracting the contribution due to $\big\{I^{[1]}_i(\bar{t}) \big\}_{i=1}^{N_2}$. Subsequently, by inverting channel matrix $H_{11}^{[N_1]}(t)$ (which can be done with probability $1$), it can determine the values of LCs $\big\{I^{[2]}_j(\bar{t}) \big\}_{j=1}^{N_1}$. Hence, as per the arguments developed earlier, R1 can decode all input symbols. The operation of R2 is similar.
\end{IEEEproof}

\begin{remark}[How is interference alignment achieved?] \label{rem: how IA is achieved dCSI 2user IC}
The signal transmitted by T1 (T2) during this phase, although it is not useful for R2 (R1), does not cause interference to R2 (R1). This is because the symbols, that T1 (T2) is supposed to transmit over this phase, have already been observed by R2 (R1), and hence, their retransmission does not cause any additional interference. In other words, the interference T1 (T2) causes at R2 (R1) over this phase gets `aligned' with the interference it has already caused to R2 (R1) over the first (second) phase. This technique, while it does not produce additional interference at R2 (R1), allows R1 (R2) to learn the remaining set of LCs that it needs to.
\end{remark}

\section{Proof of Theorem \ref{thm: MIMO_2-user IC_d-CSI_inner-bound DoF}: Case A.II} \label{sec: proof of thm: MIMO_2-user IC_d-CSI_inner-bound DoF_Case A.II}

In this case $N_1 \not= N_2$, because $N_1 = N_2 \Rightarrow M_2 < N_2 = N_1$, which contradicts with the definition of Case A. Thus, for Case A.II, the defining inequalities are
\[
M_1,N_1 > N_2 ~ \mbox{ and } ~ N_1> M_2 \geq N_2.
\]
The following lemma shows that the outer-bound is tight.

\begin{lemma} \label{lem: Case A.II d-CSI 2user IC}
Under Case A.II, bounds $L_{\{1,3\}}$ are active and the outer-bound $\mathbf{D}_{\rm{outer}}^{\rm{d-CSI}}$ is achievable. Moreover, in this case, if
\begin{enumerate}[1.]
\item $N_1 \geq M_1$, then $L_1$ is active and $\mathbf{D}^{\rm{no-CSI}} = \mathbf{D}^{\rm{d-CSI}} \subset \mathbf{D}^{\rm{p-CSI}}$;
\item $N_1 < M_1$, then $L_{\{1,3\}}$ is active and $\mathbf{D}^{\rm{no-CSI}} \subset \mathbf{D}^{\rm{d-CSI}} \subset \mathbf{D}^{\rm{p-CSI}}$.
    \end{enumerate}
\end{lemma}
\begin{IEEEproof}
Note that $L_2$ is implied by $L_3$, because for this case
\[
L_2 \equiv d_1 + d_2 \leq \min(N_1, M_1 + M_2) \mbox{ and } L_3 \equiv d_1 + d_2 \leq \min(N_1,M_1)
\]
with $\min(N_1, M_1 + M_2) \geq \min(N_1,M_1)$, which proves the first part of the lemma. For an easy reference, we restate bound $L_1$:
\[
L_1 \equiv \frac{d_1}{\min(M_1,N_1+N_2)} + \frac{d_2}{N_2} \leq 1.
\]
Again, depending upon whether $L_1$ is active and/or $L_3$ is active, we have two sub-cases:
\begin{itemize}
\item \underline{Case A.II.1: $N_1 \geq M_1 \Rightarrow N_1 \geq M_1 > N_2 \mbox{ and } N_1 > M_2 \geq N_2$:} \newline Here, $L_1 \equiv \frac{d_1}{M_1} + \frac{d_2}{N_2} \leq 1$ and $L_3 \equiv d_1 + d_2 \leq M_1$. Since $N_2 < M_1$, $L_1$ implies $L_3$, and thus $L_1$ is active. It can be verified that in this case, the outer-bound $\mathbf{D}_{\rm{outer}}^{\rm{d-CSI}}$ with $L_1$ active coincides with $\mathbf{D}^{\rm{no-CSI}}$, which yields us $\mathbf{D}^{\rm{no-CSI}} = \mathbf{D}^{\rm{d-CSI}}$ and hence the lemma.
\item \underline{Case A.II.2: $N_1 < M_1 \Rightarrow M_1 > N_1 > M_2 \geq N_2$:} \newline Here, both the bounds $L_{\{1,3\}}$ are strictly active. The typical shape of the outer-bound is as shown in Fig. \ref{fig: typical shape d-CSI 2user IC Case A}(e) from which we see that the achievability of point $P_{1,3}$ (the point of intersection of bounds $L_1$ and $L_3$) is sufficient to establish the achievability of outer-bound $\mathbf{D}_{\rm{outer}}^{\rm{d-CSI}}$. In the following, we develop a scheme to achieve point $P_{1,3}$.
\end{itemize}

\emph{\underline{An achievability scheme for point $P_{1,3}$:} }

Here,
\[
P_{1,3}\equiv \left( \frac{M_1' (N_1-N_2)}{M_1' - N_2}, ~ \frac{N_2(M_1' - N_1)}{M_1' -N_2} \right),
\]
where $M_1'=\min(M_1,N_1+N_2)$. In the following achievability scheme, we make use of only $M_1'$ antennas at T1, and thus it is assumed henceforth in this section that $M_1 = M_1'$. The basic idea behind the following achievability scheme, although it consists of two phases as opposed to three, is similar to the one presented in the last section, and therefore, to avoid repetition, we just present an outline.

It will be shown that over $M_1-N_2$ time slots, $M_1 (N_1-N_2)$ and $N_2(M_1 - N_1)$ DoF can be achieved for the two users, respectively (recall the assumption of $M_1 = M_1'$).

\underline{Phase One:} Over this phase, which comprises of $t_1 = (N_1-N_2)$ time slots, T1 transmits $M_1(N_1-N_2)$ i.i.d. complex Gaussian data symbols $\{u_{1i}(j)\}$, where $i \in [1: M_1]$ and $j \in [1:N_1-N_2]$, intended for R1 whereas T2 remains silent. At a given time $t \in [1: N_1-N_2]$, R1 receives $N_1$ LCs of input symbols $\{u_{1i}(t)\}_{i=1}^{M_1}$; and therefore, to decode them, it needs extra $M_1-N_1$ LCs of them, which, as one may expect, are observed by R2 at the first $M_1-N_1$ antennas (recall $M_1' = M_1 \leq N_1 + N_2$). Mathematically, the LCs observed by R2 at its $j^{th}$ antenna is given by
\[
I_{2j}(t) = H_{2j1}(t) \begin{bmatrix} u_{11}^*(t) & u_{12}^*(t) & \cdots & u_{1M_1}^*(t) \end{bmatrix}^* ~ \forall t \in [1:t_1],
\]
and if R1 knows the values of $I_{2[1:M_1-N_1]}(t)$ and the channel matrix $H_{21}(t)$, it can decode the data symbols $\{u_{1i}(t)\}_i$ (c.f. equation (\ref{eq: channel inversion dCSI 2user IC Case A.I.3})). Moreover, R1 knows the all channel vectors $\{H_{2[1:M_1-N_1]1}(t)\}_{t=1}^{t_1}$ at the end of this phase.

Since T1 has already transmitted the required number of data symbols intended for R1, its goal over the next phase is to provide R1 the values of LCs $\{I_{2j}(t)\}_{j,t}$, which it knows at the end of this phase. As opposed to it, T2 is yet to transmit and thus, sends all $N_2(M_1 - N_1)$ number of data symbols intended for R2 over the next phase.

\underline{Phase Two:} This is the last phase and occupies
\[
(M_1-N_2) - t_1 = (M_1-N_2) - (N_1 - N_2) = (M_1 - N_1) = t_2
\]
time slots. For a given $t \in [t_1 + 1: t_1 + t_2]$ and $t' = t-t_1$, the transmit signals are formed as follows:
\begin{eqnarray*}
X_1(t) = \begin{bmatrix} I_{2t'}(1) \\ I_{2t'}(2) \\ \vdots \\ I_{2t'}(t_1) \\ 0_{M_1-t_1} \end{bmatrix} \mbox{ and }
X_2(t)  =  \begin{bmatrix} u_{21}(t') \\ u_{22}(t') \\ \vdots \\ u_{2N_2}(t') \\ 0_{M_2-N_2}  \end{bmatrix},
\end{eqnarray*}
where $\{u_{2j}(t')\}$ with $j \in [1:N_2]$ and $t' \in [1:t_2]$ are i.i.d. complex Gaussian data symbols intended for R2. Since $t_1 = N_1-N_2$, exactly $N_1$ transmit antennas are in use (i.e., they are transmitting a non-zero signal while the rest of set to zero) at any given time during this phase.

Since R2 knows interfering symbols $\{I_{2t'}(j)\}$ for $\forall$ $j,t'$, it can subtract their contribution from its received signal (c.f. equation (\ref{eq: subtracting contribution dCSI 2user IC Case A.I.3})) and then via channel inversion can recover the desired data symbols (c.f. (\ref{eq: channel inversion dCSI 2user IC Case A.I.3})).

Further, since exactly $N_1$ antennas are in use during this phase, R1 can zero-force the interference coming from T2 to recover the signal sent by T1. Specifically, it can simply ignore the last $M_1-t_1$ and $M_2-N_2$ antennas of T1 and T2, respectively (because these are set to zero), and then can almost surely invert the square channel matrix $\begin{bmatrix} H_{11}^{[t_1]}(t) \\ H_{12}^{[N_2]}(t) \end{bmatrix}$ to itself from the first $t_1$ and the first $N_2$ transmit antennas of T1 and T2, respectively,  to recover the LCs $I_{2t'}([1:N_2])$ (and also $\{u_{2[1:N_2]}(t')\}$). Thus, at the end of this phase, it gets one LC per desired data symbol and hence can decode all of them.
\end{IEEEproof}

\section{Proof of Theorem \ref{thm: MIMO_2-user IC_d-CSI_inner-bound DoF}: Cases B.0 and B.I}
\label{sec: proof of thm: MIMO_2-user IC_d-CSI_inner-bound DoF_Case B.I}

For Case B.0, we have the following lemma.
\begin{lemma} \label{lem: Case B.0 d-CSI 2user IC}
Under Case B.0, $\mathbf{D}^{\rm{d-CSI}} = \mathbf{D}^{\rm{p-CSI}}$, which implies that bound $L_3$ is active.
\end{lemma}
\begin{IEEEproof}
Under Case B.0, the inequality $N_1 = N_2 > M_2$ holds. The lemma now follows from Remark 17 of \cite{Vaze_Dof_final}.
\end{IEEEproof}

Henceforth, in this section, we study Case B.I, for which the defining inequality is
\[
N_1 \geq M_1 > N_2 > M_2
\]
and we have the following lemma.

\begin{lemma} \label{lem: Case B.I d-CSI 2user IC}
Under Case B.I, bound $L_1$ is active and the outer-bound $\mathbf{D}_{\rm{outer}}^{\rm{d-CSI}}$ is achievable.
\end{lemma}
\begin{IEEEproof}
Here, the three outer-bounds are given as
\[
L_1 \equiv \frac{d_1}{M_1} + \frac{d_2}{N_2} \leq 1, ~ L_2 \equiv d_1 + d_2 \leq \min(N_1,M_1+M_2), ~ \mbox{ and } L_3 \equiv d_1 + d_2 \leq M_1.
\]
It can be easily verified that $L_3$ implies $L_2$ and since $N_2 < M_1$, $L_1$ implies $L_3$. Hence, bound $L_1$ is active.

The typical shape of the outer-bound is shown in Fig. \ref{fig: typical shape d-CSI 2user IC Case B}(a). It is clear that if the point
\[
P_{o2,1} \equiv \left( \frac{M_1}{N_2} (N_2-M_2), ~ M_2 \right),
\]
which is the point of intersection of the single-user bound on $d_2$ and the bound $L_1$, is known to be achievable, then the entire outer-bound can be achieved via time sharing.

\emph{\underline{A scheme to achieve point $P_{o2,1}$:} }

Over $N_2$ time slots, we show the achievability of the DoF tuple $\big( M_1(N_2-M_2), N_2 M_2 \big)$. In this scheme, R1 makes use of only $M_1$ receive antennas (even when more are available), and thus henceforth, in this section, it is assumed that $N_1 = M_1$. In this scheme, T2 at all times needs to transmit $M_2$ i.i.d. complex Gaussian data symbols intended for R2. 

Define unitary matrices $U_{12}(t)$ and $U_{22}(t)$ such that last $N_1-M_2$ and $N_2-M_2$ rows of $U_{12}(t) H_{12}(t)$ and $U_{22}(t) H_{22}(t)$, respectively, consist only of zeros. Note that these unitary matrices are functions of the channel matrices, and hence, any terminal can compute these as soon as it knows the channel matrices. Set $Y_i'(t) = U_{i2}(t) Y_i(t)$, $i=1,2$. At each time, the $i^{th}$ receiver computes $Y_i'(t)$ and uses the signal $Y_i'(t)$, instead of $Y_i(t)$, for all further decoding purposes. Note that the transmit signal $X_2(t)$ affects only the first $M_2$ entries of $Y_i'(t)$ $\forall$ $i, ~t$.

Since T2 is always transmitting $M_2$ data symbols, R1, along the first $M_2$ entries of $Y_1'(t)$, receives interference. In the following scheme, R1 simply discards the first $M_2$ entries of $Y_1'(t)$ and retains only the remaining $M_1-M_2$ entries (recall $N_1 = M_1$).

The achievability scheme, as described below, consists of two phases.

\underline{Phase One:} It lasts for $t_1 = N_2-M_2$ time slots. At time $t \in [1: t_1]$, T1 transmits $M_1$ i.i.d. complex Gaussian data symbols, $\{u_{1i}(t)\}_{i=1}^{M_1}$, intended for R1. For a given $t \in [1:t_1]$, R1 receives $M_1-M_2$ LCs of the input data symbols sent to it by T1, and requires another $M_2$ LCs of them.

Consider now the case of R2. It receives the desired data symbols along the first $M_2$ entries of $Y_2'(t)$. However, since $M_1 > N_2 > M_2$, these entries are corrupted by interference due to the signal coming from T1. More specifically, the signal received by R2 during this phase is given by
\begin{eqnarray}
Y_2'(t) = U_{22}(t) H_{22}(t) \begin{bmatrix} u_{21}^*(t) & u_{22}^*(t) & \cdots & u_{2M_2}^*(t) \end{bmatrix}^* + I_2(t) + W_2(t) \nonumber
\end{eqnarray}
where
\begin{eqnarray}
I_2(t) & = & U_{22}(t) H_{21}(t) \begin{bmatrix} u_{11}^*(t) & u_{12}^*(t) & \cdots & u_{1M_1}^*(t) \end{bmatrix}^* . \label{eq: defn interfering symbols dCSI 2user IC Case B.I}
\end{eqnarray}
Hence, if R2 knows the interfering symbols $I_{2[1:M_2]}(t)$, it can compute
\begin{equation}
Y_{2[1:M_2]}'(t) - I_{2[1:M_2]}(t) =  U_{22}^{[M_2]}(t) H_{22}(t) \begin{bmatrix} u_{21}^*(t) & u_{22}^*(t) & \cdots & u_{2M_2}^*(t) \end{bmatrix}^* + W_2(t) \label{eq: subtracting contribution simultaneous dCSI 2user IC Case B.I}
\end{equation}
and then by inverting the square matrix $U_{22}^{[M_2]}(t) H_{22}(t)$, it can recover the desired symbols. In other words, R2 needs to know the interfering symbols $\{I_{2j}(t)\}_{j=1}^{M_2}$. Moreover, since the interfering symbols $\{I_{2j}(t)\}_{j=1}^{M_2}$ are just the LCs of $\{u_{1i}(t)\}_{i=1}^{M_1}$ and there are precisely $M_2$ of these symbols, their knowledge would almost surely enable R1 to decode its desired symbols. Hence, it is beneficial to convey these interfering symbols to both the receivers, which is the goal of the next phase. Note that at the end of this phase, T1 knows $\big\{ \{I_{2j}(t)\}_{j=1}^{M_2} \big\}_{t=1}^{t_1}$ perfectly, while both the receivers know the channel-dependent coefficients that produce these interfering symbols (or the LCs).

\underline{Phase Two:} This last phase takes the remaining $M_2$ time slots. Here, T1 transmits the interfering symbols $\{I_{2j}(t)\}_{j,t}$ so that both R1 and R2 can determine these, while R2 continues to receive $M_2$ new data symbols from T2 in each slot. For a given $t \in [t_1 + 1: t_1 + M_2]$ and $t' = t- t_1$, the transmit signals are formed as follows:
\begin{eqnarray*}
X_1(t) = \begin{bmatrix} I_{2t'}(1) \\  I_{2t'}(2) \\ \vdots \\  I_{2t'}(t_1) \\ 0_{M_1-t_1} \end{bmatrix} \mbox{ and } X_2(t) = \begin{bmatrix} u_{21}(t) \\ u_{22}(t) \\ \vdots \\ u_{2M_2}(t) \end{bmatrix}.
\end{eqnarray*}
Thus, during this phase, a total $t_1 + M_2 = N_2$ transmit antennas are in use at any given time (i.e., the rest of the transmit antennas are sending nothing). Since $N_1 > N_2$, both the receivers can recover the transmit signals almost surely via channel inversion, as explained in the previous section.

Thus, R2 can learn all new desired data symbols sent during this phase as well as all the interfering symbols $\{I_{2j}(t)\}_{j,t}$, whose knowledge would enable it to decode the desired data symbols sent over Phase One. At the same time, R1 gets one interference-free LCs per desired symbol and hence can decode them.
\end{IEEEproof}

\begin{remark}[On interference alignment]
In the above achievability scheme, interference alignment is achieved by making T1 transmit the interfering symbols over Phase Two, which have already caused interference to R2. This serves three purposes: First, it allows R1 to learn the LCs it has lost due to interference. Second, it does not cause any additional interference at R2. Third, it gives R2 an opportunity to learn the interference it suffered over Phase One and thereby enabling it to decode the desired data symbols sent over the previous phase.
\end{remark}

\section{Proof of Theorem \ref{thm: MIMO_2-user IC_d-CSI_inner-bound DoF}: Case B.II} \label{sec: proof of thm: MIMO_2-user IC_d-CSI_inner-bound DoF_Case B.II}

First note from Lemma \ref{lem: eqvt form Condition 1 d-CSI 2user IC} that Condition $1$ can be expressed in an equivalent form as given by
\[
M_1 > M_1' > N_1 > N_2 > M_2 > m
\]
where $M_1' = \min(M_1,N_1+N_2-M_2)$ and $m = N_2 \frac{M_1'-N_1}{M_1'-N_2}$. Thus, the defining inequality for this case is
\[
M_1 > N_1 > N_2 > M_2 \mbox{ and either } M_1 = M_1' \mbox{ or } M_2 \leq m.
\]

Then we have the following lemma.

\begin{lemma} \label{lem: Case B.II d-CSI 2user IC}
Under Case B.II, bounds $L_{\{1,3\}}$ are active and, moreover, if
\begin{equation}
N_2 \cdot \frac{M_1-N_1}{M_1-N_2} \geq M_2 \, , \label{eq: in lem on Case B.II dCSI 2user IC}
\end{equation}
bound $L_3$ is active. Further, the outer-bound $\mathbf{D}_{\rm{outer}}^{\rm{d-CSI}}$ is achievable. More specifically, if
\begin{enumerate}[1.]
\item $M_2 \leq m$, $L_3$ is active and $\mathbf{D}^{\rm{no-CSI}} \subset \mathbf{D}^{\rm{d-CSI}} = \mathbf{D}^{\rm{p-CSI}}$;
\item $M_1 = M_1'$ and $M_2 >m$, $L_{\{1,3\}}$ is active and $\mathbf{D}^{\rm{no-CSI}} \subset \mathbf{D}^{\rm{d-CSI}} \subset \mathbf{D}^{\rm{p-CSI}}$.
\end{enumerate}
\end{lemma}

\begin{figure}[h] \centering
\includegraphics[bb=0bp 200bp 540bp 680bp,clip, height=3.2in, width=3.2in]{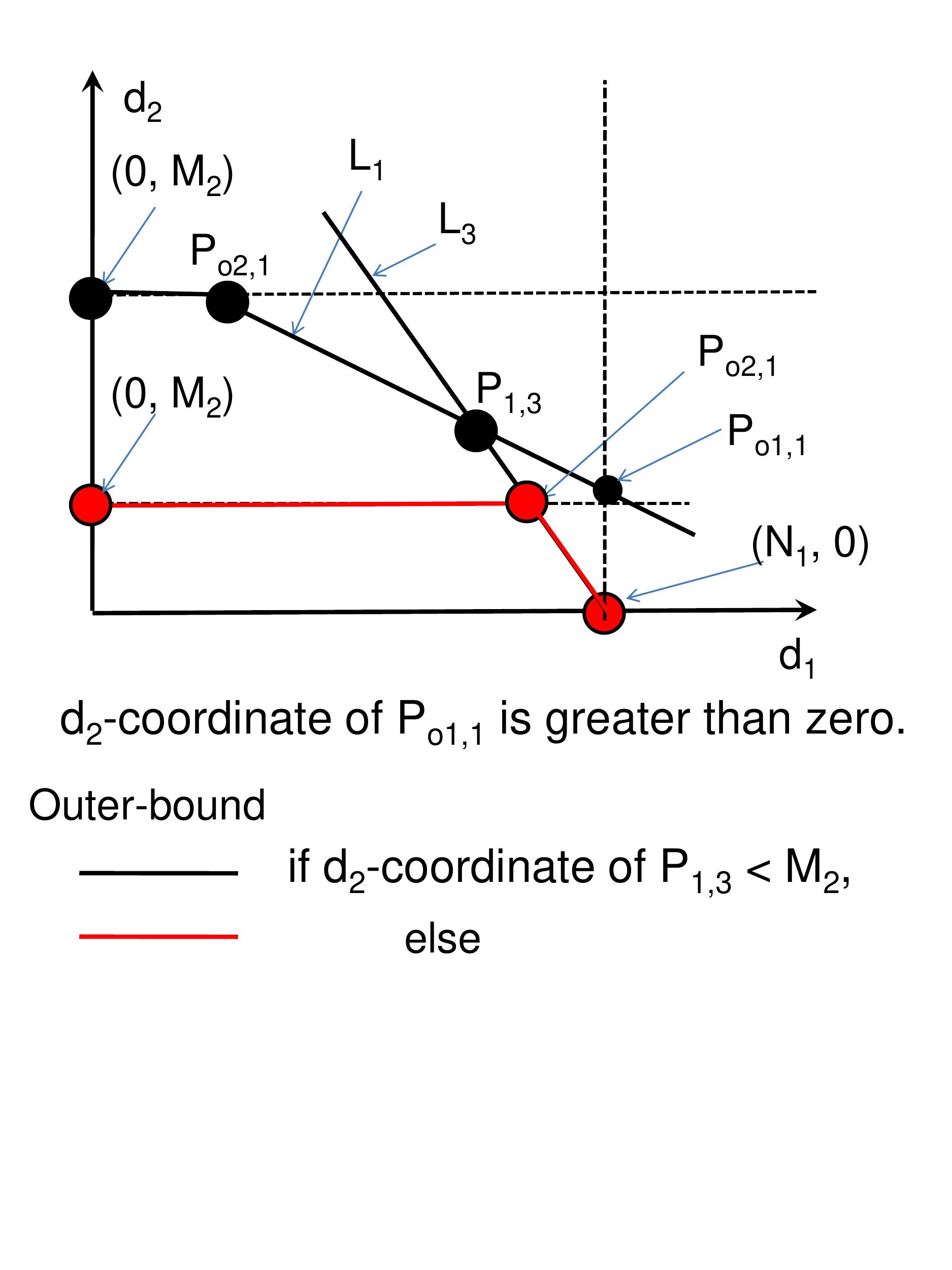}
\caption{Two Possible Shapes of the Outer-Bound: Case B.II} \label{fig: typical_shape Case B.II d-CSI 2user IC}
\end{figure}

\hspace{3pt} \emph{Proof: } A simple evaluation shows that $L_3$ implies $L_2$ (see the proof presented for Case B.I). Since the outer-bound $\mathbf{D}_{\rm{outer}}^{\rm{d-CSI}}$ does change even if $M_1$ increases beyond $N_1 + N_2$ and since all the achievability schemes developed here make use of at most $N_1 + N_2$ antennas of T1, it may be assumed, without loss of generality, that $M_1 = \min(M_1, N_1+N_2)$. Then, the bounds $L_1$ and $L_3$, which are active, are given by
\[
L_1 \equiv \frac{d_1}{M_1} + \frac{d_2}{N_2} \leq 1 \mbox{ and } L_3 \equiv d_1 + d_2 \leq N_1.
\]
The straight lines corresponding to $L_1$ and $L_3$ have unequal slopes, and hence intersect. From Fig. \ref{fig: typical_shape Case B.II d-CSI 2user IC}, one may observe that if the $d_2$-coordinate of point $P_{1,3}$ (the point of intersection of $L_1$ and $L_3$), is greater than or equal to $M_2$, only bound $L_3$ is active. Since the $d_2$-coordinate of $P_{1,3}$ is equal to $N_2 \frac{M_1-N_1}{M_1-N_2}$, bound $L_3$ is active when the inequality (\ref{eq: in lem on Case B.II dCSI 2user IC}) is true.

We now proceed to the last part of the lemma. Under Case B.II, where the inequality $M_1 > N_1 > N_2 > M_2$ always holds, Condition $1$ can not hold if either of the following two conditions are true: First, $M_2 \leq m$, and second, $\big\{ M_1 \leq N_1 + N_2 - M_2 \mbox{ and } M_2 > m \big\}$. This yields the following two subcases:
\begin{itemize}
\item \underline{Case B.II.1: $M_2 \leq m \Rightarrow M_1 > N_1 > N_2 > M_2 \mbox{ and } M_2 \leq m$:} \newline In this case, we have
    \[
    M_2 \leq m \leq N_2 \frac{M_1-N_1}{M_1-N_2},
    \]
    where the last inequality is true because for a given $N_1$ and $N_2$, $\frac{M_1-N_1}{M_1-N_2}$, as a function of $M_1$, is increasing and because $M_1 \geq M_1'$. This implies that $L_3$ is active and  $\mathbf{D}_{\rm{outer}}^{\rm{d-CSI}} = \mathbf{D}^{\rm{p-CSI}}$.
\item \underline{Case B.II.2: $\{M_1 \leq N_1 + N_2 - M_2 \mbox{ and } M_2 > m \} \Rightarrow M_1' =  M_1 > N_1 > N_2 > M_2 > m$:} \newline Here, bounds $L_{\{1,3\}}$ are (strictly) active.
\end{itemize}
For each of the above cases, we develop an achievability scheme to exhaust the outer-bound.

\subsection{Case B.II.1: $M_1 > N_1 > N_2 > M_2 ~ \mbox{ and } M_2 \leq N_2 \frac{M_1'-N_1}{M_1'-N_2}$}

It is sufficient to prove the achievability of $P_{o2,3} \equiv (N_1-M_2,M_2)$; see Fig. \ref{fig: typical shape d-CSI 2user IC Case B}(b).

\emph{\underline{A scheme to achieve point $P_{o2,3}$:} }

It will be shown that over $M_1'$ time slots, $M_1'(N_1-M_2)$ and $M_1' M_2$ DoF can be achieved for the two users, respectively. T2 needs to send $M_2$ i.i.d. complex Gaussian data symbols over each time slot. As done in the previous section, define $Y_1'(t)$ and $Y_2'(t)$ such that the transmit signal $X_2(t)$ affects only the first $M_2$ entries of these. Again, the $i^{th}$ receiver uses $Y_i'(t)$ for all decoding purposes. Further, at each time slot, R1 ignores the first $M_2$ entries of $Y_i'(t)$, which carry interference. The transmission scheme consists of two phases.

\underline{Phase One:} It consists of the first $t_1 = N_1-M_2$ time slots. At time $t \in [1: t_1]$, T1 sends $M_1'$ i.i.d. complex Gaussian data symbols, $\{u_{1i}(t)\}_{i=1}^{M_1'}$, intended for R1. Now, R1, after ignoring the first $M_2$ entries of $Y_1'(t)$, can be said to observe $N_1-M_2$ LCs of these symbols at a given time $t \in [1:t_1]$, and thus requires another $M_1' -(N_1-M_2)$ LCs of them. As done in the previous sections, these LCs required by R1 are present at the antennas of R2. With this motivation, consider the signal received by R2 at time $t \in [1:t_1]$:
\begin{eqnarray*}
Y_2'(t)  =  U_{22}(t) H_{22}(t) \begin{bmatrix} u_{21}^*(t) & \cdots & u_{2M_2}^*(t) \end{bmatrix}^* + I_2(t) + W_2(t),
\end{eqnarray*}
where
\begin{eqnarray*}
I_2(t) = U_{22}(t) H_{21}^{[M_1']}(t) \begin{bmatrix} u_{11}^*(t) & \cdots & u_{1M_1'}^*(t) \end{bmatrix}^*.
\end{eqnarray*}
It should be clear that R1 can almost surely decode the input symbols $\{u_{1i}(t)\}_i$, if it knows the values of the LCs $I_{2[1:M_2+(M_1'-N_1)]}(t)$.

Consider now the case of R2, which receives its useful signal along the first $M_2$ entries of $Y_2'(t)$. Since these entries are interfered by the LCs $I_{2[1:M_2]}(t)$, R2 needs to know these values so that it can subtract them from $Y_2'(t)$ to recover the desired data symbols (cf. (\ref{eq: subtracting contribution simultaneous dCSI 2user IC Case B.I})). Hence, the LCs $I_{2[1:M_2]}(t)$ are required by both the receivers. Further, since the signal sent by T2 affects only the first $M_2$ entries of $Y_2'(t)$, R2 knows the values of LCs $I_{2[M_2+1:M_2+(M_1'-N_1)]}(t)$, which are required by R1.

Thus, over the next phase, T1 needs to signal such that both the receivers get LCs $\{I_{2[1:M_2]}(t)\}_{t=1}^{t_1}$, and R1 additionally receives $I_{2[M_2+1:M_2+(M_1'-N_1)]}(t)$, while at the same time, R2 is able to receive $M_2$ new data symbols in each time slot from T2.

Before starting the description of the next phase, we introduce some terminology.
An element of set $\mathcal{S} = \big\{~ I_{2[1:M_2 + (M_1'-N_1)]}(t) ~ \big\}_{t=1}^{t_1}$ is referred to in the sequel as the `interfering symbol'. This set can be partitioned into two disjoint subsets: $\mathcal{S}_{\rm{R2-known}} = \big\{ ~ I_{2[M_2+1:M_2+(M_1'-N_1)]}(t) ~ \big\}_{t=1}^{t_1}$, which consists of interfering symbols that are known to R2 but are required by R1; and $\mathcal{S}_{\rm{R2-unknown}} = \big\{ ~ I_{2[1:M_2]}(t) ~ \big\}_{t=1}^{t_1}$, which consists if interfering symbols that are required by both the receivers (unknown to R2, as well).

\underline{Phase Two:} It occupies the remaining $M_1'- t_1 = t_2 = (M_1' + M_2 - N_1)$ time slots. Here, over each slot, T1 sends some of the interfering symbols. Recall that all the interfering symbols are required by R1, at which only $N_1-M_2$ dimensions are available for its useful signal. Hence, T1, at no time during this phase, should transmit more than $N_1-M_2$ interfering symbols. Moreover, some of these interfering symbols, in particular, those belonging to the set $\mathcal{S}_{\rm{R2-unknown}}$ are required by R2 as well, which implies that T1 should never transmit more than $N_2-M_2$ elements of the set $\mathcal{S}_{\rm{R2-unknown}}$ (recall, R2 needs to decode $M_2$ data symbols per time slot). Nonetheless, under these constraints, T1 should be able to transmit every element of the set $\mathcal{S}$ (at least once). Hence, the transmit signal of T1 needs to be designed to meet the above objectives. To this end, the following lemma helps.

\begin{lemma} \label{lem: partitioning Case B.II.1 d-CSI 2user IC inner-bound}
There exists a fixed, deterministic partition $\mathcal{P}$ of the set $\mathcal{S}$ into $t_2$ disjoint subsets $\{\mathcal{S}(j)\}_{j=1}^{t_2}$ of cardinality $(N_1-M_2)$ each such that no subset contains more that $(N_2-M_2)$ elements of set $\mathcal{S}_{\rm{R2-unknown}}$.
\end{lemma}
\begin{IEEEproof}
This lemma rests on the fact that $M_2 \leq m$. See Appendix \ref{app: proof of lem: partitioning Case B.II.1 d-CSI 2user IC inner-bound} for the proof.
\end{IEEEproof}

Consider a partition $\mathcal{P}$ given by the above lemma. Let us denote the elements of subset $\mathcal{S}(j)$ by $\big\{ I^{[2]}_i(j) \big\}_{i=1}^{N_1-M_2}$. At time $t \in [t_1+1:t_1+t_2]$, T1 transmits elements of set $\mathcal{S}(t-t_1)$ using the first  $N_1-M_2$ of its antennas in a one-to-one fashion, while T2 sends $M_2$ new data symbols. Thus, for a given $t \in [t_1+1: t_1 + t_2]$ and $t' = t-t_1$, T1 and T2 construct their signals as follows:
\begin{eqnarray}
X_{1i}(t) & = & I^{[2]}_i(t') ~ \forall i \in [1: N_1-M_2], ~  X_{1i}(t) = 0 ~\forall i \in [N_1-M_2+1:M_1] \mbox{ and} \nonumber  \\
X_{2j}(t) & = & u_{2j}(t) ~ \forall ~ j \in [1:M_2]. \label{eq: transmit signals1 dCSI 2user IC Case B.II.1}
\end{eqnarray}
Since the cardinality of $\mathcal{S}(t')$ can never exceed $N_1-M_2$, exactly $N_1$ transmit antennas are in use at any given time during this phase. Hence, R1 can zero-force the interference coming from T2 and recover the signal sent by T1. Thus, at the end of this phase, R1 gets one interference-free LC per desired data symbol, and hence can decode all desired data symbols.

Next, consider R2. For no $t' \in [1:t_2]$, a subset $\mathcal{S}(t')$ contains more than $N_2-M_2$ elements of the set $\mathcal{S}_{\rm{R2-unknown}}$, which implies that T1 and T2 together are transmitting at most $N_2$ symbols that are unknown to R2 (the rest are known to it). Hence, R2 can subtract the contribution due to the known interfering symbols (cf. (\ref{eq: subtracting contribution dCSI 2user IC Case A.I.3})), and then via channel inversion can recover the desired data symbols as well as the unknown interfering symbols. Once it knows the unknown interfering symbols, it can subtract the contribution due to these from the signal it has received over Phase One and then again via channel inversion can recover the desired data symbols sent to it over Phase One.

\subsection{Case B.II.2:  $N_1+N_2-M_2 \geq  M_1 > N_1 > N_2 > M_2 > N_2 \frac{M_1-N_1}{M_1-N_2}$ }

In this case, bounds $L_{\{1,3\}}$ are active and a typical shape of the outer-bound is as shown in Fig. \ref{fig: typical shape d-CSI 2user IC Case B}(c). We need to establish the achievability of points $P_{o2,1}$ and $P_{1,3}$. We develop two separate achievability schemes for these two points.

\emph{\underline{1) An achievability scheme for point $P_{o2,1}$:} }

Here, point $P_{o2,1}$ is given by
\[
P_{o2,1} \equiv \Big( \frac{M_1}{N_2}(N_2-M_2), ~ M_2 \Big).
\]
Recall that in the context of Case B.I, we developed an achievability scheme for point $P_{o2,1}$ in the last section. Here, we modify this scheme such that it works for the present case. In the following, we will just point out the important differences. In the scheme of the present section, R1 makes use of all its antennas, unlike the one of the last section where it uses only $M_1$ of its antennas even if $N_1 \geq M_1$ (which is the case in the last section).

The objective is to achieve a DoF tuple $\Big( M_1 (N_2-M_2), M_2 N_2 \Big)$ over $N_2$ time slots. Recall T2 needs to transmit $M_2$ i.i.d. data symbols in each time slot.

\underline{Phase One:} It again consists of $t_1 = N_2-M_2$ time slots. T1 transmits $M_1$ i.i.d. complex data symbols in each time slot. R1 experiences interference at $M_2$ of its receive dimensions and thus observes $N_1-M_2$ interference-free LCs per time slot, which also implies that it needs another $M_1+M_2-N_1$ LCs. Note that up to this point, the scheme is exactly identical to the one of the last section, except that we let R1 use all $N_1$ antennas and have accounted for the fact that $M_1$ is now greater than $N_1$.

Now, R2 receives its useful signal along the first $M_2$ entries of $Y_2'(t)$ (which is obtained from $Y_2(t)$ after a suitable unitary transformation). Hence, as argued in the last section, R2 can decode its desired signal if it knows the interfering symbols $I_{2[1:M_2]}(t)$ (see equation (\ref{eq: defn interfering symbols dCSI 2user IC Case B.I}) of the last section).

On the other hand, R1 can decode the desired data symbols sent at time $t$ if it knows the values of LCs $I_{2[1:M_1+M_2-N_1]}(t)$ ($M_1+M_2-N_1 \leq N_2$). Recall that in the scheme of the previous section, it was sufficient for R1 to know just the values of $I_{2[1:M_2]}(t)$; whereas in the present case it needs to learn more -- a difference due to $M_1 > N_1$ here, whereas in the last section, we had $M_1 \leq N_1$. Also note that since $X_2(t)$ affects only the first $M_2$ entries of $Y_2'(t)$, the values of LCs $I_{2[M_2+1:M_1+M_2-N_1]}(t)$ are known to R2.

Thus over the next phase, T1 needs to signal such that the values of LCs $\{ I_{2[1:M_2]}(t) \}_{t=1}^{t_1}$ are conveyed to both the receivers, and additionally, those of $\{ I_{2[M_2+1:M_1+M_2-N_1]}(t)\}_{t=1}^{t_1}$ are delivered to R1, and simultaneously R2 is able to decode $M_2$ new i.i.d. data symbols coming from T2 in each slot.

Define $\mathcal{S}_{\rm{R2-known}} = \big\{ ~ I_{2[M_2+1:M_2+M_1-N_1]}(t) ~ \big\}_{t=1}^{t_1}$, $\mathcal{S}_{\rm{R2-unknown}} = \big\{ ~ I_{2[1:M_2]}(t) ~ \big\}_{t=1}^{t_1}$, and $\mathcal{S} = \mathcal{S}_{\rm{R2-unknown}} \cup \mathcal{S}_{\rm{R2-known}}$ -- the elements of which are called the interfering symbols. For a set $\mathcal{S}$, $| \mathcal{S} |$ denotes its cardinality.

Partition the set $\mathcal{S}_{\rm{R2-known}}$ into $t_2 = N_2 - t_1 = M_2$ disjoint subsets $\mathcal{S}_{\rm{R2-known}}(j)$, where $j \in [1:t_2]$, such that no subset has cardinality more than $N_1-N_2$, and as per the following lemma, this is feasible.
\begin{lemma}
$\big| \mathcal{S}_{\rm{R2-known}} \big| = (M_1-N_1) (N_2-M_2) \leq t_2 (N_1-N_2)$.
\end{lemma}
\begin{IEEEproof}
The first equality follows from the definition. Suppose the second inequality is not true. Then we have
\begin{eqnarray*}
(M_1-N_1) (N_2-M_2) > M_2 (N_1-N_2) \Rightarrow \frac{N_2-M_2}{M_2} > \frac{N_1-N_2}{M_1-N_1} \Rightarrow \frac{N_2}{M_2} > \frac{M_1-N_2}{M_1-N_1},
\end{eqnarray*}
which contradicts with the defining inequality of Case B.II.2. Hence, the second inequality is true.
\end{IEEEproof}

Let the elements of $\mathcal{S}_{\rm{R2-known}}(j)$ be $I^{[2]}_i(j)$, $i \in [1:k_j]$, where $k_j$ is some integer such that $1 \leq k_j \leq (N_1-N_2)$ $\forall$ $j$.

\underline{Phase Two:} T1 transmits interfering symbols over this phase. At time $t \in [t_1+1: t_1+t_2]$ with $t' = t-t_1$, the transmit signals are formed as follows:
\begin{eqnarray*}
X_1(t) & = & \begin{bmatrix} I_{2t'}^*(1) & \cdots & I_{2t'}^*(N_2-M_2) & I^{[2]}_1(t') & \cdots & I^{[2]}_{k_{t'}}(t') & 0_{M_1-(N_2-M_2+k_{t'})} \end{bmatrix}^*\mbox{ and}\\
X_2(t) & = & \begin{bmatrix} u_{21}^*(t) & \cdots & u_{2M_2}^*(t) \end{bmatrix}.
\end{eqnarray*}
Since $(N_2-M_2) + k_{t'} + M_2 \leq N_1$ $\forall$ $t'$, not more than $N_1$ transmit antennas are in use at any time during this phase. Hence, R1 can recover the interfering symbols after zero-forcing the interference coming from T2. Further, since R2 knows all elements of set $\mathcal{S}_{\rm{R2-known}}$, exactly $N_2$ symbols unknown to R2 are being transmitted at any given time. Hence, R2, after subtracting the contribution due to known interfering symbols, can recover the desired data symbols as well as the unknown interfering symbols.

Thus, at the end of each phase, each receiver get one interference-free LC per desired data symbol and hence each of them can decode the desired symbols.

\emph{\underline{2) An achievability Scheme for Point $P_{1,3}$:} }

We have
\[
P_{1,3} \equiv \left( \frac{M_1(N_1-N_2)}{M_1-N_2}, ~ \frac{N_2(M_1-N_1)}{M_1-N_2} \right).
\]
We now describe below a generic IA-based achievability scheme, $\achsch$, which is defined in terms of certain set of parameters. This scheme with appropriate choice for parameters can be used to achieve point $P_{1,3}$ above. Moreover, it is useful in the next section as well.

Let $\mathbb{N}$ and $\mathbb{N}_0$ be the sets of positive and non-negative integers, respectively. Consider the achievability scheme $\achsch \Big\{(M_1,M_2,N_1,N_2), (d_1^{\star},d_2^{\star}), (T,t_1,t_2), \{m_t\}_{t=1}^{t_1}, \{n_t\}_{t=1}^{t_1} \Big\}$, where $d_1^{\star}$, $d_2^{\star}$, $T$, $t_1$, $t_2 \in \mathbb{N}$; $m_t$, $n_t \in \mathbb{N}_0$ $\forall$ $t \in [1:t_1]$; $t_1 < T$, $t_2 = T- t_1$; $N_1 \leq m_t \leq M_1$, $n_t \leq M_2$, $m_t + n_t \leq N_1 + N_2$ $\forall$ $t \in [1:t_1]$; and $\sum_{t=1}^{t_1} m_t = d_1^{\star}$. We will design this coding scheme such that over $T$ time slots, $d_1^{\star}$ and $d_2^{\star}$ DoF can be achieved respectively for the two users, provided the parameters $t_1$, $\{m_t\}$, $\{n_t\}$ are chosen appropriately. Thus, using this scheme, it is possible to achieve the DoF pair $\Big(\frac{d_1^{\star}}{T},\frac{d_2^{\star}}{T} \big)$.

The scheme takes a total of $T$ time slots and consists of two phases. \newline
\underline{Phase 1:} This phase constitutes the initial $t_1$ time slots. At time $t \in [1:t_1]$, T1 transmits $m_t$ i.i.d. complex Gaussian data symbols, $\{u_{1j}(t)\}_{j=1}^{m_t}$, intended for R1; whereas T2 transmits $n_t$ i.i.d. complex Gaussian data symbols, $\{u_{2j}(t)\}_{j=1}^{n_t}$, intended for R2. Note that this is feasible since by definition $m_t \leq M_1$ and $n_t \leq M_2$.

Consider now the scenario at the receivers. It may be assumed, without loss of generality, that the signal transmitted by T2 at time $t \in [1 : t_1]$ affects only the first $n_t$ entries of $Y_1(t)$ and $Y_2(t)$ (cf. the discussion in Section \ref{sec: proof of thm: MIMO_2-user IC_d-CSI_inner-bound DoF_Case B.I} on the achievability of point $P_{o2,1}$). Thus, R1 observes $N_1-n_t$ LCs of input symbols $\{u_{1j}(t)\}_{j=1}^{m_t}$ without any interference, and therefore, needs another $m_t-(N_1-n_t)$ LCs to decode these symbols. The additional LCs required by R1 are present at (some of the) antennas of R2 because $m_t \leq N_1 + N_2 - n_t$ $\forall$ $t \in [1:t_1]$. More precisely, the signal received by R2 at time $t$ during this phase is given by
\begin{eqnarray*}
Y_2(t) & = & H_{22}^{[n_t]}(t) \begin{bmatrix} u_{21}^*(t) & \cdots & u_{2n_t}^*(t) \end{bmatrix} + I_2(t) + W_2(t), \mbox{ where} \\
I_2(t) & =  & H_{21}^{[m_t]}(t) \begin{bmatrix} u_{11}^*(t) & \cdots u_{1m_t}^*(t) \end{bmatrix}^*.
\end{eqnarray*}
Then R1 can decode the desired data symbols $\{u_{1j}(t)\}_{j=1}^{m_t}$, if it knows the values of LCs $\big\{ I_{2[1:n_t + (m_t-N_1)]}(t) \big\}$. Now, R2, at time $t$, receives the desired symbols along the first $n_t$ entries of $Y_2(t)$, which experience interference due to $I_{2[1:n_t]}(t)$. Thus, R2 can decode its desired data symbols, provided it knows the values of $I_{2[1:n_t]}(t)$. It must be noted that R2 knows the values of $\{I_{2[n_t+1:n_t + (m_t-N_1)]}(t) \}$.

Thus, over the next phase, we have to accomplish the following objectives: \begin{inparaenum}[\itshape a\upshape)] \item communicate the values of LCs $\big\{ ~ I_{2[1:n_t]}(t) ~ \big\}_{t=1}^{t_1}$ to both the receivers, \item additionally, deliver those of $\big\{ ~ I_{2[n_t+1 : n_t + (m_t-N_1)]}(t) ~ \big\}$ to R1, and \item transmit $n' = d_2^{\star} - \sum_{t=1}^{t_1} n_t $ number of new data symbols to R2. \end{inparaenum}

Before starting the description of the next phase, we introduce some terminology and a lemma that allows us to design the transmission strategy over the next phase.

Let $\mathcal{S}_{\rm{IS}} = \big\{ I_{2[1: n_t + (m_t-N_1)]}(t) \big\}_{t=1}^{t_1}$ and, as before, an element of it is referred to as the interfering symbol. Further define $\mathcal{S}_{\rm{IS-R2unknown}} = \big\{ I_{2[1: n_t]}(t) \big\}_{t=1}^{t_1}$ and $\mathcal{S}_{\rm{IS-R2known}} = \big\{ I_{2[n_t + 1: n_t + (m_t-N_1)]}(t) \big\}_{t=1}^{t_1}$ so that $\mathcal{S}_{\rm{IS}} = \mathcal{S}_{\rm{IS-R2unknown}} \cup \mathcal{S}_{\rm{IS-R2known}} $. Let $\{u_{2j}\}_{j=1}^{n'}$ be the i.i.d. complex Gaussian data symbols to be sent by T2 over the next phase and set $\mathcal{S}_{\rm{DS}} = \{u_{2j}\}_{j=1}^{n'}$ and $\mathcal{S} = \mathcal{S}_{\rm{IS}} \cup \mathcal{S}_{\rm{DS}}$.

Consider the following lemma about partitioning of set $\mathcal{S}$.
\begin{lemma} \label{lem: partitioning Case B.II.2 d-CSI 2user IC inner-bound}
There exists a fixed, deterministic partition $\mathcal{P}$ of set $\mathcal{S}$ into $t_2 $ disjoint subsets $\mathcal{S}(j)$, $j \in [1: t_2]$, of cardinality at most $N_1$ such that each resulting subset $\mathcal{S}(j)$ satisfies the following properties:
\begin{enumerate}
\item It does not contain more than $M_2$ elements of set $\mathcal{S}_{\rm{DS}}$.
\item It does not contain more than $N_2$ elements of set $\mathcal{S}_{\rm{IS-R2unknown}} \cup \mathcal{S}_{\rm{DS}}$.
\end{enumerate}
\end{lemma}
\begin{IEEEproof}
This lemma needs to be proved whenever the achievability scheme $\achsch$ is to be used and a particular choice for the parameters is made. In the proof of this lemma, the following facts are useful:
\begin{eqnarray}
| \mathcal{S}_{\rm{DS}} | & = & n' = d_2^{\star} - \sum_{t=1}^{t_1} n_t \label{eq: 1 proof of lem: partitioning Case B.II.2 d-CSI 2user IC inner-bound} \\
| \mathcal{S}_{\rm{IS-R2unknown}} | & = & \sum_{t=1}^{t_1} n_t \label{eq: 2 proof of lem: partitioning Case B.II.2 d-CSI 2user IC inner-bound} \\
| \mathcal{S}_{\rm{IS-R2known}} | & = & \sum_{t=1}^{t_1} m_t - N_1 t_1 = d_1^{\star} - N_1 t_1. \label{eq: 3 proof of lem: partitioning Case B.II.2 d-CSI 2user IC inner-bound}
\end{eqnarray}
From the proof of Lemma \ref{lem: partitioning Case B.II.1 d-CSI 2user IC inner-bound} presented Appendix \ref{app: proof of lem: partitioning Case B.II.1 d-CSI 2user IC inner-bound}, it is easy to see that the present lemma holds if the following three inequalities are true: \begin{inparaenum}[(a)] \item $M_2 t_2 \geq |\mathcal{S}_{\rm{DS}}| = n'$; \item $N_2 t_2 - |\mathcal{S}_{\rm{DS}}| \geq |\mathcal{S}_{\rm{IS-R2unknown}}|$; and \item $N_1 t_2 - |\mathcal{S}_{\rm{DS}}| - |\mathcal{S}_{\rm{IS-R2unknown}}| \geq |\mathcal{S}_{\rm{IS-R2known}}|$. \end{inparaenum} These inequalities are shown whenever the lemma needs to be proved.
\end{IEEEproof}
Assuming that this lemma is true, we design the next phase of the scheme.

\underline{Phase Two:} This is the last phase and takes the remaining $t_2$ time slots. At time $t \in [t_1+1 : t_1+ t_2]$ with $t' = t-t_1$, a subset $\mathcal{S}(t')$, obtained from the earlier lemma, is chosen. T1 transmits all the interfering symbols belonging to this set (i.e., the elements of set $\mathcal{S}(t') \cap \mathcal{S}_{\rm{IS}})$, whereas T2 transmits all data symbols in this set (i.e., the elements of set $\mathcal{S}(t') \cap \mathcal{S}_{\rm{DS}})$. As far as T1 is concerned, this is feasible because for no $t'$ is the cardinality of the subset $\mathcal{S}(t')$ more than $N_1$, which itself is smaller than $M_1$. Further, for no $t' \in [1:t_2]$ does the subset $\mathcal{S}(t')$ contain more than $M_2$ data symbols, and hence, T2 at time $t$ can transmit the data symbols belonging to the set $\mathcal{S}(t')$.

For a $t \in [t_1+1:t_1+t_2]$, T1 and T2 together need to transmit at most $N_1$ symbols. These symbols are transmitted in a one-to-one fashion (c.f. equation (\ref{eq: transmit signals1 dCSI 2user IC Case B.II.1})). Therefore, not more than $N_1$ transmit antennas are in use at any given time during this phase. This implies that R1 via simple channel inversion can recover the required interfering symbols. After recovering the interfering symbols, R1 has one interference-free LC per data symbol and can hence decode the message.

As for R2, the second property of Lemma \ref{lem: partitioning Case B.II.2 d-CSI 2user IC inner-bound} implies that at most $N_2$ symbols that are unknown to R2 are transmitted at any given time during this phase. Hence, R2 can subtract the contribution due to the known interfering symbols and then recover the desired data symbols as well as the unknown interfering symbols. Thus, it can also decode all the desired data symbols sent to it over the two phases.

In other words, at the end of $T$ time slots, the $i^{th}$ receiver can decode $d_i^{\star}$ data symbols which are intended for it (note, $n' + \sum_{t=1}^{t_1} n_t = d_2^{\star}$), provided Lemma \ref{lem: partitioning Case B.II.2 d-CSI 2user IC inner-bound} holds.

Thus, if the achievability scheme $\achsch$ is used with parameters which are such that Lemma \ref{lem: partitioning Case B.II.2 d-CSI 2user IC inner-bound} holds, then we know that the DoF tuple $\Big( \frac{d_1^{\star}}{T}, \frac{d_1^{\star}}{T} \Big)$ is achievable over the $(M_1,M_2,N_1,N_2)$ MIMO IC.

Now, returning to the achievability of point $P_{1,3}$ under Case B.II.2, we use achievability scheme $\achsch$ with the following choice of parameters:
$T= M_1 - N_2$; $t_1 = N_1 - N_2$; which implies $t_2 = M_1-N_1$; $d_1^{\star} = M_1 (N_1 - N_2)$; $d_2^{\star} = N_2 (M_1 - N_2)$; $m_t = M_1$ $\forall$ $t \in [1:t_1]$ so that $\sum_{t=1}^{t_1} m_t = d_1^{\star}$; $n_t$ are deterministic integers such that
\[
\sum_{t=1}^{t_1} n_t = \min \big\{ M_2(N_1-N_2), N_2(M_1-N_1) \big\}.
\]
It is now sufficient to prove that Lemma \ref{lem: partitioning Case B.II.2 d-CSI 2user IC inner-bound} holds with these choices of parameters, which is done in Appendix \ref{app: proof of lem: partitioning Case B.II.2 d-CSI 2user IC inner-bound}. Hence, the DoF pair $\Big( \frac{d_1^{\star}}{T}, \frac{d_1^{\star}}{T} \Big)$, which coincides with $P_{1,3}$ is achievable.

This concludes the proof of Lemma \ref{lem: Case B.II d-CSI 2user IC}. \IEEEQED

\section{Proof of Theorem \ref{thm: MIMO_2-user IC_d-CSI_inner-bound DoF}: Case B.III} \label{sec: proof of thm: MIMO_2-user IC_d-CSI_inner-bound DoF_Case B.III}

\begin{figure}[h] \centering
\includegraphics[bb=0bp 45bp 540bp 700bp,clip, height=5in, width=4in]{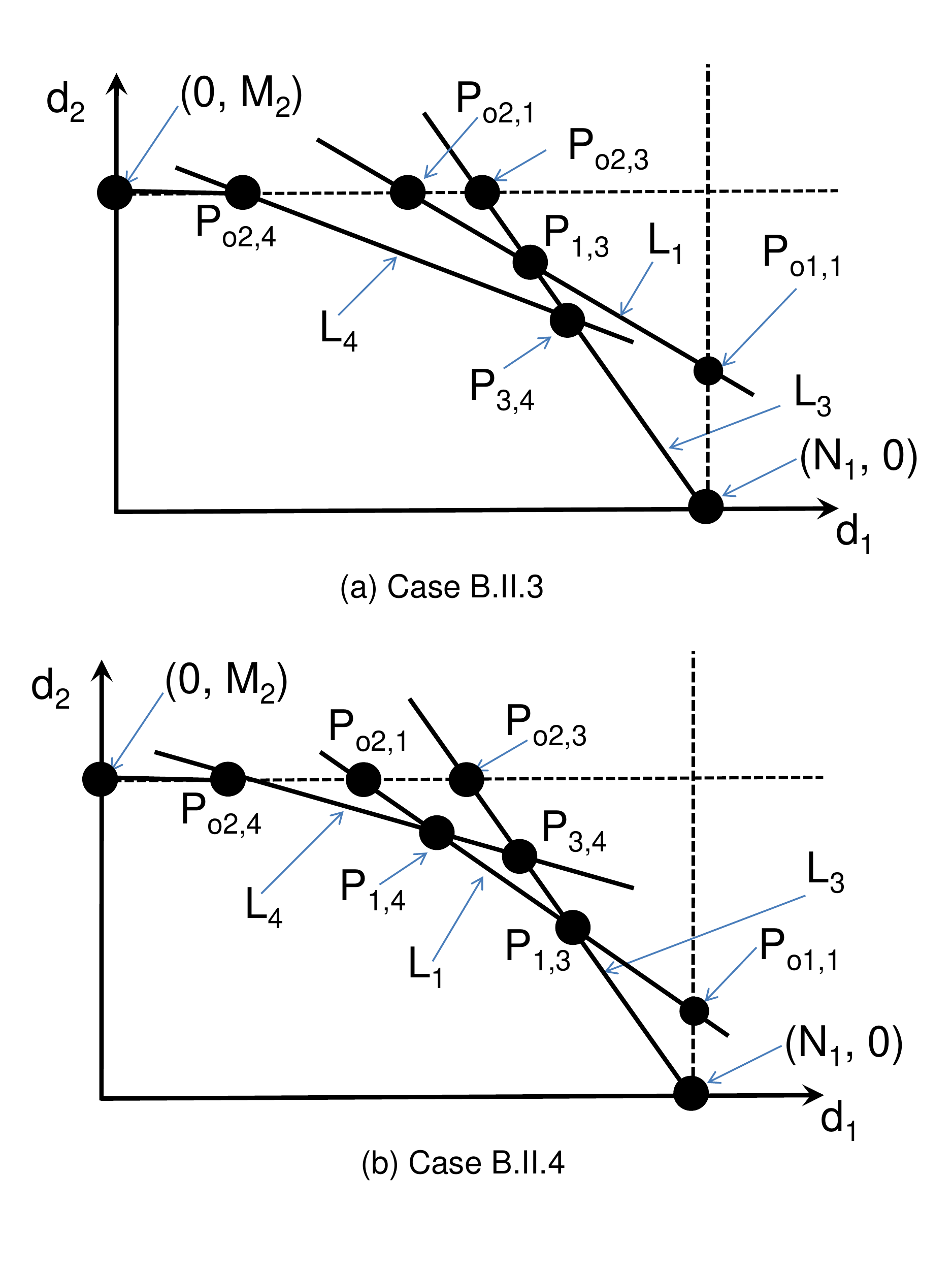}
\caption{Two Possible Shapes of the Outer-Bound: Case B.III} \label{fig: typical shape dCSI MIMO IC B.II.3,4 part2}
\end{figure}

Using Lemma \ref{lem: eqvt form Condition 1 d-CSI 2user IC}, the defining inequality of this case is given by
\[
M_1 > M_1' > N_1 > N_2 > M_2 > m.
\]
We have the following lemma for this case.
\begin{lemma} \label{lem: Case B.III d-CSI 2user IC}
Under Case B.III, bounds $L_{\{1,3,4\}}$ are active, the outer-bound $\mathbf{D}_{\rm{outer}}^{\rm{d-CSI}}$ achievable, and $\mathbf{D}^{\rm{no-CSI}}  \subset \mathbf{D}^{\rm{d-CSI}} \subset \mathbf{D}^{\rm{p-CSI}}$. Moreover, \newline
1. if $M_1 \geq N_1 + N_2 - m$, then bounds $L_{\{3,4\}}$ are active; and \newline
2. if $M_1 < N_1 + N_2 - m$, then bounds $L_{\{1,3,4\}}$ are active.
\end{lemma}
\begin{IEEEproof}
We first determine the typical shape of the outer-bound. To this end, note from the previous section that whenever $M_1 > N_1> N_2 > M_2$, the bounds $L_{\{1,3,4\}}$ are active. Let $d_i(P_{A,B})$ denote the $d_i$-coordinate of point $P_{A,B}$. It can be verified that $d_1(P_{o2,4}) \leq d_1(P_{o2,1}), d_1(P_{o2,3})$ (see Fig. \ref{fig: typical shape dCSI MIMO IC B.II.3,4 part2}). In other words, at $d_2 = M_2$, bound $L_4$ is active. Then it is not difficult to see that if $d_2(P_{1,3}) \geq d_2 (P_{1,4})$, then bounds $L_{\{3,4\}}$ are active (see Fig. \ref{fig: typical shape dCSI MIMO IC B.II.3,4 part2}). It turns out that
\[
d_2(P_{1,3}) \geq d_2 (P_{1,4}) \Leftrightarrow M_1 \geq N_1 + N_2 - m
\]
and hence, under Case B.III.1, bounds $L_{\{3,4\}}$ are active (see Fig. \ref{fig: typical shape dCSI MIMO IC B.II.3,4 part2}(a)). This also implies that under Case B.III.2, bounds $L_{\{1,3,4\}}$ are active (see Fig. \ref{fig: typical shape dCSI MIMO IC B.II.3,4 part2}(b)).

\subsection{Case B.III.1: Condition $1$ holds and $M_1 \geq N_1 + N_2 - m$}    \label{subsec: proof of thm: MIMO_2-user IC_d-CSI_inner-bound DoF part2 B.II.3}

From Fig. \ref{fig: typical shape dCSI MIMO IC B.II.3,4 part2}(a), one may deduce that the entire outer-bound can be achieved via time sharing, provided the points $P_{o2,4}$ and $P_{3,4}$ can be achieved, which is the topic of the remainder of this subsection.

\emph{\underline{1) An achievability scheme for $P_{o2,4}$:} } We have
\[
P_{o2,4} \equiv \left( \frac{M_1' (N_2-M_2)}{N_2}, M_2 \right) \mbox{ with } M_1' = \min \{ M_1,N_1 + N_2-M_2\} = N_1 + N_2 -M_2.
\]
Note that $(M_1',M_2,N_1,N_2)$ MIMO IC belongs to Case B.II.2. Therefore, the achievability of point $P_{o2,1}$, proved in the context of Case B.II.2, implies the achievability of point $P_{o2,4}$ for Case B.III.1 (see previous section).

\emph{\underline{2) An achievability scheme for $P_{3,4}$:} } We have
\[
P_{3,4} \equiv \left( N_1 - \frac{N_2^2}{M_1'}, \frac{N_2^2}{M_1'} \right).
\]
Note that the point $P_{3,4}$ remains invariant even if $M_1$ increases beyond $N_1 + N_2$. Moreover, the scheme that we propose here to achieve $P_{3,4}$ makes use of at most $N_1 + N_2$ antennas at R1 (even if more are available). Thus, it can be assumed without loss of generality that $M_1 \leq N_1 + N_2$.

We first introduce some notation. Let $\lceil N_1+N_2 - m \rceil$ ($\lfloor N_1 + N_2 - m\rfloor$) be the smallest (largest) integer greater (smaller) than or equal to $N_1 + N_2 - m$. Note that $N_1 + N_2 - m \leq M_1, N_1 + N_2$ and $N_1 + N_2 - m \geq N_1 + N_2 - M_2$.

We will use the scheme $\achsch$ with the following choice of parameters: $T = M_1'$; $t_1 = N_1 -M_2$, which imples $t_2 = T - t_1 = N_2$; $d_1^{\star} = N_1 M_1' - N_2^2$; and $d^{\star}_2 = N_2^2$. Choose $m_t$ such that $m_{t} \in \big\{ ~ \lceil N_1+N_2 - m \rceil,\lfloor N_1 + N_2 - m\rfloor ~ \big\}$ $\forall$ $t \in [1:t_1]$ and $\sum_{t=1}^{t_1} m_{t} = d_1^{\star} = N_1 M_1' - N_2^2$ and such a choice can be made because $(N_1 + N_2 - m) t_1 = N_1 M_1'-N_2^2$. Next, set $n_t = N_1 + N_2 - m_{t}$ $\forall$ $t \in [1:t_1]$. This choice ensures that $m_t \leq M_1$ and $n_t \leq M_2$ $\forall$ $t \in [1:t_1]$. Note that $\Big( \frac{d_1^{\star}}{T}, \frac{d_1^{\star}}{T} \Big)$ is equal to $P_{3,4}$.

With this choice of parameters, Lemma \ref{lem: partitioning Case B.II.2 d-CSI 2user IC inner-bound} is proved below.

\emph{Proof of Lemma \ref{lem: partitioning Case B.II.2 d-CSI 2user IC inner-bound} for point $P_{3,4}$ under Case B.III.1:} Since $n_t = N_1 + N_2 - m_t$, one can easily compute
\[
\sum_{t=1}^{t_1} n_t = N_2 (N_2 - M_2),
\]
which yields $n' = N_2 M_2$. Using these facts and equations (\ref{eq: 1 proof of lem: partitioning Case B.II.2 d-CSI 2user IC inner-bound}), (\ref{eq: 1 proof of lem: partitioning Case B.II.2 d-CSI 2user IC inner-bound}), and (\ref{eq: 1 proof of lem: partitioning Case B.II.2 d-CSI 2user IC inner-bound}), it can be verified that the inequalities (a), (b), and (c) (stated just after equation (\ref{eq: 1 proof of lem: partitioning Case B.II.2 d-CSI 2user IC inner-bound})) hold. \IEEEQED

Since Lemma \ref{lem: partitioning Case B.II.2 d-CSI 2user IC inner-bound} holds, we know that the point $P_{3,4}$ is achievable.

\subsection{Case B.III.2: Condition $1$ holds and $M_1 < N_1 + N_2 - m$}    \label{subsec: proof of thm: MIMO_2-user IC_d-CSI_inner-bound DoF part2 B.II.4}
From Fig. \ref{fig: typical shape dCSI MIMO IC B.II.3,4 part2}(b), we observe that the entire outer-bound can be achieved via time sharing, provided the points $P_{o2,4}$, $P_{1,4}$, and $P_{1,3}$ can be achieved, which is the topic of the remainder of this subsection.

The achievability of point $P_{o2,4}$ can be proved in the same manner as done in the previous subsection.

\emph{\underline{1) An achievability scheme for $P_{1,3}$:} } We have
\[
P_{1,3} \equiv \left( \frac{M_1 (N_1-N_2)}{M_1 - N_2}, \frac{N_2(M_1-N_1)}{M_1 - N_2} \right) .
\]
Let $M_2' = N_1 + N_2 - M_1 < M_2$. We use the achievability scheme $\achsch$ with the following choice of parameters: $T = M_1 - N_2$; $t_1 = N_1 - N_2$; $t_2 = T = t_1 = (M_1 - N_1)$; $d_1^{\star} = M_1 (N_1-N_2)$; $d_2^{\star} = N_2(M_1-N_1)$; $m_t = M_1$ $\forall$ $t \in [1:t_1]$ so that $\sum m_t = d_1^{\star}$; choose $n_t$ as deterministic integers such that $n_t \leq M_2'$ $\forall$ $t \in [1:t_1]$ and $\sum_{t=1}^{t_1} n_t = \min \{ M_2' t_1, d_2^{\star} \}$. The required lemma, namely, Lemma \ref{lem: partitioning Case B.II.2 d-CSI 2user IC inner-bound} has been proved in Appendix \ref{app: proof of lem: partitioning Case B.II.2 d-CSI 2user IC inner-bound part2 Case B.III.2 P13}.

\emph{\underline{2) An achievability scheme for $P_{1,4}$:} } We have
\[
P_{1,4} \equiv \left( \frac{M_1 ( M_1' - N_1) }{M_1' + N_2 - M_1}, N_2 \frac{N_1 + N_2 - M_1}{M_1'+N_2 - M_1} \right).
\]
We use the scheme $\achsch$ with the following choice of parameters: $T = M_1' + N_2 - M_1$; $t_1 = M_1'-N_1$; $t_2 = T - t_1 = M_2'$; $d_1^{\star} = M_1 ( M_1' - N_1)$; $d_2^{\star} =  N_2 (N_1 + N_2 - M_1)$; $m_t = M_1$ and $n_t = M_2'$ $\forall$ $t \in [1:t_1]$. Lemma \ref{lem: partitioning Case B.II.2 d-CSI 2user IC inner-bound} is proved in Appendix \ref{app: proof of lem: partitioning Case B.II.2 d-CSI 2user IC inner-bound part2 Case B.III.2 P14}.
\end{IEEEproof}

\section{Conclusion}          \label{sec:conclusions}
In this paper, we study the $2$-user MIMO IC with delayed CSI. Inner and outer-bounds to its DoF region are obtained and are shown to coincide for all possible values of the $4$-tuple $(M_1,M_2,N_1,N_2)$. To derive an outer-bound, the property of statistical equivalence of the channel outputs is used, whereas to obtain an inner-bound, interference alignment based achievability schemes, that require only delayed CSIT, are developed.

\newpage
\appendices

\section{Lemmas Useful for Determining the Shape of the DoF Region} \label{app: lemmas to determine the shape d-CSI 2user IC}
Consider the following lemma.
\begin{lemma} \label{lem: d1max then d2=0 d-CSI 2user IC}
For the IC with $N_1 \geq N_2$ and $M_1 > N_2$,
\[
d_1 = \min(M_1,N_1) \Rightarrow d_2 = 0,
\]
for the cases of delayed and no CSI.
\end{lemma}
\begin{IEEEproof}
Since $\mathbf{D}^{\rm{no-CSI}} \subseteq \mathbf{D}^{\rm{d-CSI}}$, it is sufficient to prove the lemma for the case of delayed CSI, which is considered below. If $N_1 \geq M_2$, bound $L_3$ implies that
\begin{eqnarray*}
d_1 + d_2 & \leq & \min \big\{ M_1 + M_2 , ~ N_1 + N_2, ~ \max(M_1,N_2), ~ \max(M_2,N_1) \big\} \\
& = & \min \big\{ M_1+M_2, ~ N_1+N_2, ~ M_1, N_1\big\} = \min (M_1,N_1).
\end{eqnarray*}
Hence, the lemma follows. Now, when $N_1 < M_2$, the bounds $L_1$ and $L_3$ are given by
\[
L_1 \equiv \frac{d_1}{\min(M_1,N_1+N_2)} + \frac{d_2}{N_2} \leq 1 \mbox{ and } L_2 \equiv \frac{d_1}{N_1} + \frac{d_2}{\min(N_1+N_2,M_2)} \leq 1.
\]
If $N_1 \leq M_1$, $L_2$ implies that $d_2 = 0 $ whenever $d_1 = \min(M_1,N_1) = N_1$; whereas $N_1 > M_1$, the same is implied by $L_1$.
\end{IEEEproof}

\begin{lemma} \label{lem: d2max then d1=0 d-CSI 2user IC}
Consider the MIMO IC with delayed CSI for which $N_1 \geq N_2$ and $M_1 > N_2$. Under Case A, where $M_2 \geq N_2$, we have
\[
d_2 = \min(M_2,N_2) \Rightarrow d_1 = 0,
\]
which is not true with Case B, where $M_2 < N_2$.
\end{lemma}
\begin{IEEEproof}
Whenever $N_1 \geq N_2$ and $M_1 > N_2$, bound $L_1$ is given by
\[
L_1 \equiv \frac{d_1}{\min(M_1,N_1+N_2)} + \frac{d_2}{N_2} \leq 1.
\]
Therefore, under Case A, if $d_2 = \min(M_2,N_2) = N_2$, $d_1$ can not be more than zero, which proves the first part of the lemma. Under Case B, when $d_2 = \min(M_2,N_2) = M_2$, one can achieve $d_1 = N_2 - M_2$, since $N_1, M_1 \geq N_2$, even without CSI \cite{Vaze_Dof_final}. This proves the second part of the lemma.
\end{IEEEproof}

\begin{lemma} \label{lem: eqvt form Condition 1 d-CSI 2user IC}
Condition $1$ can equivalently be stated as
\[
M_1 > M_1' > N_1 > N_2 > M_2 > m,
\]
where $M_1' = \min(M_1,N_1+N_2-M_2)$ and $m = N_2 \frac{M_1'-N_1}{M_1'-N_2}$.
\end{lemma}
\begin{IEEEproof}
Under Condition $1$, $M_1' = N_1 + N_2 - M_2$, which implies that $M_1' - N_1 = N_2-M_2$ and $M_1'-N_2 = N_1-M_2$. Hence, Condition $1$ implies the inequality stated in this lemma. Further, if the inequality stated in the lemma is true, then $M_1 > N_1 + N_2-M_2$, which then implies Condition $1$.
\end{IEEEproof}

\section{Proof of Lemma \ref{lem: partitioning Case B.II.1 d-CSI 2user IC inner-bound}}  \label{app: proof of lem: partitioning Case B.II.1 d-CSI 2user IC inner-bound}


As a first step towards obtaining the desired partition of the set $\mathcal{S}$, we partition the set $\mathcal{S}_{\rm{R2-unknown}}$ into $t_2$ subsets $\big\{ \mathcal{S}_{\rm{R2-unknown}}(j) \big\}_{j=1}^{t_2}$ of cardinality at most $N_2-M_2$. This is feasible as per the following lemma.
\begin{lemma}
$\big|\mathcal{S}_{\rm{R2-unknown}}\big| = M_2 (N_1-M_2) \leq (N_2-M_2) t_2$.
\end{lemma}
\begin{IEEEproof}
The fact that $\big|\mathcal{S}_{\rm{R2-unknown}}\big| = M_2 (N_1-M_2)$ can be easily verified by noting that $t_1 = N_1-M_2$. Suppose the inequality stated in the lemma is not true, i.e., the inequality $M_2 (N_1-M_2) > (N_2-M_2) (M_1' +M_2-N_2)$ holds (recall $t_2 = (M_1' +M_2-N_2)$). This yields us
\begin{eqnarray*}
\lefteqn{ M_2 (N_1-M_2) > (N_2-M_2) (M_1' + M_2-N_2)  \Leftrightarrow  \frac{N_1 - M_2}{M_1'+M_2-N_1} > \frac{N_2 - M_2}{M_2} } \\
&& {} \Leftrightarrow  \frac{M_1'}{M_1' + M_2 - N_1} > \frac{N_2}{M_2}  \Leftrightarrow  \frac{M_2}{N_2} > \frac{M_2 + (M_1' - N_1)}{N_2 + (M_1'-N_2)}  \Leftrightarrow  M_2 > N_2 \frac{M_1' - N_1}{M_1'-N_2} = m,
\end{eqnarray*}
which contradicts the defining inequality of Case B.II.1. Hence, the inequality stated in lemma holds.
\end{IEEEproof}

Consider now the following algorithm that yields the required partition $\mathcal{P}$.
\begin{enumerate}
\item Set $j = 1$ and $x = \big| \mathcal{S}_{\rm{R2-unknown}}(j) \big|$.
\item Compute $x' = (N_1-M_2) - x$. Pick $x'$ distinct elements of set $\mathcal{S}_{\rm{R2-known}}$ and let $s$ be the set of chosen elements.
    Set $\mathcal{S}(j) = \mathcal{S}_{\rm{R2-unknown}}(j) \cup s$ and $\mathcal{S}_{\rm{R2-known}} = \mathcal{S}_{\rm{R2-known}} \backslash s$.
\item If $j = t_2$, stop; else increment $j$ by $1$ and go to Step 1 above.
\end{enumerate}
Note here that a given element of set $\mathcal{S}_{\rm{R2-known}}$ can belong to a subset $\{\mathcal{S}(j)\}_j$ for at most one $j$ because at every step distinct elements are chosen and once an element is chosen it is removed from further consideration. Hence, the resulting subsets are disjoint. Further, the algorithm can accommodate
\begin{eqnarray*}
\sum_{j=1}^{t_2} \Big\{(N_1-M_2) - \big| \mathcal{S}_{\rm{R2-unknown}}(j) \big| \Big\} & = & t_2 (N_1-M_2) - \big| \mathcal{S}_{\rm{R2-unknown}} \big| \\
& = & t_2 (N_1-M_2) - M_2(N_1-M_2) = (t_2-M_2) (M_1'-N_1) \\
& = & (M_1'-N_2)(M_1'-N_1)
\end{eqnarray*}
elements of set $\mathcal{S}_{\rm{R2-known}}$. Since $\big| \mathcal{S}_{\rm{R2-known}} \big| = (N_1-M_2) (M_1'-N_1)$ and since the subsets $\{\mathcal{S}(j)\}_j$ have been shown to be disjoint, and every element of set $\mathcal{S}_{\rm{R2-known}}$ belongs to $\{\mathcal{S}(j)\}$ for exactly one $j \in [1:t_2]$. Hence, the algorithm yields a partition with the required properties.

\section{Proof of Lemma \ref{lem: partitioning Case B.II.2 d-CSI 2user IC inner-bound} for Point $P_{1,3}$ under Case B.II.2} \label{app: proof of lem: partitioning Case B.II.2 d-CSI 2user IC inner-bound}

We consider two cases separately: first, when $M_2(N_1-N_2) \geq N_2(M_1-N_1)$, and second when $M_2(N_1-N_2) < N_2(M_1-N_1)$.

\subsection{$M_2(N_1-N_2) \geq N_2(M_1-N_1)$:} \label{sub-app: proof of lem: partitioning Case B.II.2 d-CSI 2user IC inner-bound n'=0}

In this case, $\sum_{t=1}^{t_1} n_t = N_2 (M_1-N_1)$, i.e., $n' = 0$ or no data symbols needs to be sent to R2 over Phase Two. Thus, the first property stated in the lemma is satisfied trivially. Now, we have
\begin{eqnarray*}
\big| \mathcal{S}_{\rm{IS-R2unknown}} \big| & = & \sum_{t=1}^{t_1} n_t = N_2(M_1-N_1) = N_2 t_2, \\
\big| \mathcal{S}_{\rm{IS-R2known}} \big| & = & (M_1-N_1) ( N_1-N_2), \mbox{ and } \big| \mathcal{S}_{\rm{DS}} \big| = 0.
\end{eqnarray*}
The second property stated in the lemma dictates that at most $N_2 t_2$ elements of the set $\mathcal{S}_{\rm{IS-R2unknown}}$ can be accommodated into subsets $\{\mathcal{S}(j)\}_{j=1}^{t_2}$ (given that $\big| \mathcal{S}_{\rm{DS}} \big| = 0$). Since $\big| \mathcal{S}_{\rm{IS-R2unknown}} \big| = N_2 t_2$, the elements of the set $\mathcal{S}_{\rm{IS-R2unknown}}$ can be distributed into subsets $\{\mathcal{S}(j)\}_j$ such that every element of it belongs to $\mathcal{S}(j)$ for exactly one $j$. After this step, the subsets $\{\mathcal{S}(j)\}_j$, each of which is of cardinality at most $N_1$, can accommodate at most $N_1 t_2 - N_2 t_2 = (N_1 - N_2) t_2$ elements of set $ \mathcal{S}_{\rm{IS-R2known}}$, which equals the cardinality of this set itself. Hence, the partition $\mathcal{P}$ with the required properties can be obtained. See also the algorithm presented in Appendix \ref{app: proof of lem: partitioning Case B.II.1 d-CSI 2user IC inner-bound}, which can be suitably modified to apply here.

\subsection{$M_2(N_1-N_2) < N_2(M_1-N_1)$:}

Here, $n' > 0$, i.e., T2 needs to transmit new data symbols during Phase Two. Let us first focus on the first property, which says that at most $M_2 t_2$ elements of set $\mathcal{S}_{\rm{DS}}$ can be accommodated in the subsets resulting $\{\mathcal{S}(j)\}_j$. But,
\begin{eqnarray*}
M_2 t_2 = M_2 ( M_1-N_1) & = & M_2 ( M_1-N_2) - (N_1-N_2) M_2 \\
          & > & N_2(M_1-N_1) - (N_1-N_2) M_2 = \big| \mathcal{S}_{\rm{DS}} \big|,
\end{eqnarray*}
where the inequality follows from the fact that under Case B.II.2, $M_2 > N_2 \frac{M_1-N_1}{M_1-N_2}$. Hence, as a first step, the elements of set $\mathcal{S}_{\rm{DS}}$ are distributed into subsets $\{\mathcal{S}(j)\}_{j=1}^{t_2}$ such that no subset contains more than $M_2$ elements and every element belongs to exactly one subset.

Let us now focus on the second property stated in the lemma, which says that at most $N_2 t_2$ elements of the set $\mathcal{S}_{\rm{IS-R2unknown}} \cup \mathcal{S}_{\rm{DS}}$ can be accommodated into subsets $\{\mathcal{S}(j)\}_{j=1}^{t_2}$. After having performed the first step, the subsets $\{\mathcal{S}(j)\}$ can contain a total of \begin{eqnarray*}
N_2 t_2 - \big| \mathcal{S}_{\rm{DS}} \big|    & = & N_2 (M_1-N_1) - \{ N_2(M_1-N_1) - (N_1-N_2) M_2 \} \\
& = & M_2 ( N_1 - N_2),
\end{eqnarray*}
elements of $\mathcal{S}_{\rm{IS-R2unknown}}$, which equals its cardinality. Having performed the first step above, the elements of the set $\mathcal{S}_{\rm{IS-R2unknown}}$ can be distributed into subsets $\{\mathcal{S}(j)\}$ such that the two properties stated in the lemma hold and every element of $\mathcal{S}_{\rm{IS-R2unknown}} \cup \mathcal{S}_{\rm{DS}}$ belongs to exactly one subset.

At the final step, the elements of $\mathcal{S}_{\rm{IS-R2known}}$ are distributed into subsets $\{\mathcal{S}(j)\}$ such that every element belongs to exactly one subset. This can be done because the resulting subsets can accommodate at most $N_1 t_2 - \{|\mathcal{S}_{\rm{DS}}| + |\mathcal{S}_{\rm{IS-R2unknown}}|\} = N_1 t_2 - N_2 t_2 = (N_1-N_2) t_2 = (M_1-N_1) (N_1-N-2) = |\mathcal{S}_{\rm{IS-R2known}}|$ elements of the set $\mathcal{S}_{\rm{IS-R2known}}$, i.e., all the elements of $\mathcal{S}_{\rm{IS-R2known}}$ can be accommodated.


\section{Proof of Lemma \ref{lem: partitioning Case B.II.2 d-CSI 2user IC inner-bound} for Case B.III.2}

\subsection{For Point $P_{1,3}$} \label{app: proof of lem: partitioning Case B.II.2 d-CSI 2user IC inner-bound part2 Case B.III.2 P13}

If $M_2' (N_1-N_2) \geq N_2 (M_1-N_1)$ (i.e., $n' = 0$), the proof given in Appendix \ref{sub-app: proof of lem: partitioning Case B.II.2 d-CSI 2user IC inner-bound n'=0} holds without any change. Consider thus the case of $n' > 0$. First, we prove the inequality (a), i.e., $M_2 t_2 \geq n'$. Suppose the inequality does not hold. Then we get
\begin{eqnarray}
\lefteqn{  M_2 (M_1 - N_1) < n' = N_2(M_1-N_1) - (N_1 + N_2 - M_1)(M_1-N_1) } \nonumber \\
&& {} \Rightarrow (N_2 - M_2) (M_1 - N_1) > (N_1+N_2 - M_1) (N_1-N_2) \nonumber \\
&& {} \Rightarrow N_2 ( N_1 - N_2) < (M_1 - N_1)(N_1-M_2) \nonumber \\
&& {} \Rightarrow N_2 ( N_1 - N_2) < (N_1 - M_2) (N_2 - N_2 \frac{N_2 - M_2}{N_1 - M_2} ) \label{eq: key step (a) sub-app: lem: partitioning of set S part2 Case B.II.4 P13} \\
&& {} \Rightarrow N_2 ( N_1 - N_2) < N_2( N_1- N_2), \nonumber
\end{eqnarray}
where the inequality (\ref{eq: key step (a) sub-app: lem: partitioning of set S part2 Case B.II.4 P13}) holds since under Case B.III.2, we have $M_1 < N_1 + N_2 - m$. The last inequality yields us a contradiction, which implies that the desired inequality (a) is true. The remaining two inequalities, namely, (b) and (c) can be easily proved and the details have been omitted.

\subsection{For Point $P_{1,4}$} \label{app: proof of lem: partitioning Case B.II.2 d-CSI 2user IC inner-bound part2 Case B.III.2 P14}

We can compute $|\mathcal{S}_{\rm{DS}}| = N_2 M_2' - M_2' (M_1'-N_1) = M_2 M_2' = M_2 t_2$, which implies nequality (a). Now, $|\mathcal{S}_{\rm{IS-R2unknown}}| = M_2' t_1 = M_2' (M_1' - N_1) = M_2' (N_2 - M_2) = N_2 t_2 - |\mathcal{S}_{\rm{DS}}|$, which imples inequality (b). To prove the last inequality, we need to show that $|\mathcal{S}_{\rm{IS-R2known}}| = (N_2-M_2') (M_1'-N_1) \leq M_2' (N_1 - N_2)$. Suppose assume the contrary. Then we have
\begin{eqnarray*}
\lefteqn{ \frac{N_1 - N_2}{M_1 - N_1} < \frac{N_2-M_2}{N_1 + N_2 - M_1}  \Rightarrow \frac{M_1 - N_2}{M_1 - N_1} < \frac{N_1 + 2N_2 - M_1 - M_2}{N_1 + N_2 - M_1} }
\\
&& {} \Rightarrow \frac{N_1 + N_2 - M_1}{M_1 - N_1} < \frac{N_1 + 2N_2 - M_1 - M_2}{M_1 - N_2} \Rightarrow \frac{N_1 + N_2 - M_2}{M_1 - N_2} > \frac{N_2}{M_1 - N_1} \\
&& {} \Rightarrow \frac{M_1- N_1}{M_1 - N_2} > \frac{N_2}{N_1 + N_2 - M_2} \Rightarrow M_1 > N_1 + N_2 \frac{N_1- N_2}{N_1-M_2} \\
&& {} \Rightarrow M_1 > N_1 + N_2 \frac{N_1 - M_2 - [N_2-M_2]}{N_1-M_2} \Rightarrow M_1 > N_1 + N_2 - m,
\end{eqnarray*}
which is a contradiction since under Case B.III.2, we have $M_1 < N_1 + N_2 - m$. Hence, inequality (c) holds.

\bibliographystyle{IEEEtran}
\bibliography{references_DPC_DoF}
\end{document}